\documentclass[epj]{svjour}
\usepackage{latexsym}
\usepackage{graphics}
\usepackage{amsmath}
\usepackage[utf8]{inputenc}
\usepackage{morefloats}
\usepackage{hyperref}
\hypersetup{
    colorlinks=True,       
    linkcolor=blue,          
    citecolor=blue,        
    filecolor=blue,      
    urlcolor=blue           
}
\usepackage{xcolor}
\usepackage{lipsum}
\newcommand{\red}[1]{{\bf\textcolor{red}{#1}}}
\newcommand{\rou}[1]{{\bf\textcolor{red}{#1}}}
\newcommand{\blu}[1]{{\bf\textcolor{blue}{#1}}}
\newcommand{\ora}[1]{{\bf\textcolor{orange}{#1}}}
\newcommand{\vio}[1]{{\bf\textcolor{violet}{#1}}}
\newcommand{\gre}[1]{{\bf\textcolor{green}{#1}}}
\newcommand{\gra}[1]{{\bf\textcolor{gray}{#1}}}
\newcommand{\bro}[1]{{\bf\textcolor{brown}{#1}}}
\newcommand{\cya}[1]{{\bf\textcolor{cyan}{#1}}}
\newcommand{\yel}[1]{{\bf\textcolor[RGB]{247, 214, 0}{#1}}}

\begin{document}
\title{World influence and interactions of  universities from Wikipedia networks}
\author{
C\'elestin Coquid\'e\inst{1}\thanks{\emph{email address:} celestin.coquide@utinam.cnrs.fr}
\and
Jos\'e Lages\inst{1}\thanks{\emph{email address:} jose.lages@utinam.cnrs.fr}
\and
Dima L. Shepelyansky\inst{2}\thanks{\emph{email address:} dima@irsamc.ups-tlse.fr}
}
%
%
\institute{
Institut UTINAM, Observatoire des Sciences de l'Univers THETA, CNRS, 
Universit\'e de Bourgogne Franche-Comt\'e, Besançon, France
\and
Laboratoire de Physique Th\'eorique, IRSAMC, Universit\'e de 
Toulouse, CNRS, UPS, 31062 Toulouse, France
}
\date{Received:  / Revised version: date}
%
\abstract{
We present Wikipedia Ranking of World Universities (WRWU)
based on analysis of networks of 24 Wikipedia editions
collected in May 2017. With PageRank and CheiRank
algorithms we determine ranking of universities 
averaged over cultural views of these editions.
The comparison with the Shanghai ranking 
gives overlap of 60\% for top 100 universities
showing  that WRWU gives more significance to their
historical development. We show that 
the new reduced Google matrix algorithm
allows to determine interactions between 
leading universities on a scale of ten centuries.
This approach also determines the influence of 
specific universities on world countries.
We also compare different cultural views of Wikipedia
editions on significance and influence of universities.
\PACS{
      {89.75.Fb}{Structures and organization in complex systems}   \and
      {89.75.Hc}{Networks and genealogical trees} \and
      {89.20.Hh}{World Wide Web, Internet}
     } 
} 
\maketitle

\section{Introduction}
\label{Introduction}
The importance of universities for progress of humanity
is broadly recognized world wide.
Thus in 2017 UNESCO emphasizes the role
of universities and higher education institutes 
in fostering sustainable development and empowering
learners \cite{unesco2017}. The efficiency
of university education gained a high political importance
in many world countries.  Various tools have been developed to
measure quantitatively this efficiency among which the ranking
of universities gained significant importance
as reviewed in \cite{hazelkorn}. Thus the Academic Ranking of World
Universities (ARWU), 
compiled by Shanghai Jiao Tong University since 2003 
(Shanghai ranking) \cite{shanghai}, generated a significant political impact on
evaluation of higher education efficiency in many countries \cite{hazelkorn}.
 For example, the ARWU stimulated 
the emergence of LABEX, IDEX projects in France \cite{freducation}
and the Russian Academic Excellence Project  \cite{ru5top100}
with allocation of significant financial supports. In addition
to ARWU other international ranking systems
of universities appeared (see e.g. \cite{times,umultirank,ireg}).
Various strong and weak features of ranking methodology are reviewed 
in \cite{jons2013,eua2013,docampo2014}. Of course,
the ranking systems are based on different specific criteria
with different cultural preferences of rather larger groups 
realizing these rankings. Already, the presence of many
ranking systems indicates the presence of bias in each 
of above ranking systems.

Another purely mathematical
and statistical approach to ranking of world universities
has been developed in \cite{wikizzs,wikievol,wrwu2013}
on the basis of Google matrix analysis of Wikipedia networks.
For each Wikipedia language edition a network is
composed by all Wikipedia articles with directed links
between them generated by mutual quotations 
of a given article to other articles.
In \cite{wikizzs} the analysis was performed only for
English Wikipedia (ENWIKI) of year 2009, other years for ENWIKI
were considered in \cite{wikievol},
while in \cite{wrwu2013}
this analysis was done with 24 language Wikipedia editions of 2013
that allowed to reduce significantly cultural bias
(these 24 networks had been collected and analyzed for historical figures
in \cite{eomwiki24} in the frame of EC FET Open project NADINE \cite{nadine}). 
The Google matrix analysis \cite{wikizzs,wikievol,wrwu2013}
is based on the PageRank algorithm \cite{brin} which detailed description is
given in \cite{meyer}. Some additional characteristics have been also used for
description of network nodes (Wikipedia articles), like CheiRank and 2DRank, as
described in \cite{rmp2015,gmscholar}. 
Thus the Wikipedia Ranking of World Universities (WRWU2013) 
from 24 Wikipedia networks was introduced in \cite{wrwu2013}
and it was shown that its top 100 universities has 62\% overlap 
with ARWU. In addition WRWU2013 attracted a significant interest
world wide (see \cite{pagel} and various press highlights listed at
\cite{wrwu2013page}). Other research groups
also start to apply Wikipedia ranking in Wikiometrics \cite{rokach}.
We also note the growing interest to
scientific analysis of several language editions of Wikipedia \cite{perc}.

In this work we extend the WRWU studies started in \cite{wrwu2013}.
The new elements are: we use 24 Wikipedia editions collected in May 2017
\cite{24wiki2017} and also we apply the recently invented reduced Google matrix (REGOMAX)
method \cite{greduced}. This new method allows to
determine effective interactions between a selected relatively small subset of
network nodes taking into account all pathways between them via the global huge 
network with millions of nodes. The efficiency of the REGOMAX method
has been demonstrated on examples of analysis of interactions
of political leaders \cite{politwiki}, terror networks \cite{spgroups}
and protein-protein interactions in cancer networks \cite{proteinplos}.
Here, using the REGOMAX method we obtain effective 
interactions between a group of selected universities and
determine their influence on world countries.
The new ranking of universities from Wikipedia 2017 editions 
is compared with those of 2013.

The paper is composed as follows: Section 2 gives description of 
network datasets; Section 3 describes Google matrix
construction and PageRank, CheiRank algorithms with 
overview of the reduced Google matrix approach;
Section 4 presents results on global ranking of universities from
24 editions and their distribution over world countries;
Section 5 provides REGOMAX results for English edition
determining the world influence of specific universities;
the interactions between top 20 universities are analyzed
in Section 6 comparing  views of  English, French, German and Russian editions;
in Section 7 we obtained the reduced Google matrix of top 100
universities averaged over 24 editions and analyze the 
interactions between universities on the scale of 10 centuries
and all continents; discussion of the results is given in Section 8.
All detailed ranking results of WRWU2017 are
available at \cite{wrwu2017page} and complementary figures and tables are in Supplementary Information (see from p.~\pageref{SI}).

\section{Datasets}

We use the datasets of 24 Wikipedia editions extracted in May 2017
\cite{24wiki2017} (see also \cite{wrwu2017page}). The size of each network
is given in Table~\ref{tab:24editions}. Compared to
2013 discussed in \cite{eomwiki24} there is a significant 
size increase for each edition, especially for Swedish (SV)
where a part of articles is now computer generated.
The number of links is given at \cite{wrwu2017page}.
On average there are about 20 links per node.
Self-citation links are not considered 
(references on the article inside the same article 
are eliminated).

\begin{table}[b]
\caption{
Wikipedia directed networks of 2017 from  24 considered language editions;
here $N$ is the number of articles. Wikipedia data were collected in
May 2017.}

\resizebox{\columnwidth}{!}{
\begin{tabular}{llr|llr}
\hline
Edition&Language&$N$&Edition&Language&$N$\\
\hline\hline
EN &English&5416537&ZH&Chinese&939625 \\
SV &Swedish&3786455&FA&Persian&539926 \\
DE &German&2057898&AR&Arabic&519714 \\
NL &Dutch&1900222&HU&Hungarian&409297 \\
FR &French&1866546&KO&Korean&380086 \\
RU &Russian&1391225&TR&Turkish&291873 \\
IT &Italian&1353276&MS&Malaysian&289234 \\
ES &Spanish&1287834&DA&Danish&225523 \\
PL &Polish&1219733&HE&Hebrew&205411 \\
VI &Vietnamese&1155932&EL&Greek&130429 \\
JP &Japanese&1058950&HI&Hindi&121503 \\
PT &Portuguese&967162&TH&Thai&116495 \\
\hline
\end{tabular}
}
\label{tab:24editions}
\end{table}

\section{Description of algorithms and methods}
\subsection{Google matrix, PageRank and CheiRank algorithms}
The mathematical grounds of this study are based on Markov chain theory and, in particular, 
on the Google matrix analysis initially introduced in 1998 by Google's co-founders, 
Brin and Page \cite{brin}, for hypertext analysis of the World Wide Web. 
Let us consider the network of the $N$ articles of a given Wikipedia edition. 
The network adjacency matrix element $A_{ij}$ is equal to $1$ if article $j$ 
quotes article $i$ and equal to $0$ otherwise. The Google matrix element 
$G_{ij}=\alpha S_{ij}+(1-\alpha)/N$ gives a transition probability 
that a random reader jumps from article $j$ to article $i$. 
The stochastic matrix element $S_{ij}$ is $S_{ij}=A_{ij}/\sum_{i=1}^NA_{ij}$
 if article $j$ quotes at least one other article, otherwise $S_{ij}=1/N$. 
The second term in $G$  proportional to $(1-\alpha)$, where $0.5<\alpha<1$ is the damping factor, 
allows to a random reader 
to escape from isolated sets of articles. More details can be found in \cite{meyer}.
Here we use the value $\alpha=0.85$ 
typical for WWW studies \cite{meyer}. The Google matrix $G$, constructed as described above, 
belongs to the class of Perron-Frobenius operators \cite{meyer}. 
The eigenvector $\mathbf{P}$  with the largest  eigenvalue $\lambda=1$ is the solution of equation 
$G\mathbf{P}=\mathbf{P}$. This PageRank vector $\mathbf{P}$ has positive or zero components and 
describes the steady-state probability distribution  
of the Markov process encoded in the Google matrix $G$. 
Assuming an infinite random process, the vector component $P_{i}$ is proportional 
to the number of times a random reader reaches an article $i$. 
It is convenient to sort the vector components $P_1,\dots,P_N$ in descending order: 
the article associated to the highest (lowest) vector component has 
the top (last) rank index $K=1$ ($K=N$).
The PageRank algorithm measures the relative influence of articles. 
Recursively, more an article is quoted by influent articles, more high is its probability.

As proposed in \cite{cheirank} we also
consider the same network of articles but with inverted links, i.e., 
article $j$ points toward article $i$ if article $j$ is quoted by article $i$. 
This inverted network is defined by the adjacency matrix elements, $A_{ij}^*=A_{ji}$,
which can be used to build successively the corresponding stochastic matrix elements, $S_{ij}^*$, 
and the corresponding Google matrix elements, $G^*_{ij}$. The CheiRank vector $\mathbf{P}^*$ 
is then defined such as $G^*\mathbf{P}^*=\mathbf{P}^*$ and 
the CheiRank is constructed similarly to the PageRank \cite{cheirank,wikizzs,rmp2015}. 
The CheiRank algorithm measures 
the relative communicative ability of the articles. Recursively, 
the more an articles quotes very communicative articles, the more it is communicative.

The properties of the Google matrix spectrum and eigenstates and their various
applications are discussed in detail in \cite{meyer,rmp2015,gmscholar}.

\subsection{The reduced Google matrix}

The concept of reduced Google matrix (REGOMAX) was introduced in 
\cite{greduced} and tested with Wikipedia networks in \cite{politwiki,spgroups}
and protein-protein networks \cite{proteinplos}.
The method is based on the construction
of a Google matrix for a relatively small subset of nodes embedded into a much larger
network taking into account all indirect interactions
between subset nodes via the remaining huge part of the network.

Let us consider a small subset $\mathcal{S}_\mathrm{r}$ of $n_\mathrm{r}\ll N$ articles, 
and the complementary subset $\mathcal{S}_\mathrm{s}$ of the $n_\mathrm{s}=N-n_\mathrm{r}\simeq N$ 
remaining articles.
For convenience, the Google matrix can be rewritten as
\begin{equation}
G=\left(
\begin{array}{cc}
G_\mathrm{rr}&G_\mathrm{rs}\\
G_\mathrm{sr}&G_\mathrm{ss}
\end{array}
\right)
\end{equation}
where
the submatrix $G_\mathrm{rr}$, of size  $n_\mathrm{r}\times n_\mathrm{r}$, 
encodes the transitions between articles 
of the subset $\mathcal{S}_\mathrm{r}$, the  submatrix $G_\mathrm{ss}$, 
of size $n_\mathrm{s}\times n_\mathrm{s}$,
encodes the transitions between articles of the subset $\mathcal{S}_\mathrm{s}$,
the submatrix $G_\mathrm{rs}$, of size   $n_\mathrm{r}\times n_\mathrm{s}$, 
encodes the transitions from articles 
of the subset $\mathcal{S}_\mathrm{s}$ toward articles of the subset $\mathcal{S}_\mathrm{r}$,
the  submatrix $G_\mathrm{sr}$, of size $n_\mathrm{s}\times n_\mathrm{r}$, 
encodes the transitions from articles 
of the subset $\mathcal{S}_\mathrm{r}$ toward articles of the subset $\mathcal{S}_\mathrm{s}$.
Since $G\mathbf{P}=\mathbf{P}$,
the PageRank vector can be rewritten as
\begin{equation}
\mathbf{P}=\left(
\begin{array}{c}
\mathbf{P}_\mathrm{r}\\
\mathbf{P}_\mathrm{s}
\end{array}
\right)
\end{equation}
where the vector $\mathbf{P}_\mathrm{r}$ ($\mathbf{P}_\mathrm{s}$) of size $n_\mathrm{r}$ ($n_\mathrm{s}$) 
contains the PageRank vector components associated to articles of 
the $\mathcal{S}_\mathrm{r}$ ($\mathcal{S}_\mathrm{s}$) subset. The reduced Google matrix $G_\mathrm{R}$ 
associated to articles of the subset $\mathcal{S}_\mathrm{r}$ is the $n_\mathrm{r}\times n_\mathrm{r}$ 
matrix defined implicitly by the following relation $G_\mathrm{R}\mathbf{P}_\mathrm{r}=\mathbf{P}_\mathrm{r}$. 
After some algebra, the reduced Google matrix can be written as \cite{greduced,politwiki,proteinplos}
\begin{equation}
G_\mathrm{R}=G_\mathrm{rr}+G_{\mathrm{ind}}\mbox{ where }G_{\mathrm{ind}}=
G_\mathrm{rs}\left(1_\mathrm{s}-G_\mathrm{ss}\right)^{-1}G_\mathrm{sr}.
\end{equation}
Here $1_\mathrm{s}$ is the $n_\mathrm{s}\times n_\mathrm{s}$ identity matrix. The reduced Google matrix 
$G_\mathrm{R}$ is composed by the $G_\mathrm{R}$-submatrix $G_\mathrm{rr}$ which encodes 
the direct links (direct quotations) between the $n_\mathrm{r}$ articles of the $\mathcal{S}_\mathrm{r}$ 
subset and by an additional scattering term $G_{\mathrm{ind}}$ 
which quantifies the indirect links between articles. 
If there is no direct link from article $j\in\mathcal{S}_\mathrm{r}$ to article $i\in\mathcal{S}_\mathrm{r}$, 
i.e., $A_{ij}=0$, then the corresponding $G_\mathrm{rr}$ element will be minimum 
($G_{\mathrm{rr}_{ij}}\sim 1/N\sim10^{-7}$ for the May 2017 English Wikipedia network). 
Conversely, the corresponding $G_{\mathrm{ind}}$ element can be very high highlighting the fact 
that two articles can be strongly indirectly linked through successive direct 
links between articles of the $\mathcal{S}_\mathrm{s}$ subset 
(eg, $j\in\mathcal{S}_\mathrm{r}\rightarrow k_1\in\mathcal{S}_\mathrm{s}\rightarrow k_2
\in\mathcal{S}_\mathrm{s}\rightarrow \dots\rightarrow k_n\in\mathcal{S}_\mathrm{s}\rightarrow i
\in\mathcal{S}_\mathrm{r}$). 
The PageRank vector of $G_\mathrm{R}$ has the same components of $n_r$ nodes
as in the global matrix $G$ (up to a constant normalization factor).
The reduced Google matrix $G_\mathrm{R}$, which conserves 
the global Google matrix PageRank hierarchy between the $n_\mathrm{r}$ 
articles of the $\mathcal{S}_\mathrm{r}$ subset, encodes direct links and effective indirect 
links between articles. The direct calculation of $G_\mathrm{ind}$ converges very slowly since 
the matrix $\left(1_\mathrm{s}-G_\mathrm{ss}\right)^{-1}$ is almost singular, indeed as 
$n_\mathrm{r}\ll n_\mathrm{s}$, $G_\mathrm{ss}\sim G$, the leading eigenvalue of 
$G_\mathrm{ss}$ is $\lambda_c\sim1$. Let us associate to the eigenvalue $\lambda_c$ 
the right eigenvector $\mathbf{\Psi}_\mathrm{R}$ and the left eigenvector $\mathbf{\Psi}_L$ 
such as $G_\mathrm{ss}\mathbf{\Psi}_\mathrm{R}=\lambda_c\mathbf{\Psi}_\mathrm{R}$ and 
$\mathbf{\Psi}_L^T\mathbf{\Psi}_\mathrm{R}=1$. 
To speed up calculations, we follow the same procedure as in 
\cite{greduced,politwiki,proteinplos}, splitting the term $\left(1_\mathrm{s}-G_\mathrm{ss}\right)^{-1}$ 
in a term $\mathbf{\Psi}_\mathrm{R}\mathbf{\Psi}_L^T\left(1-\lambda_c\right)^{-1}$ which is 
a projection onto the subspace associated to $\lambda_c$ and 
a term $(1_\mathrm{s}-\mathbf{\Psi}_\mathrm{R}\mathbf{\Psi}_L^T)\left(1_\mathrm{s}-G_\mathrm{ss}\right)^{-1}$ 
which is a projection onto the complementary subspace. 
This procedure enables us to rewrite the reduced Google matrix as
\begin{equation}
G_\mathrm{R}=G_\mathrm{rr}+G_{\mathrm{pr}}+G_{\mathrm{qr}}
\label{eq:GR}
\end{equation}
where $G_{\mathrm{pr}}=G_\mathrm{rs}\mathbf{\Psi}_\mathrm{R}\mathbf{\Psi}_L^TG_\mathrm{sr}\left(1-\lambda_c\right)^{-1}$ 
encodes essentially already known information concerning 
the PageRank (since $\mathbf{\Psi}_\mathrm{R}\sim \mathbf{P}$) and 
$G_{\mathrm{qr}}=G_\mathrm{rs}(1_\mathrm{s}-\mathbf{\Psi}_\mathrm{R}\mathbf{\Psi}_L^T)\left(1_\mathrm{s}-G_\mathrm{ss}\right)^{-1}G_\mathrm{sr}$ encodes hidden interactions between articles which appear 
due to indirect links via the global network \cite{greduced,politwiki,proteinplos}.
In the following we perform analysis of the three components 
present in (\ref{eq:GR}), we also consider the matrix component $G_{\mathrm{qrnd}}$
obtained from $G_{\mathrm{qr}}$ by taking out the diagonal terms
since the self-citations are not very interesting.

\section{Wikipedia Ranking of World Universities from 24 Wikipedia editions of 2017}

Once the articles of  24 considered editions are ranked using PageRank and 
CheiRank algorithms, we extract for each edition the top 100 articles devoted to institutions 
of higher education and research. We also consider 2DRank which is a combination
of PageRank and CheiRank (see \cite{wikizzs,wrwu2013}; 2DRank results 
are available at \cite{wrwu2017page}). 
As in \cite{eomwiki24,wrwu2013}, from these 24 top 100 listings we obtain 
the following cumulative score for a given university $U$
\begin{equation}
\Theta_{U,A}=\sum_{E}\left(101-R_{U,E,A}\right)
\label{eq:theta}
\end{equation}
where $A$ denotes the algorithm used for the ranking (PageRank or CheiRank or 2DRank), 
$E$ the Wikipedia edition, $R_{U,E,A}$ the rank of university $U$ in 
the top 100 universities obtained using algorithm $A$ from edition $E$ of Wikipedia. 
If an university $U'$ is absent from the top 100 universities obtained from an edition 
$E'$ with an algorithm $A'$ then we artificially set $R_{U',E',A'}=101$.
We use ISO 3166-1 alpha-2 country codes \cite{isowiki} 
(all the used country codes are available at \cite{wrwu2017page}).

\begin{figure}[h]
\centering
\resizebox{\columnwidth}{!}{%
\includegraphics{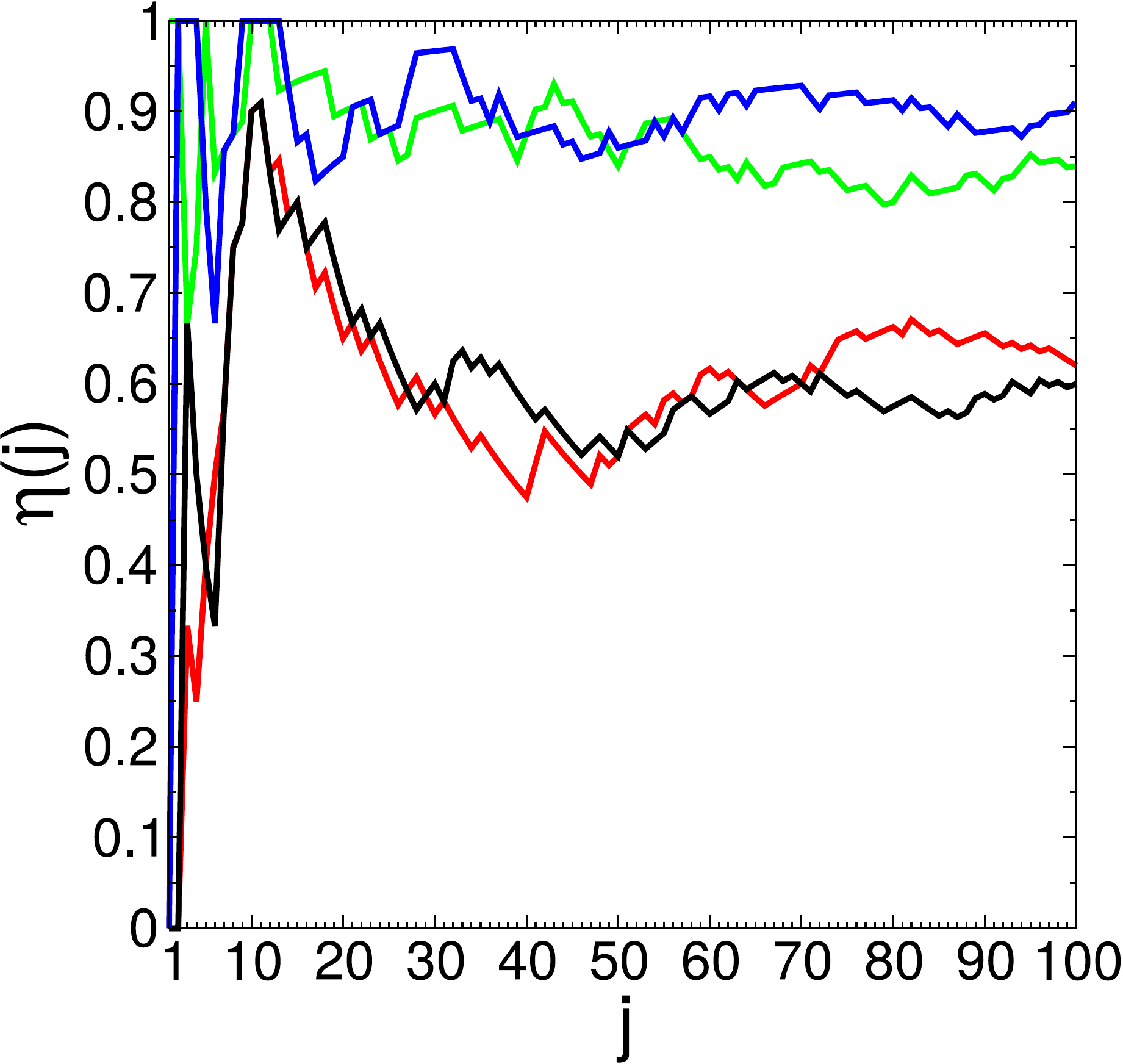}
\includegraphics{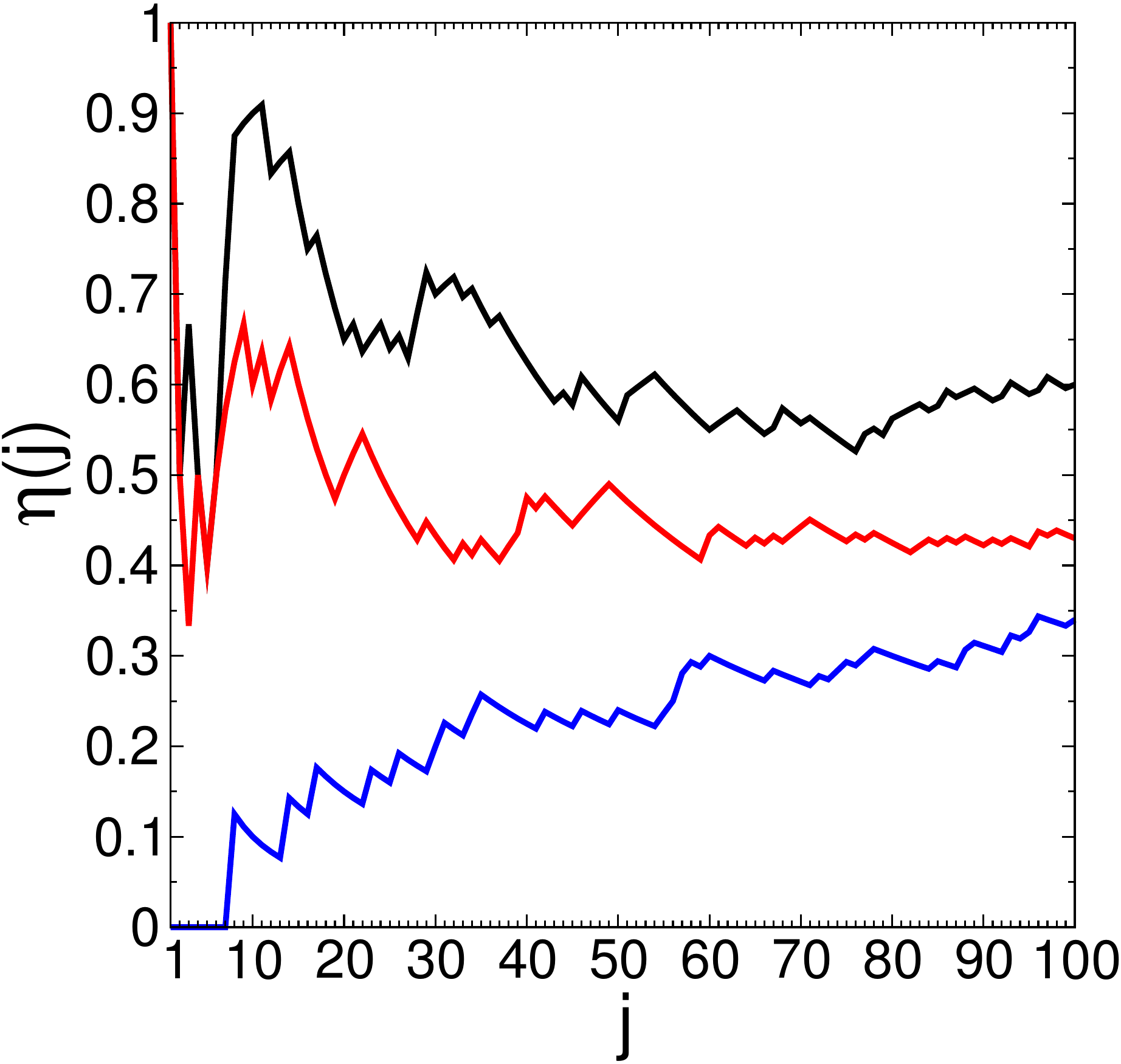}
}
\caption{Left panel: Overlap $\eta(j)=j_c/j$ of ARWU2017 with WRWU2017 
as a function of the rank $j$ of WRWU2017. 
Here $j_c$ gives the number of common universities among the first 
$j$ universities of the two rankings. 
The color curves show the overlap of ARWU2017 with WRWU2017 (black curve), 
of ARWU2013 with WRWU2013 (red curve), 
of ARWU2013 with ARWU2017 (green curve), and of WRWU2013 with WRWU2017 (blue curve). 
Right panel: Overlap of ARWU2017 with ENWRWU2017 (black curve), 
of ARWU2017 with FRWRWU2017 (red curve) 
and of ARWU2017 with DEWRWU2017 (blue curve). 
The horizontal axis label $j$ is the rank of ARWU2017.}
\label{fig:overlap}
\end{figure}

\begin{table}[b]
\caption{List of the first 10 universities of the 2017 Wikipedia Ranking of World Universities 
using PageRank algorithm. For a given university, the score $\Theta_{PR}$ is defined
by (\ref{eq:theta}), $N_a$ is the number of appearances in the top 100 lists of 24 Wikipedia editions, 
CC is the country code, LC is the language code, and FC is the foundation century.}
\centering
\resizebox{\columnwidth}{!}{
\begin{tabular}{lrrlllr}
\hline
Rank&$\Theta_{PR}$&$N_a$&University&CC&LC&FC\\
\hline
\hline
1st&2281&24&University of Oxford&UK&EN&11\\
2nd&2278&24&University of Cambridge&UK&EN&13\\
3rd&2277&24&Harvard University&US&EN&17\\
4th&2099&24&Columbia University&US&EN&18\\
5th&1959&23&Yale University&US&EN&18\\
6th&1917&24&University of Chicago&US&EN&19\\
7th&1858&23&Princeton University&US&EN&18\\
8th&1825&21&Stanford University&US&EN&19\\
9th&1804&21&Massachusetts Institute of Technology&US&EN&19\\
10th&1693&20&University of California, Berkeley&US&EN&19\\
\hline
\end{tabular}
}
\label{tab:top10WRWU}
\end{table}

\begin{table}[b]
\caption{
List of the first 10 universities of ARWU2017 \cite{shanghai}.
The last columns show the difference between ARWU2017 rank and WRWU2017 rank.} 
\resizebox{\columnwidth}{!}{
\begin{tabular}{llrrr}
\hline
Rank&ARWU17&WRWU17\\
\hline
\hline
1st & Harvard University & -2\\
2nd & Stanford University & -6\\
3rd & University of Cambridge& +1\\
4th & Massachusetts Institute of Technology & -5\\
5th & University of California, Berkeley  & -5\\
6th & Princeton University & -1\\
7th & University of Oxford & +6\\
8th & Columbia University & +4\\
9th & California Institute of Technology & -13\\
10th & University of Chicago & +4\\
\hline
\end{tabular}
}
\label{tab:top10ARWU}
\end{table}

The top 10 universities from WRWU2017 with PageRank algorithm and from ARWU2017  
are given in Tab.~\ref{tab:top10WRWU} and Tab.~\ref{tab:top10ARWU} respectively.
The top 3 places of WRWU are occupied by Oxford, Cambridge and Harvard
while for ARWU it is Harvard, Stanford and Cambridge. The universities 
with significantly lower positions in WRWU (compared to ARWU)
are MIT, Berkeley  and Caltech, while Oxford significantly improves its
position at WRWU going to the first place from 7th at ARWU.

The overlap between different rankings are presented in Fig.~\ref{fig:overlap}.
For the top 100 universities we have 60\% overlap between
WRWU PageRank list and ARWU list in 2017. For 2013 this overlap was slightly higher at 62\%.
The overlap between WRWU2017 list and  WRWU2013 list is 91\% 
and between ARWU2017 list and ARWU2013 list is 84\%.
WRWU appears to be stable, top10s in 2013 (Tab.~\ref{tab:top10WRWU}) 
and 2017 (Tab.~3 in \cite{wrwu2013}) contain the same universities but with some changes in places:
Oxbridge keep the two first places but Oxford supersedes 
Cambridge at the first place; Yale ($9\rightarrow5$) and 
Chicago ($7\rightarrow6$) improve their ranking whereas 
Princeton ($5\rightarrow7$) and MIT ($6\rightarrow9$) recede; 
Harvard ($3\rightarrow3$), Columbia ($4\rightarrow4$), Stanford ($8\rightarrow8$), 
Berkeley ($10\rightarrow10$) keep their positions.
Between 2013 \cite{wrwu2013page} and 2017 \cite{wrwu2017page}, 
only 9 universities went out from top100
(IPSA /
Karolinska Institutet /
Rockefeller /
Rutgers  /
Tsinghua  /
Amsterdam /
Hamburg /
Strasbourg /
Wroc\l{}aw)
and 9 new universities enter in the top100
(Seoul National University /
TU Munich /
UC, San Diego /
Boulder /
Freiburg /
Kiel /
Marburg /
Salamanca /
Sydney).

The above numbers and comparisons show 
that WRWU approach gives a reliable ranking which remains relatively close to ARWU at different years.
At the same time WRWU has about 40\% of different universities compared to ARWU.
The origin of this difference is based on variety of cultural views
well present in 24 editions. Thus the right panel of Fig.~\ref{fig:overlap}
shows a spectacular difference between EN, FR and DE editions:
for top 10 universities the German edition
has only about 10\% overlap with ARWU while EN and FR have about 50\%.
For top 100 this difference still remains significant 
being approximately 34\% for DE, 43\% for FR and 60\% for EN.
Thus the case of German edition demonstrates rather different cultural
view on importance of their universities. As discussed in \cite{wrwu2013}
there are clear historical grounds for this difference related to
the  world dominance of German and Italian universities before 19th century
as it is clearly seen in Fig.~10 in \cite{wrwu2013}. 

\begin{figure}[t]
\centering
\resizebox{0.9\columnwidth}{!}{%
\includegraphics{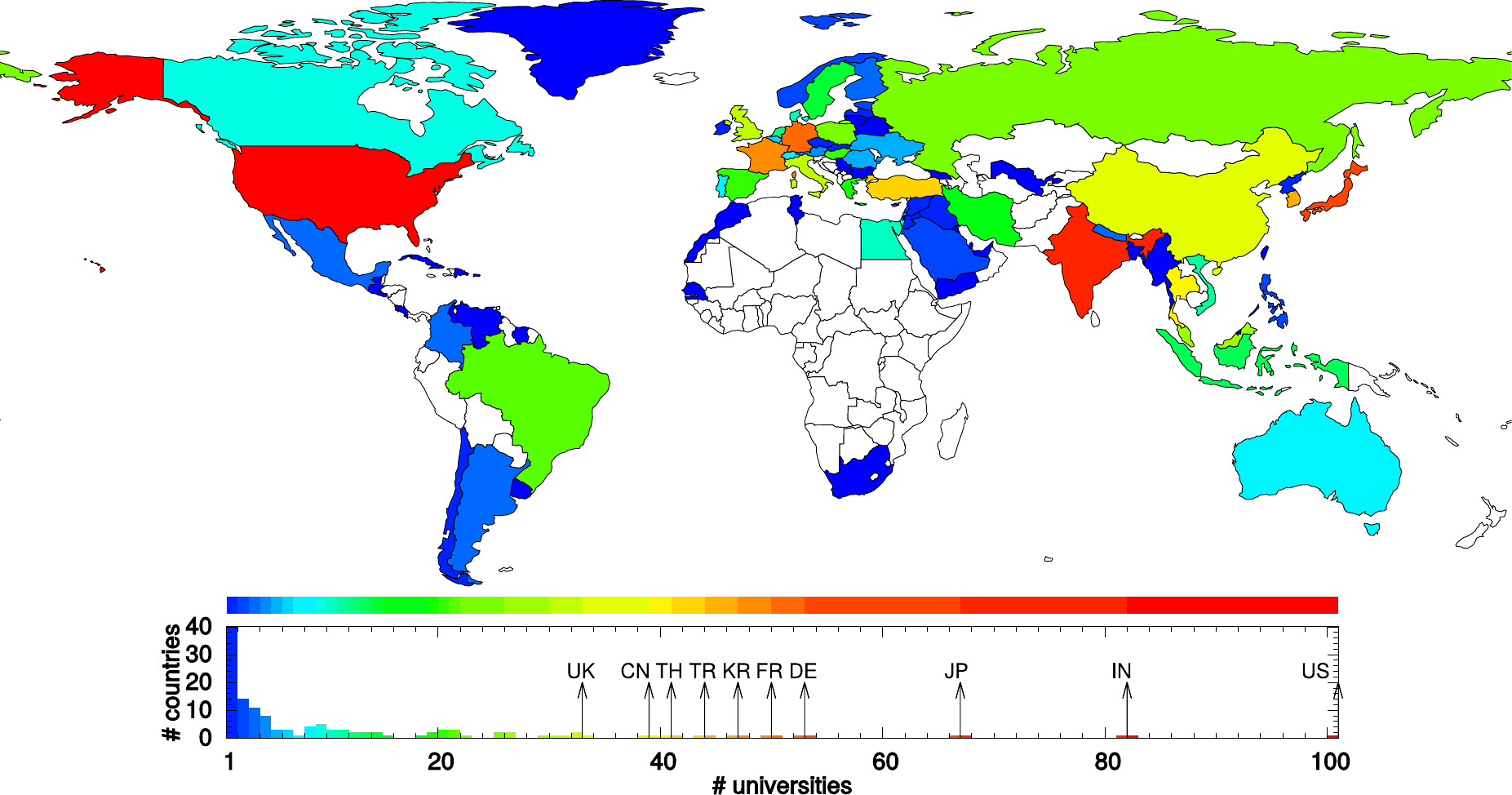}
}
\resizebox{0.9\columnwidth}{!}{%
\includegraphics{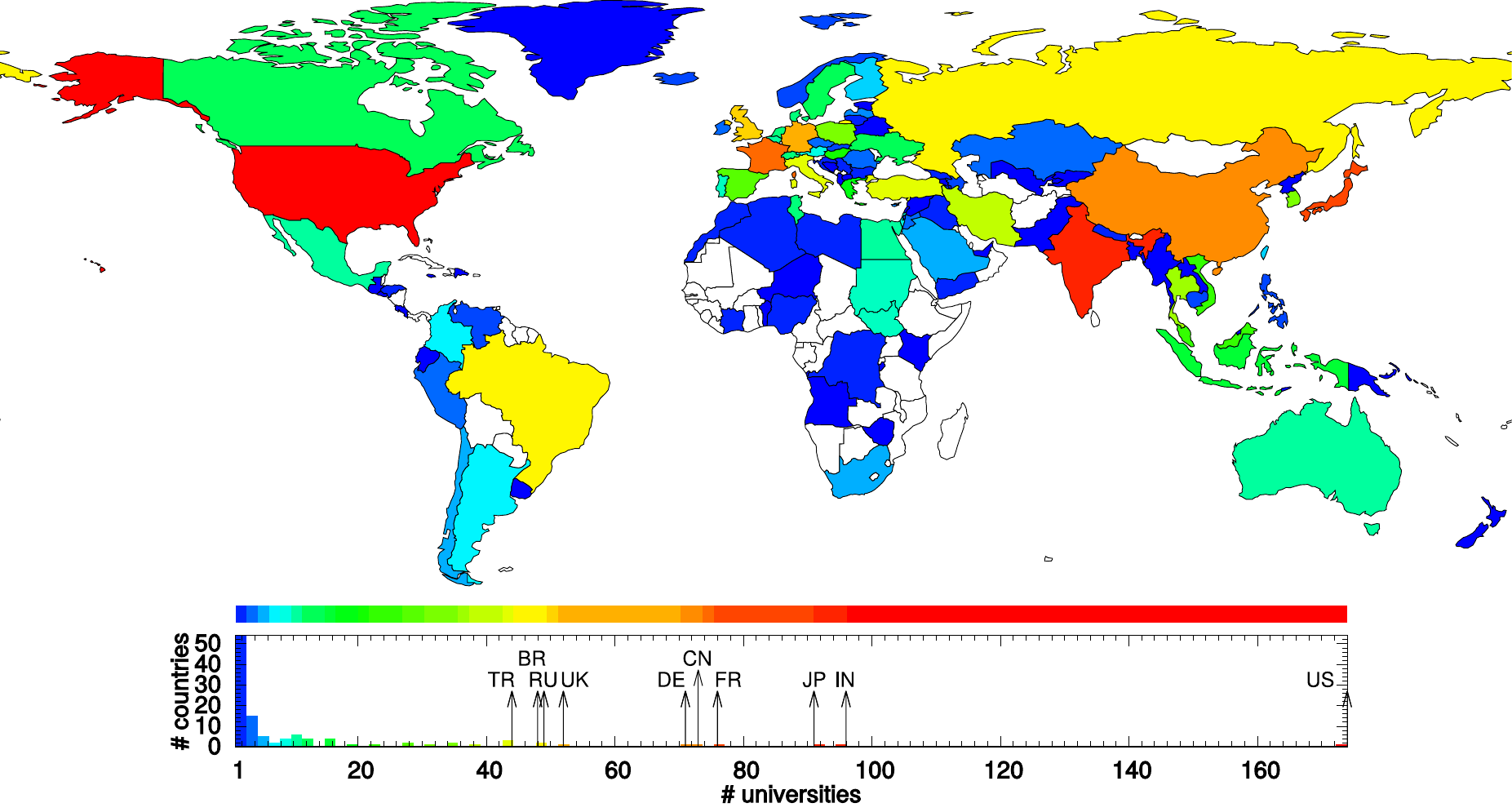}
}
\caption{Geographical distribution of the universities entering the 2017 Wikipedia 
Ranking of World Universities using PageRank  (top panel) and CheiRank algorithms (bottom panel).
The total number of universities is 1011 (1464) for WRWU using PageRank (CheiRank) algorithm. 
US universities are the most numerous: 101 (174) universities for 
WRWU using PageRank (CheiRank) algorithm.
Countries with white color have no universities in the top 100 edition lists.
Here and below the color categories are obtained using the Jenks natural 
breaks classification method \cite{jenkswiki}.}
\label{fig:geoWPRWU}
\end{figure}

In total for all 24 editions we find 1011 and 1464 different universities with PageRank
and CheiRank algorithms respectively. Their geographical distribution
over the country world map is shown in Fig.~\ref{fig:geoWPRWU}.
The largest number of top universities is in US but we see a significant numbers
also for India, Japan, Germany and France for PageRank which characterizes the 
university influence.
The communicativity is highlighted by CheiRank with top countries being US, India, 
Japan, France and China.
Of course, Hindi, Japan, Chinese editions give certain preference to their 
own universities but in global the high positions of these universities
and countries reflect significant efforts in higher education
performed by these countries. 

\begin{figure}[t]
\centering
\resizebox{0.9\columnwidth}{!}{%
\includegraphics{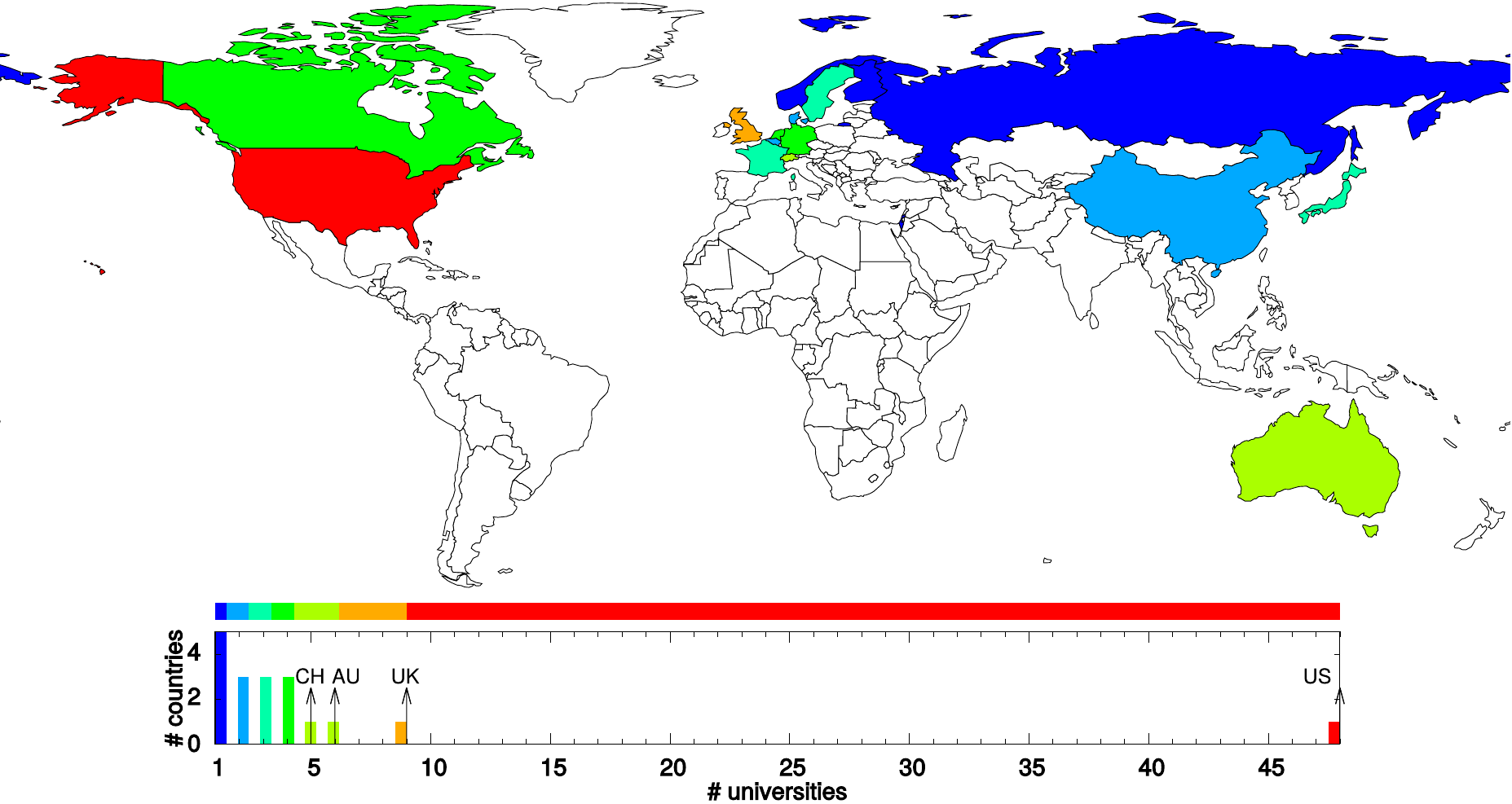}
}
\resizebox{0.9\columnwidth}{!}{%
\includegraphics{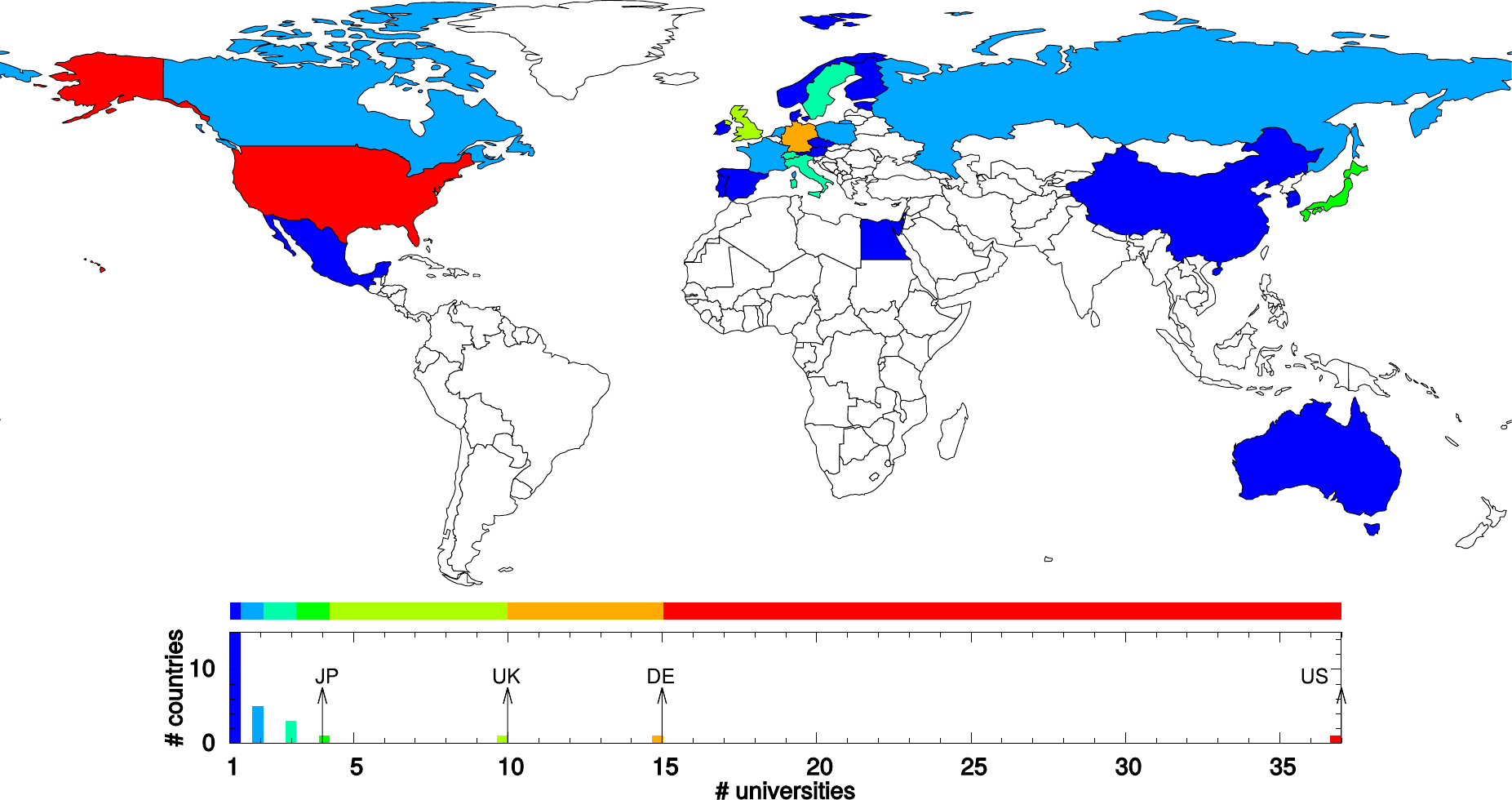}
}
\caption{Geographical distribution of the first 100 universities from ARWU2017 (top panel) and 
WRWU2017 from PageRank (bottom panel). US universities are the most numerous, 
48 universities for ARWU2017 and 37 universities for WRWU2017.
}
\label{fig:geotop100}
\end{figure}

The geographical distributions of top 100 universities
of ARWU2017 and WRWU2017 are presented in Fig.~\ref{fig:geotop100}.
For ARWU the top countries are US, UK, Australia and China
while for WRWU we find US, Germany, UK and Japan.
It is clear that ARWU gives too high significance to Anglo-saxon and Chinese universities
while WRWU provides more balanced historical view
taking into account a significant role played e.g. by German universities.

\begin{figure}[h]
\centering
\resizebox{\columnwidth}{!}{%
\includegraphics{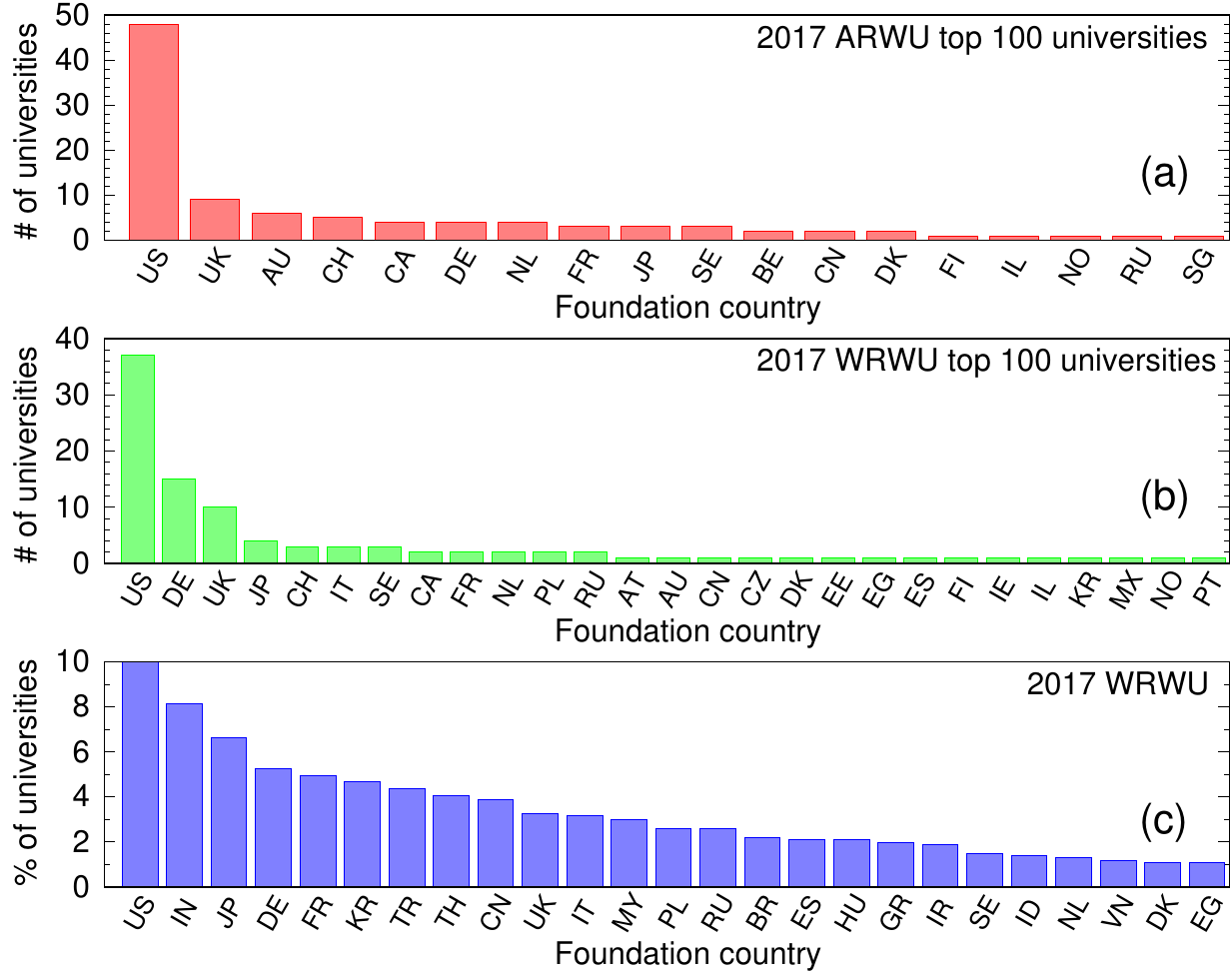}
}
\caption{Distribution over countries of (a) 2017 ARWU top 100 universities and of 
(b) 2017 WRWU top 100 universities. Panel (c) gives the percentage per country of universities 
among the 1011 universities listed in 2017 WRWU, countries with less than 10\% are not shown. 
Countries with equal number of universities are sorted by alphabetic order.}
\label{fig:histograms}
\end{figure}

A more detailed view on the universities distribution over countries for
ARWU and WRWU is shown in Fig.~\ref{fig:histograms}.
The ranking of WRWU universities 
inside each country is given in \cite{wrwu2017page}.

\section{Influence of world universities on countries from English Wikipedia edition}

In this Section we use the REGOMAX approach to analyze the influence 
of universities on world countries. With this aim 
the reduced Google matrix is constructed for the subset of articles devoted to 
the PageRank top 20 universities of ENWRWU 
(see Tab.~\ref{tab:top20ENWRWU}) and articles devoted to the 85 countries 
to which belong universities appearing in WRWU (see Tab.~\ref{tab:WRWUcountries}).
Thus the total size of the reduced Google matrix is $n_r=105$, to be compared to
the global ENWIKI network size which is about 5.4 millions articles.

\begin{table}[b]
\caption{List of the PageRank top  20 universities of English edition WRWU2017. 
The color code corresponds to the regional location of universities: red for 
US west coast, orange for US central region, blue for US east coast, and violet for UK.} 
\resizebox{\columnwidth}{!}{
\begin{tabular}{rl|rl}
\hline
Rank&University&Rank&University\\
\hline
\hline
1st&\blu{Harvard University}&11th&\ora{University of Michigan}\\
2nd&\vio{University of Oxford}&12th&\blu{Cornell University}\\
3rd&\vio{University of Cambridge}&13th&\rou{University of California, Los Angeles}\\
4th&\blu{Columbia University}&14th&\blu{University of Pennsylvania}\\
5th&\blu{Yale University}&15th&\blu{New York University}\\
6th&\rou{Stanford University}&16th&\ora{University of Texas at Austin}\\
7th&\blu{Massachusetts Institute of Technology}&17th&\blu{University of Florida}\\
8th&\rou{University of California, Berkeley}&18th&\vio{University of Edinburgh}\\
9th&\blu{Princeton University}&19th&\ora{University of Wisconsin-Madison}\\
10th&\ora{University of Chicago}&20th&\rou{University of Southern California}\\
\hline
\end{tabular}
}
\label{tab:top20ENWRWU}
\end{table}

\begin{table}[b]
\caption{List of the countries associated to the universities appearing in WRWU2017. 
Countries are ranked from 2017 Wikipedia English edition using PageRank algorithm.} 
\resizebox{\columnwidth}{!}{
\begin{tabular}{rlc|rlc}
\hline
Rank&University&CC&Rank&University&CC\\
\hline
\hline
1&United States&US&44&Chile&CL\\
2&France&FR&45&Republic of Ireland&IE\\
3&Germany&DE&46&Singapore&SG\\
4&United Kingdom&UK&47&Serbia&RS\\
5&Iran&IR&48&Vietnam&VN\\
6&India&IN&49&Nepal&NP\\
7&Canada&CA&50&Estonia&EE\\
8&Australia&AU&51&Iraq&IQ\\
9&China&CN&52&Bangladesh&BD\\
10&Italy&IT&53&Syria&SY\\
11&Japan&JP&54&Myanmar&MM\\
12&Russia&RU&55&Slovakia&SK\\
13&Brazil&BR&56&Venezuela&VE\\
14&Spain&ES&57&Morocco&MA\\
15&Netherlands&NL&58&Cuba&CU\\
16&Poland&PL&59&Puerto Rico&PR\\
17&Sweden&SE&60&Saudi Arabia&SA\\
18&Mexico&MX&61&Lithuania&LT\\
19&Turkey&TR&62&Lebanon&LB\\
20&Romania&RO&63&Cyprus&CY\\
21&South Africa&ZA&64&Latvia&LV\\
22&Norway&NO&65&Belarus&BY\\
23&Switzerland&CH&66&United Arab Emirates&AE\\
24&Philippines&PH&67&Uruguay&UY\\
25&Austria&AT&68&North Korea&KP\\
26&Belgium&BE&69&Yemen&YE\\
27&Argentina&AR&70&Costa Rica&CR\\
28&Indonesia&ID&71&Tunisia&TN\\
29&Greece&GR&72&Jordan&JO\\
30&Denmark&DK&73&Guatemala&GT\\
31&South Korea&KR&74&Greenland&GL\\
32&Israel&IL&75&Dominican Republic&DO\\
33&Hungary&HU&76&Uzbekistan&UZ\\
34&Finland&FI&77&Kuwait&KW\\
35&Egypt&EG&78&Qatar&QA\\
36&Portugal&PT&79&Senegal&SN\\
37&Taiwan&TW&80&El Salvador&SV\\
38&Ukraine&UA&81&Suriname&SR\\
39&Czech Republic&CZ&82&Faroe Islands&FO\\
40&Malaysia&MY&83&Brunei&BN\\
41&Thailand&TH&84&Palestine&PS\\
42&Colombia&CO&85&Georgia&GE\\
43&Bulgaria&BG&&&\\
\hline
\end{tabular}
}
\label{tab:WRWUcountries}
\end{table}

\begin{figure}[h]
\centering
\resizebox{\columnwidth}{!}{%
\includegraphics{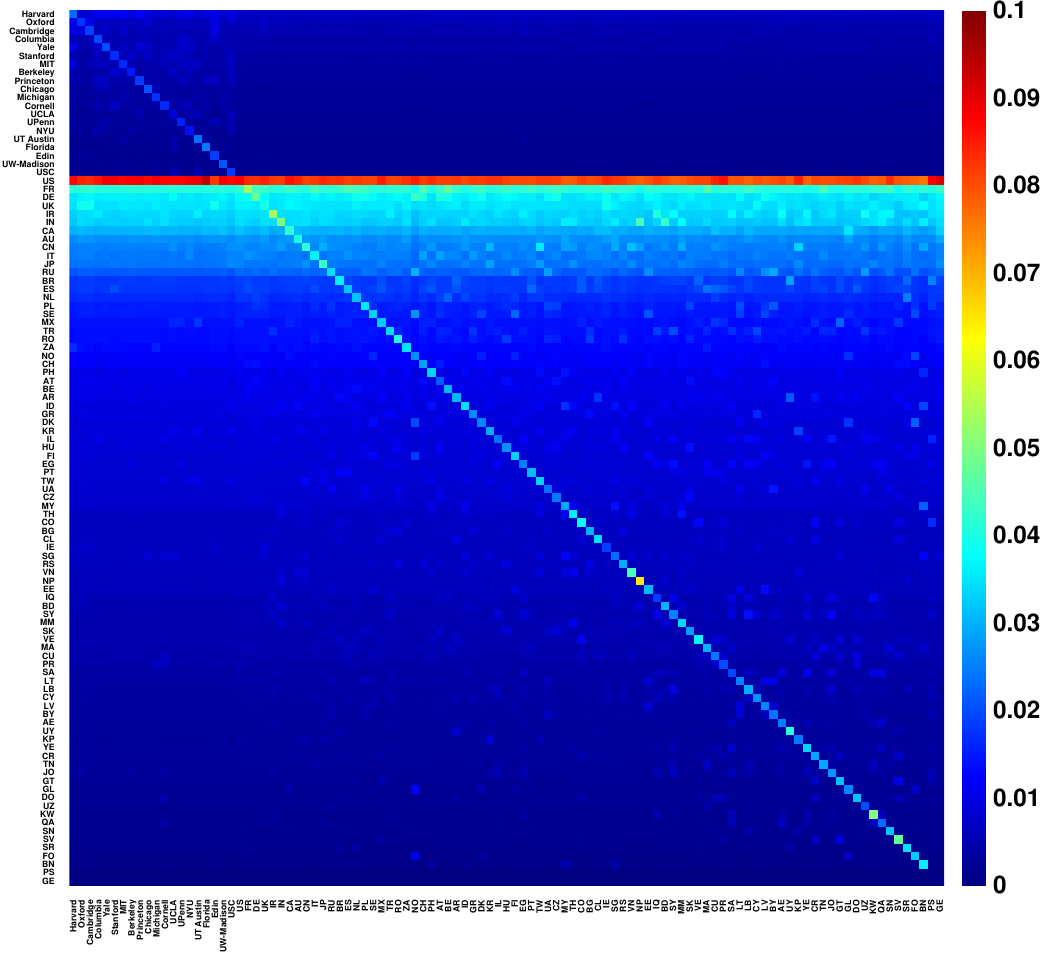}
\includegraphics{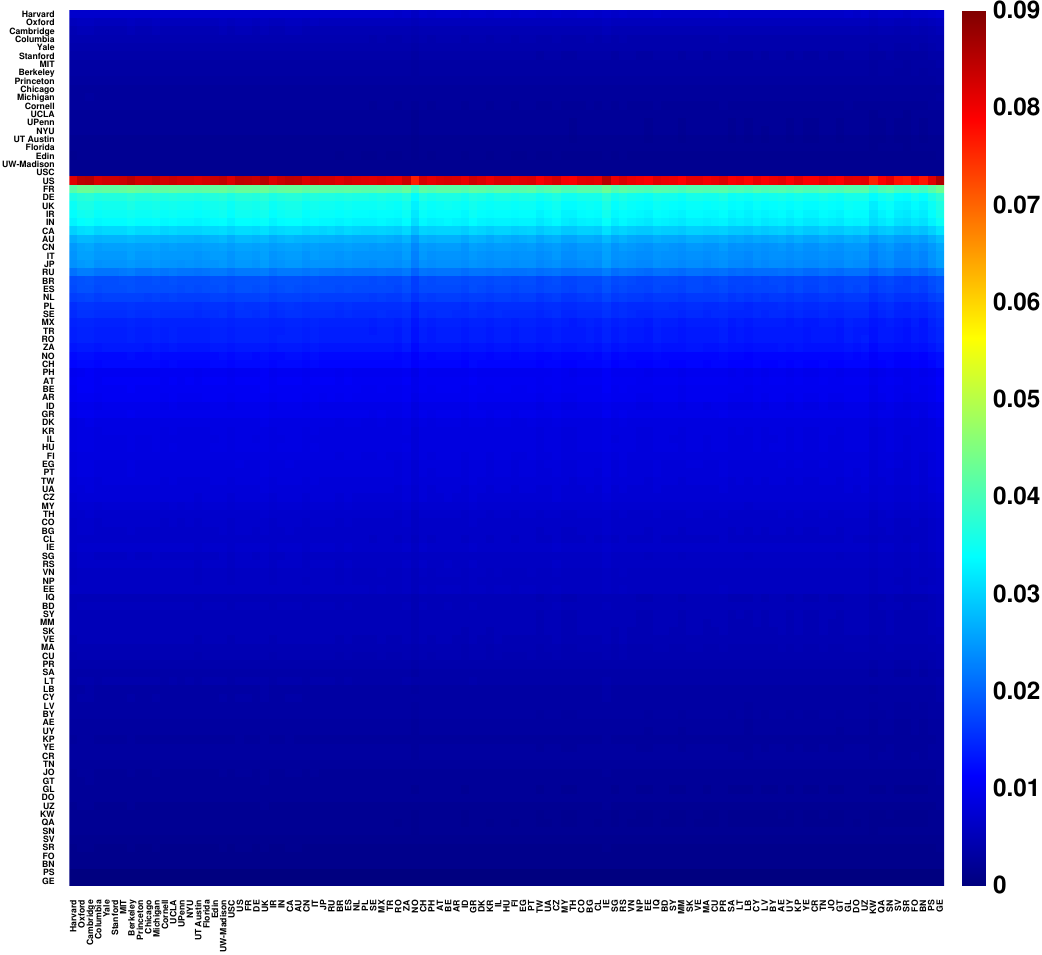}
}
\resizebox{\columnwidth}{!}{%
\includegraphics{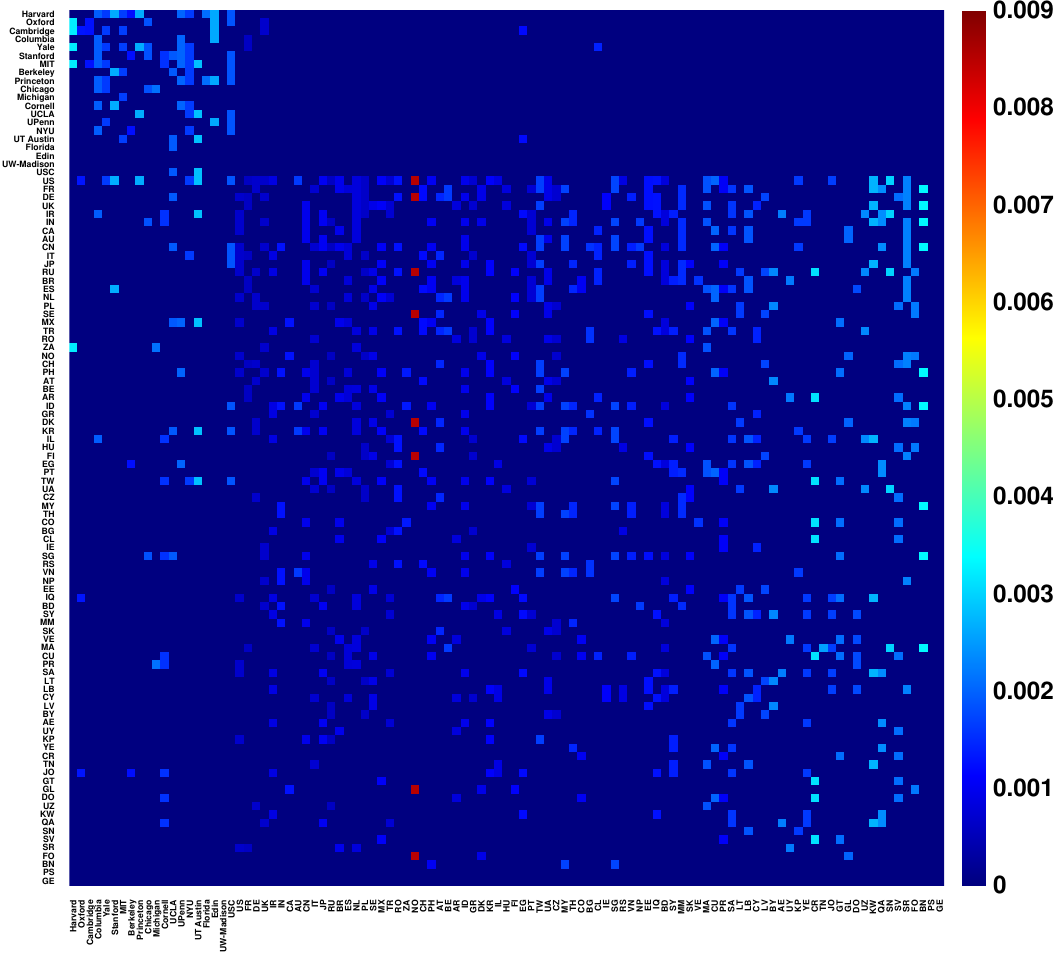}
\includegraphics{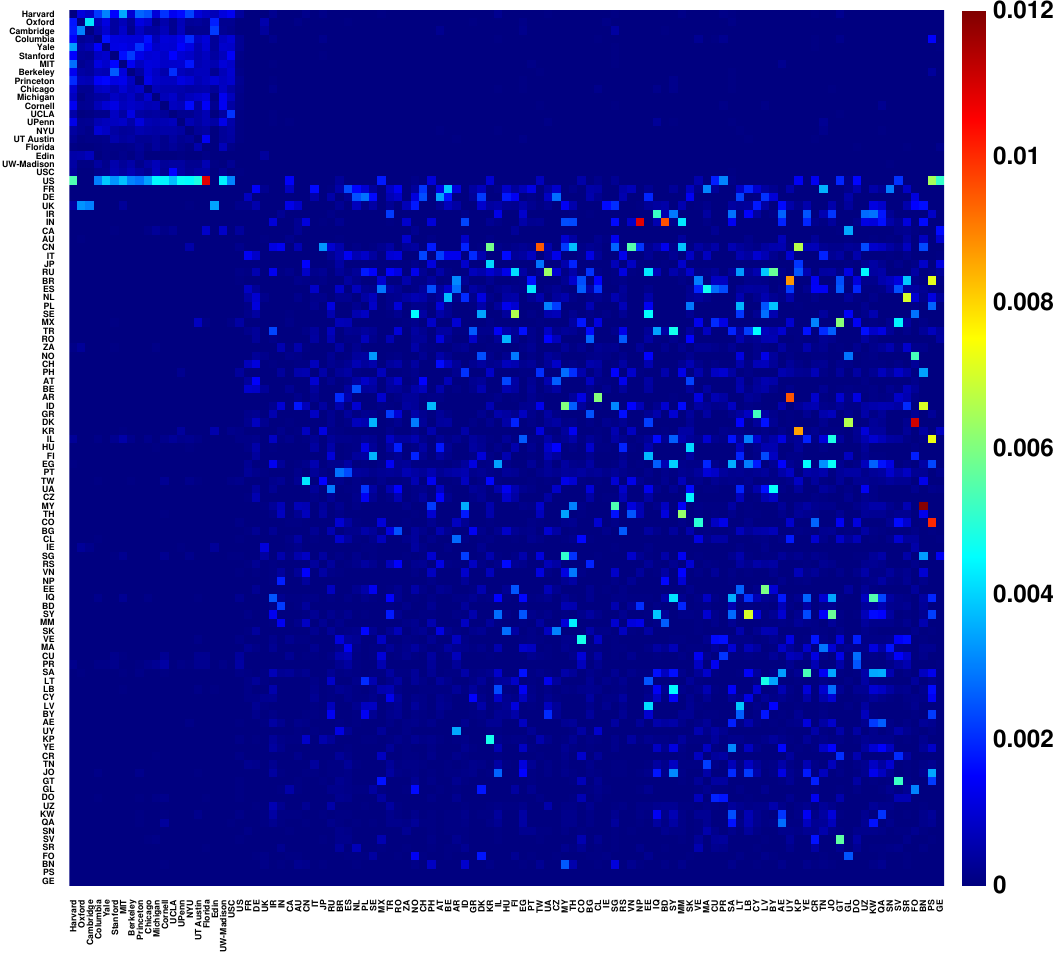}
}
\caption{Reduced Google matrix $G_\mathrm{R}$ for the first 20 universities ranked in 
ENWRWU (Tab.~\ref{tab:top20ENWRWU}) and the 85 countries to which universities 
from WRWU belong (Tab.~\ref{tab:WRWUcountries}). The full reduced Google matrix $G_\mathrm{R}$ 
is presented in top left panel, $G_{\mathrm{pr}}$ in top right panel, 
$G_{\mathrm{rr}}$ in bottom left panel, and $G_{\mathrm{qrnd}}$ in bottom right panel. 
The weights of $G_\mathrm{R}$ 
matrix components are $W_\mathrm{R}=1$, 
$W_{\mathrm{pr}}=0.948273$, $W_{\mathrm{rr}}=0.0144137$, 
and $W_{\mathrm{qr}}=0.0373132$. The node order presents first 20 universities 
in order of Tab.~\ref{tab:top20ENWRWU} and then 85 countries in order of
Tab.~\ref{tab:WRWUcountries}.
}
\label{fig:ENWIKIGRU20C85}
\end{figure}

The images of the corresponding reduced Google matrix $G_\mathrm{R}$ and its
components $G_{\mathrm{rr}}$, $G_{\mathrm{pr}}$, and $G_{\mathrm{qr}}$, 
are shown in Fig.~\ref{fig:ENWIKIGRU20C85}.
As discussed above and in \cite{politwiki}
the $G_{\mathrm{pr}}$ component is rather close to
the matrix composed by identical columns of
PageRank vector of $n_r$ nodes,
the direct links are presented by the component
 $G_{\mathrm{rr}}$ and indirect links by   $G_{\mathrm{qr}}$
(and related $G_{\mathrm{qrnd}}$). The weights of these three
components (sum of elements of all columns divided by 
matrix size $n_r$) are respectively 
$W_\mathrm{R}=1$, 
$W_{\mathrm{pr}}=0.948273$, $W_{\mathrm{rr}}=0.0144137$, 
and $W_{\mathrm{qr}}=0.0373132$. The weights of
the components  $G_{\mathrm{rr}}$ and  $G_{\mathrm{qr}}$
are small, compared to those of $G_{\mathrm{pr}}$,
but these two components provide new important information
on interactions between nodes. The weight of indirect links
is larger than the direct ones $W_{\mathrm{qr}} > W_{\mathrm{rr}}$.

The knowledge of all matrix elements of  $G_\mathrm{R}$ 
allows us to determine the influence or sensitivity
of a given university $u$ on a given country $c$.
To measure the sensitivity we change the matrix element
$G_\mathrm{R}(u \rightarrow c)$ by a factor $(1+\delta)$ with $\delta\ll1$,
we renormalize to $1$ the sum of the column associated to university $u$, 
and we compute the logarithmic derivative
of PageRank probability $P(c)$ associated to country $c$:
$D(u \rightarrow c, c) = d \ln P(c)/d\delta$ 
(diagonal sensitivity). 
It is also possible to consider the nondiagonal
sensitivity $D(u\rightarrow c, c') =  d \ln P(c')/d\delta $
when the variation is done for the link from
$u$ to $c$ and the derivative
of PageRank probability is computed
for another country $c'$.
This approach was already used in
\cite{spgroups,painters} showing its efficiency.

\begin{figure}[h]
\centering
\resizebox{0.9\columnwidth}{!}{%
\includegraphics{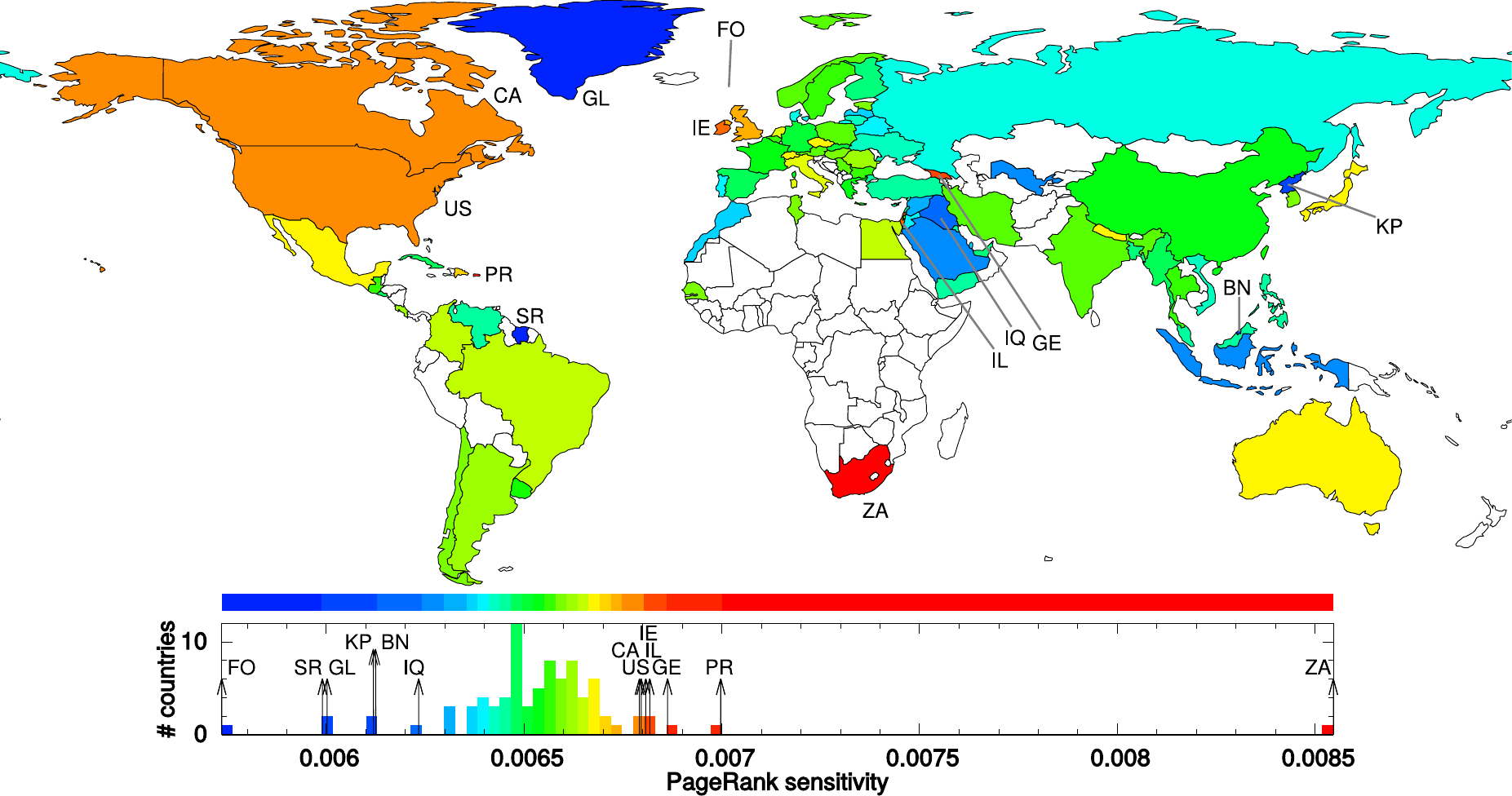}
}
\resizebox{0.9\columnwidth}{!}{%
\includegraphics{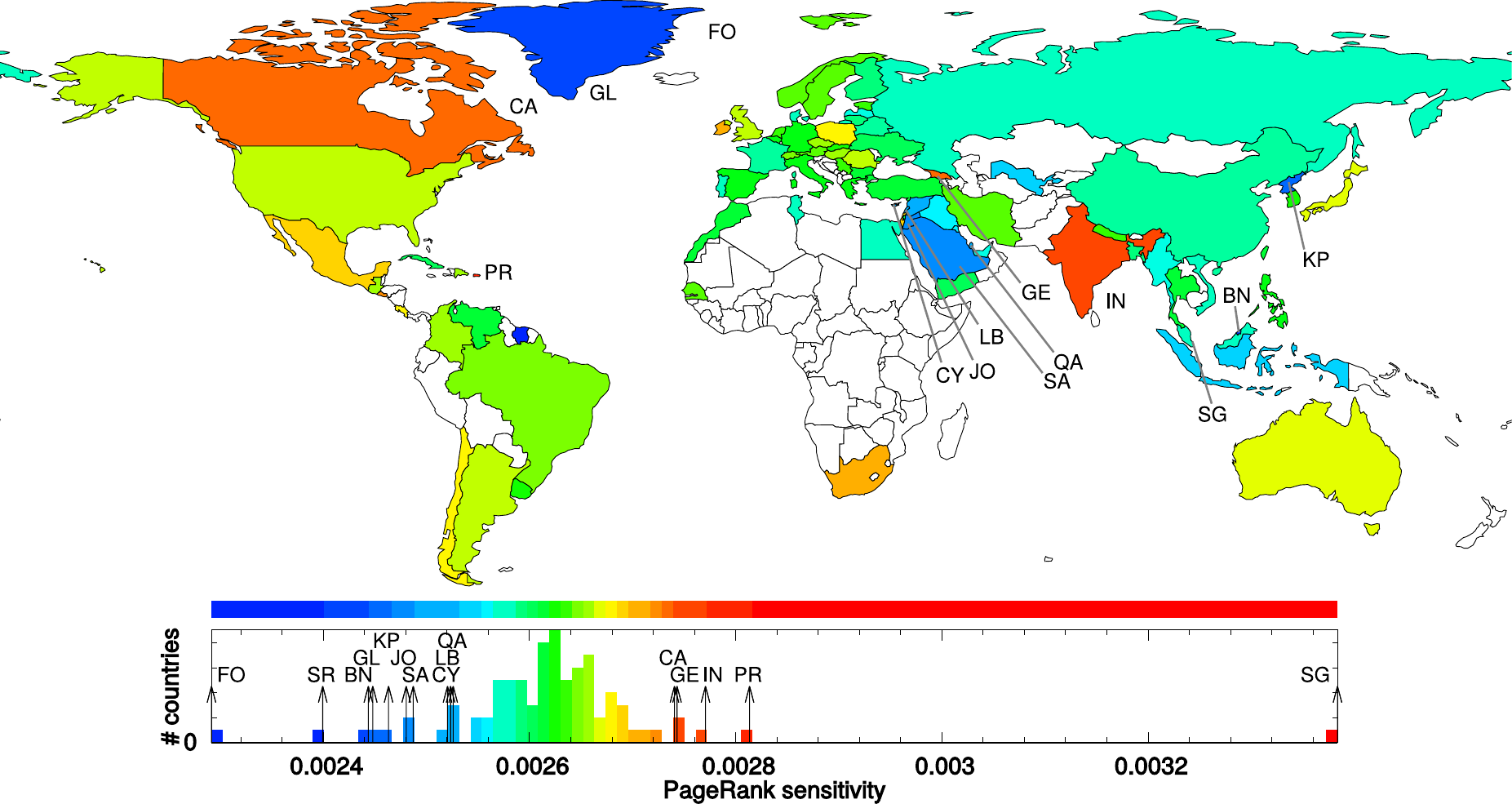}
}
\resizebox{0.9\columnwidth}{!}{%
\includegraphics{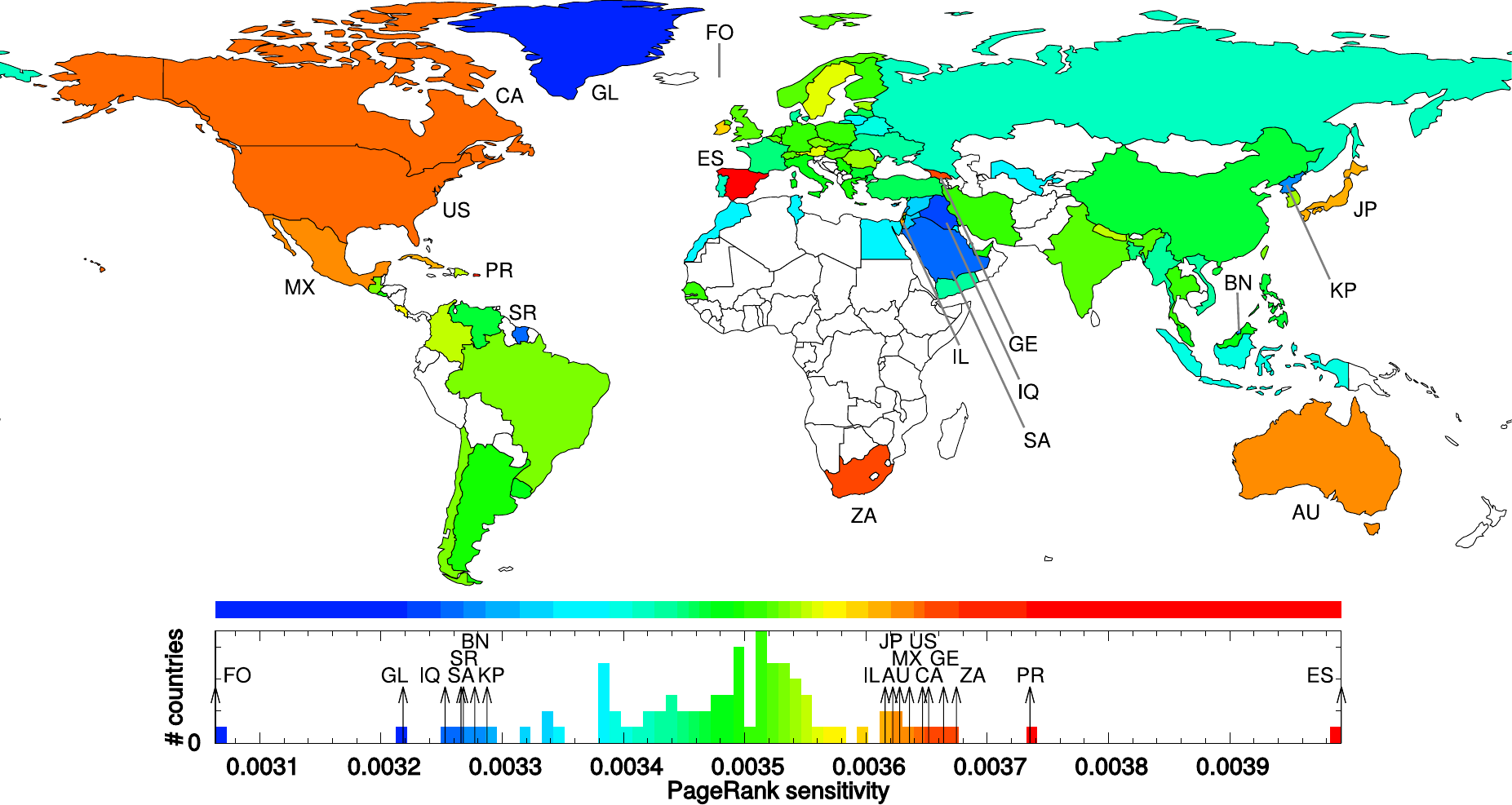}
}
\resizebox{0.9\columnwidth}{!}{%
\includegraphics{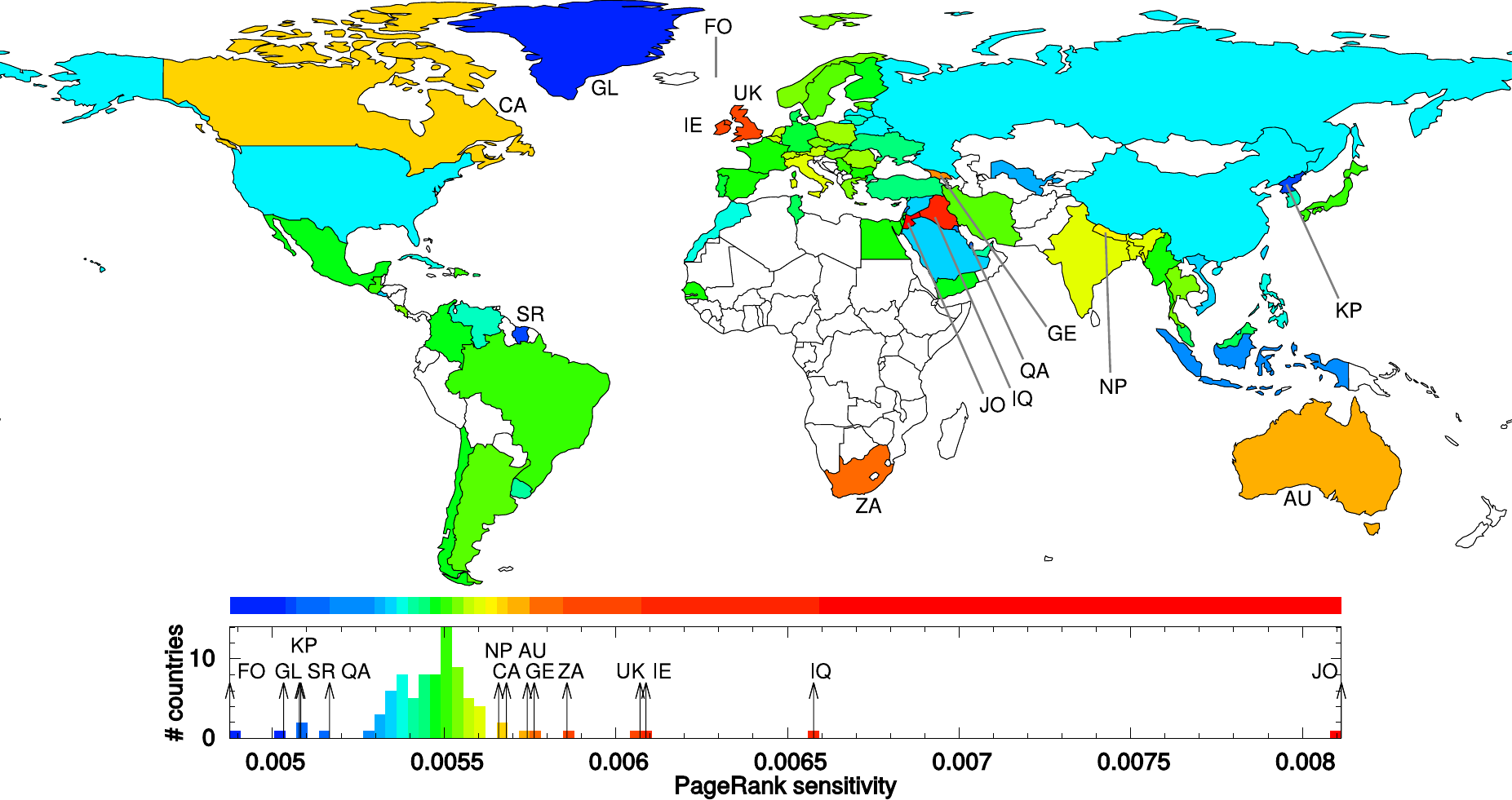}
}
\caption{World map of diagonal sensitivity of world countries $D(u \rightarrow c, c)$ 
to the change of the reduced Google matrix 
link university $u$ $\rightarrow$ country $c$. 
The cases of 4 universities are shown from top to bottom: 
 Harvard University, University of Chicago, Stanford University, and University of Oxford.
}
\label{fig:sensenwikidiago}
\end{figure}

The world maps of university influence on countries, 
expressed by the diagonal sensitivity  $D(u \rightarrow c, c)$
for 4 selected universities, are shown in Fig.~\ref{fig:sensenwikidiago}.

For Harvard the most sensitive country is South Africa (ZA) due to the
well known scandal linked to Harvard investments 
in apartheid ZA pointed on the Harvard wi\-ki\-page.
Puerto Rico also appears on this wikipage
in relation with oldest universities in the America.
However, next influenced countries are Georgia (GE), Israel (IL)
and Ireland (IR) which are not present on the wikipage
which appearance we attribute to indirect links.

For Chicago the most influenced countries are Singapore (SG),
Puerto Rico (PR) and India (IN). The first two countries SG, PR are not present
on the wikipage showing that indirect links play an important role for them.
India has a direct link related to the following facts: Chicago campus opened in India and
a faculty member was erstwhile governor of India central bank.

For Stanford the top countries are Spain (ES), present on the wikipage,
PR and ZA appearing due to indirect links.

For Oxford the top three countries are
Jordan (JO), appearing of wikipage since 
Abdullah II of JO has been educated at Oxford;
Iraq (IQ), also appearing on wikipage 
since T.~E.~Lawrence educated at Oxford  
played a major role in establishing and administering the modern state of IQ;
IR which is not present on wikipage but has close links with UK.

\begin{figure}[h]
\centering
\resizebox{0.9\columnwidth}{!}{%
\includegraphics{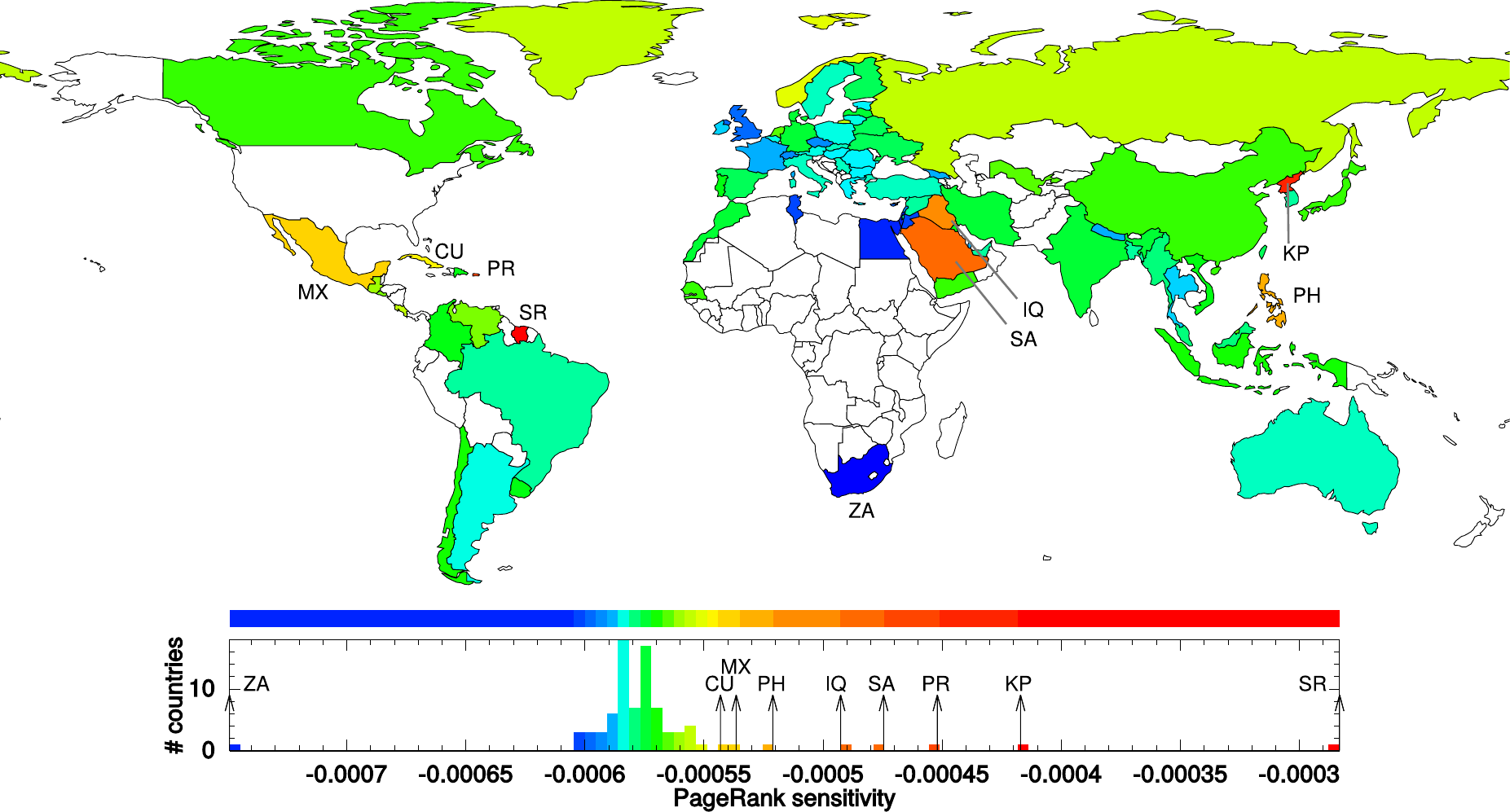}
}
\resizebox{0.9\columnwidth}{!}{%
\includegraphics{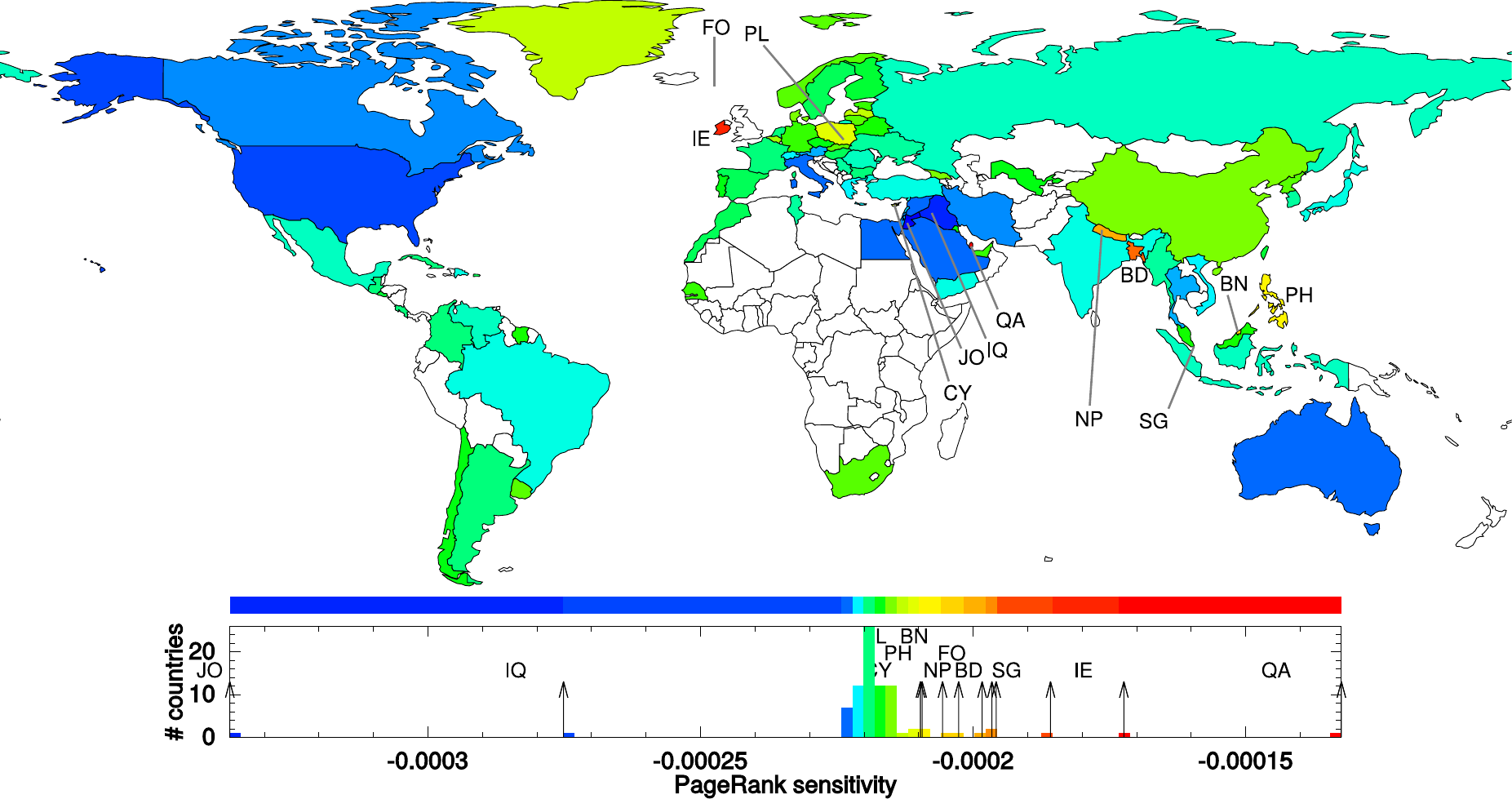}
}
\caption{World map of nondiagonal sensitivity $D(u \rightarrow c, c')$
of country $c'$
to the change of the reduced Google matrix link 
Harvard$\rightarrow$US (top panel) and Oxford$\rightarrow$UK (bottom panel). 
Red (blue) color corresponds to the greatest (lowest) absolute value.
Diagonal values  $D(u \rightarrow c, c'=c)$ are not shown.
}
\label{fig:sensenwikinondiago}
\end{figure}

Examples of nondiagonal sensitivity
are shown in Fig.~\ref{fig:sensenwikinondiago} for the link variation Harvard University to US
and University of Oxford to UK.
For the link Harvard$\rightarrow$US the most influenced countries are
Suriname (SR), People's Republic of Korea (KP) and
Puerto Rico (PR). For the link variation Oxford$\rightarrow$UK
the most influenced countries are Qatar (QA),
Ireland (IR) and Singapore (SG).
By definition the nondiagonal sensitivity
contains indirect effects and it is not
so easy to find the pathways of links which
are responsible for this dominant influence.
These examples show the strength
of reduced Google matrix approach.

\begin{figure}[h]
\centering
\resizebox{0.9\columnwidth}{!}{%
\includegraphics{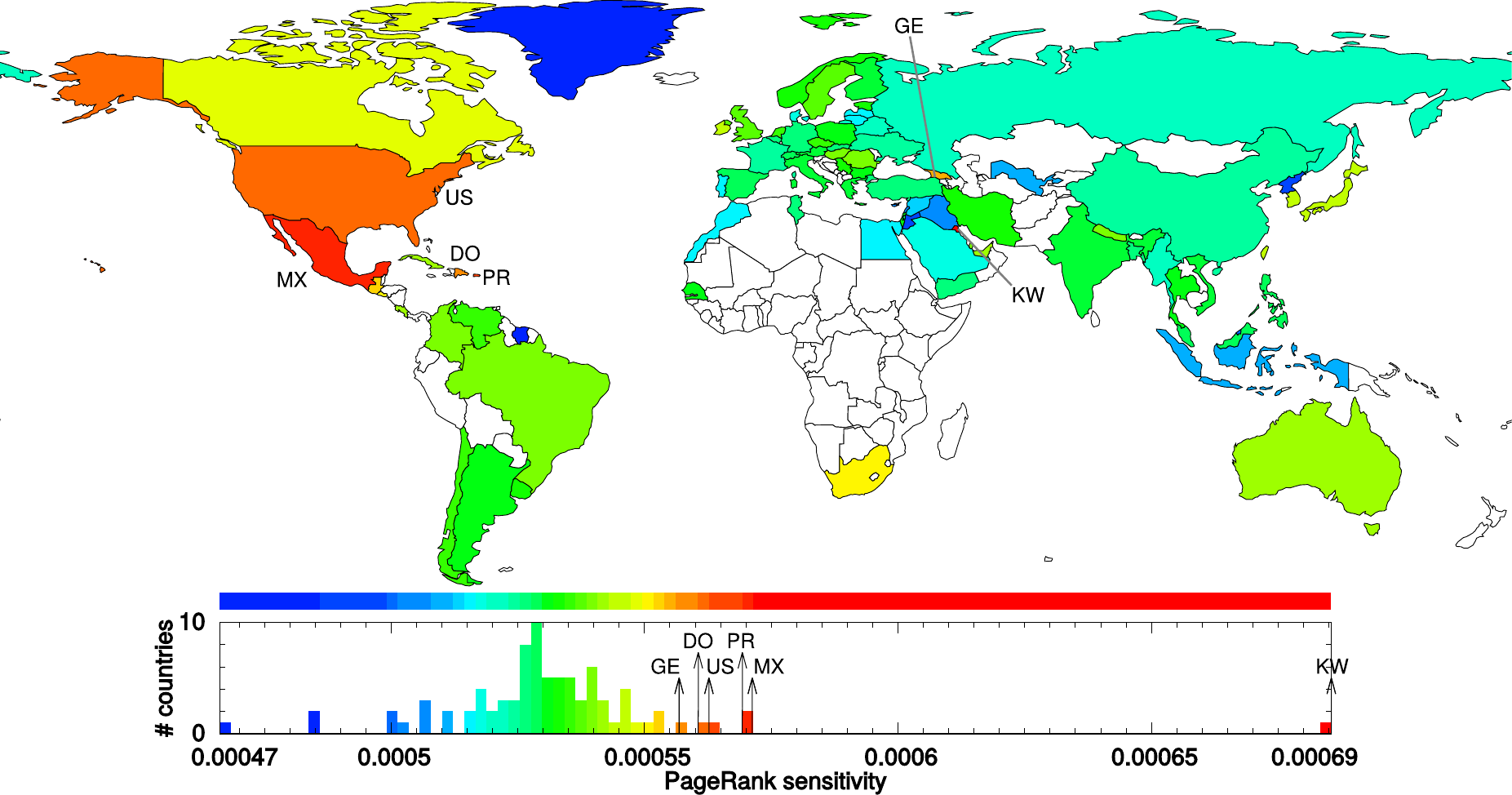}
}
\caption{Same as in Fig.~\ref{fig:sensenwikidiago}  with diagonal 
sensitivity $D(\mbox{``Rice University''} \rightarrow c, c)$ 
to the change of the reduced Google matrix 
link Rice University $\rightarrow$ country $c$.
}
\label{fig:rice}
\end{figure}

Finally we discuss an example of Rice University.
It has rank 74 in ARWU2017 being at position 37 inside USA,
but according to WRWU2017 its PageRank position  is only 357 and 56 inside USA.
This shows that the Wikipedia article of Rice University
is not sufficiently developed and its visibility via Wikipedia
is rather low and can be improved by more skillful
organization of its wikipage.
We note that Wikipedia visibility of a university
plays rather important role since very often it is
on top lines of Google search and
many language editions spread the wikipage content
world wide free of charge.
Also the WRWU position 357 is only due to appearance
of Rice in top 100 universities of FA, KO, and TH language editions
probably due to wikipage creation and information
given by Rice alumni speaking these languages. This indicates
an importance of links between university and its alumni.

The world influence of Rice University,
expressed via its diagonal sensitivity,
is shown in Fig.~\ref{fig:rice}.
The most influenced countries are
Kuwait (KW), Mexico (MX) and Puerto Rico (PR).
These countries are not present on the wikipage
of Rice University 
and their appearance is related to indirect links.


\section{Reduced networks of world universities}

In this Section we analyze the interactions between top 20 universities
obtained from the REGOMAX approach.
We consider the different cultural views from EN, FR, DE, RU  editions
and make a comparative analysis.
For EN edition we also consider the links of selected universities with
countries.


\subsection{Top 20 universities in English Wikipedia edition}

Top 20 PageRank universities of  ENWRWU2017 are 
given in Tab.~\ref{tab:top20ENWRWU}. They are
either US or UK universities. 
We define 4 regional groups with their PageRank leaders: Stanford University for US west coast, 
University of Chicago for US central region, Harvard University for US east coast, 
and University of Oxford for UK; each group is marked by color in Tab.~\ref{tab:top20ENWRWU}.


\begin{figure}[t]
\centering
\resizebox{\columnwidth}{!}{%
\includegraphics{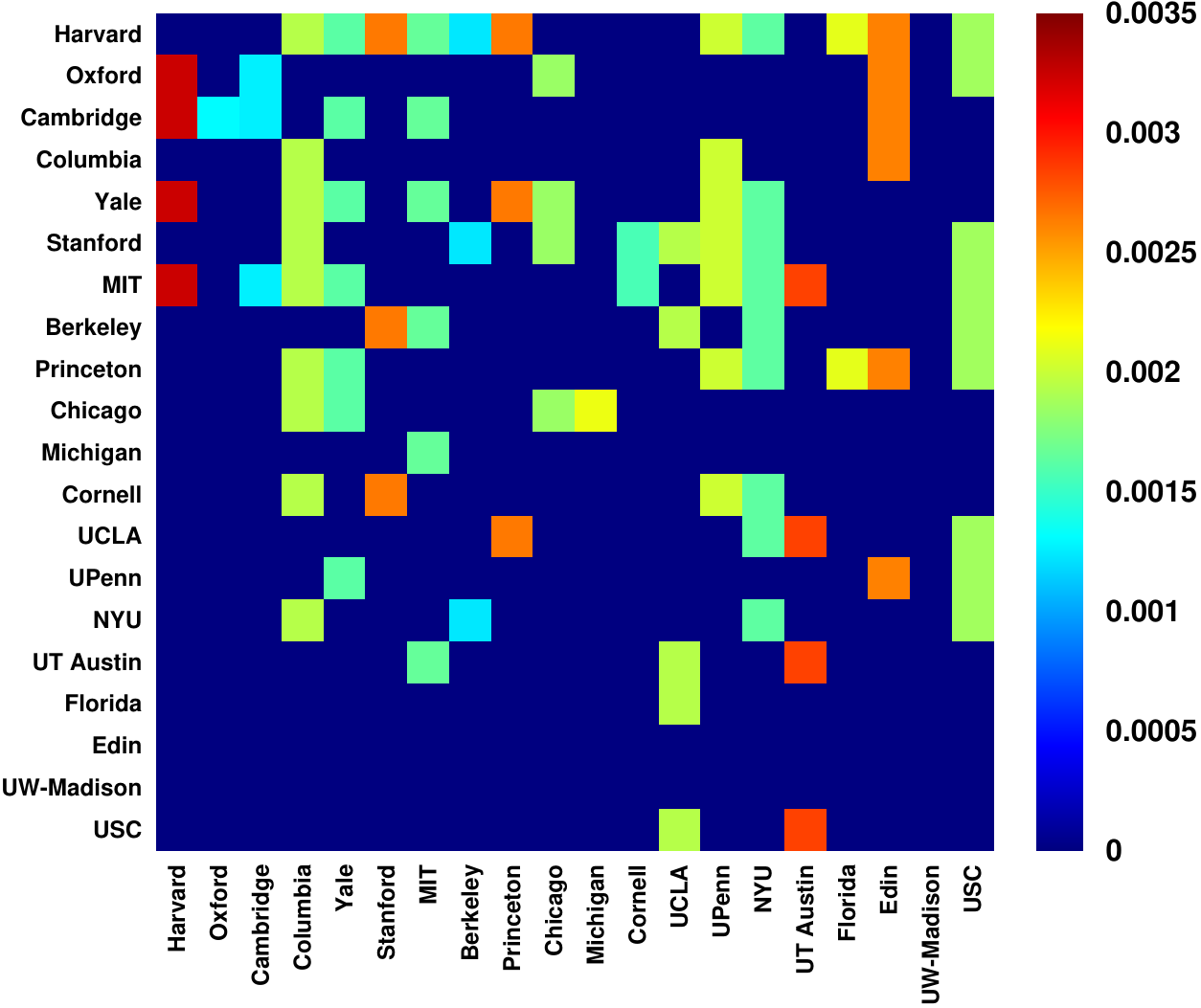}
\includegraphics{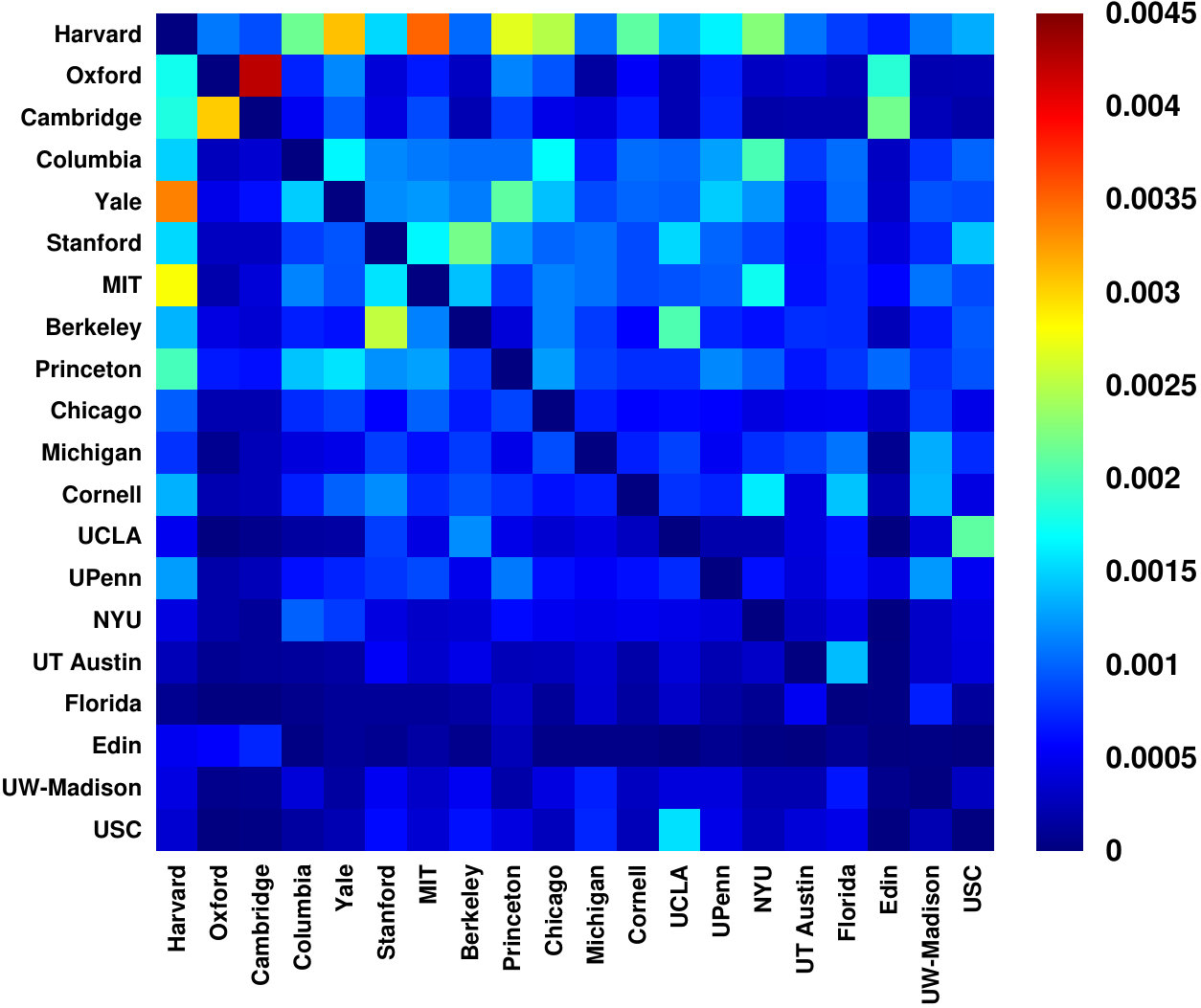}
}
\caption{Matrices $G_{\mathrm{rr}}$ (left panel) and $G_{\mathrm{qrnd}}$ (right panel) for top20 ENWRWU 
(Tab.~\ref{tab:top20ENWRWU}). The universities are ordered by their PageRank index as in Tab.~\ref{tab:top20ENWRWU}.
The matrix weights are $W_{\mathrm{rr}} = 0.00877$ and $W_{\mathrm{qrnd}} = 0.01381$.
Color shows the strength of matrix elements.
The same components for top 20 universities of FRWIKI, DEWIKI, RUWIKI
are available in Figs.~\ref{fig:FRU20Grr_Gqr}, \ref{fig:DEU20Grr_Gqr}, \ref{fig:RUU20Grr_Gqr}.
}
\label{fig:ENU20Grr_Gqr}
\end{figure}

The components of reduced Google matrix describing direct links $G_{\mathrm{rr}}$ 
and indirect links  $G_{\mathrm{qrnd}}$ are shown in Fig.~\ref{fig:ENU20Grr_Gqr}
(only nondiagonal links are shown for  $G_{\mathrm{qr}}$). The weight of indirect nondiagonal links is
 about 50\% percent stronger than the weight of direct links. This shows the importance of indirect 
interactions between top 20 universities.

\begin{figure}
\centering
\resizebox{\columnwidth}{!}{%
\includegraphics{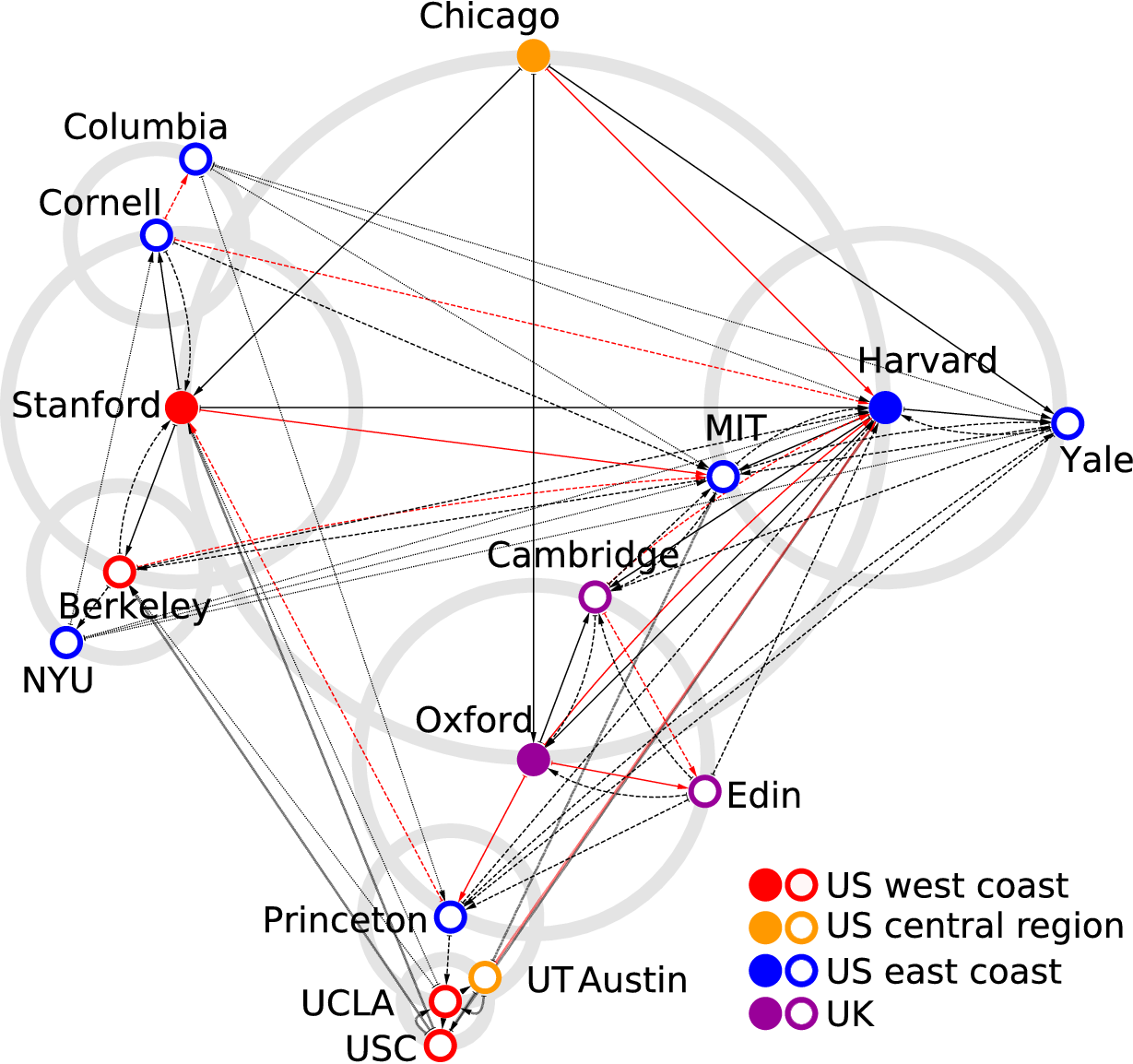}
}
\caption{Network of friends of top 20 PageRank universities of ENWRWU 
obtained from matrix of direct and indirect links $G_{\mathrm{rr}}+G_{\mathrm{qrnd}}$.
Color filled nodes are regional leaders. Red links are purely hidden links, i.e., 
no corresponding adjacency matrix entry. We obtain 4 friendship levels (gray circles). 
Links originating from 1st level universities are presented by solid lines, 
from 2nd level by dashed lines, from 3rd level by doted lines, and 
from 4th level by ``\textbackslash'' symbol lines.
}
\label{fig:ENU20netw}
\end{figure}

\begin{figure}[h]
\centering
\resizebox{\columnwidth}{!}{%
\includegraphics{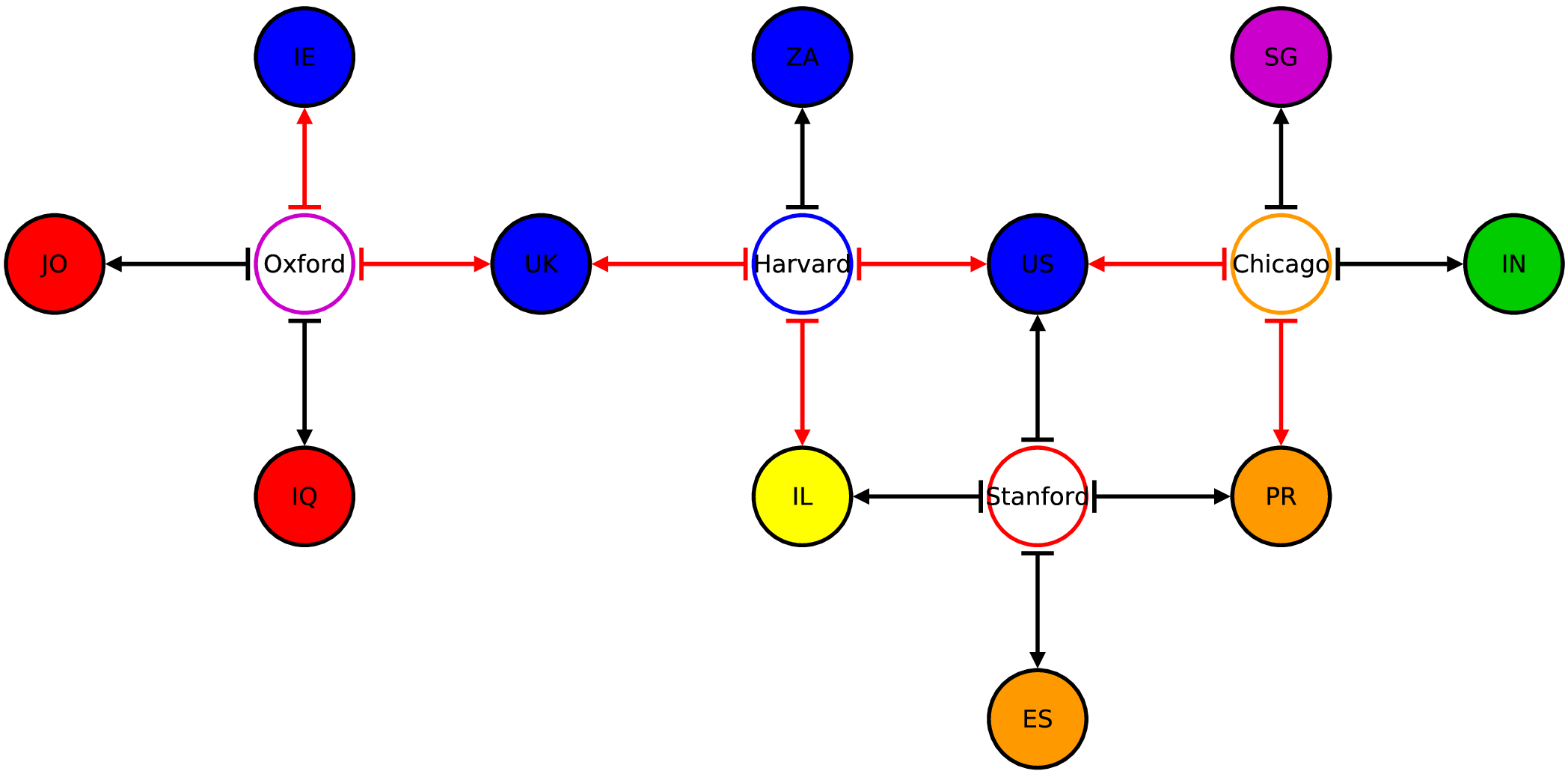}
}
\caption{Network of friends from $G_{\mathrm{rr}}+G_{\mathrm{qrnd}}$ 
associated to the top20 ENWRWU and 85 countries listed in Tab.~\ref{tab:WRWUcountries}.
For each regional leaders, Stanford University, University of Chicago, 
Harvard University, University of Oxford, the four strongest links to one 
of the 85 countries listed in Tab.~\ref{tab:WRWUcountries} are presented. 
Universities (countries) are represented by empty (full) nodes.
The color code for countries depends on the main spoken language: 
\blu{blue} for English, \red{red} for Arabic, \ora{orange} for Spanish, 
\vio{violet} for Chinese, \gre{green} for Hindi, and \yel{yellow} for Hebrew. 
Red links are purely indirect links and black ones are from direct links.
}
\label{fig:ENWIKIGsumU20C85}
\end{figure}

To characterize the importance of direct and indirect links
we consider the sum of two matrix components given by $G_{\mathrm{rr}}+G_{\mathrm{qrnd}}$.
From this matrix we construct the network of friends of regional leaders shown in 
Fig.~\ref{fig:ENU20netw} (the drawings of networks have been produced using Cytoscape \cite{shannon03}).
First we place the regional leader on a circle (1st level of possible friendship).
From a regional leader we look at the four biggest outgoing
links in $G_{\mathrm{rr}}+G_{\mathrm{qrnd}}$, these four links define the four best friends of the regional leader.
If these friends are not present in the network of friends, i.e., they are not themselves regional leaders, then they
are placed on the circle around the regional leader (2nd level of friendship). If several regional leaders share
the same friend, by preference, the friend is placed in the circle around the leader of its region.
Then, from each friends of regional leaders, we define in the same way four new friends. Each
new friend is either already placed in the friendship network or not. If the new friend is not
present, it is placed in the circle around the corresponding friend of regional leader (3rd level of friendship). If a new friend is shared by several friends of regional leaders, the new friend is placed by preference on the circle around the friend of regional leader belonging to its region.
In the same manner we then define the 4th level of friendship and
so on. The procedure continues until no new friends can be placed on the friendship network (because already placed on it). For
the 2017 ENWRWU top20 the procedure stops after four levels of friendship. A red arrow represents a pure hidden link, i.e., a link from university $u$ to university $u'$ with a null adjacency matrix entry, $A_{u'u}=0$, or otherwise stated, with a minimal value in $G_{\mathrm{rr}}$, $\left(G_{\mathrm{rr}}\right)_{u'u}=(1-\alpha)/N$.

\begin{table}[b]
\caption{List of top 20 PageRank universities of French edition WRWU2017. 
The color code corresponds to the country location of universities:
 blue for US, violet for UK, red for FR, green for CA, and yellow for BE.} 
\resizebox{\columnwidth}{!}{
\begin{tabular}{rl|rl}
\hline
Rank&University&Rank&University\\
\hline
\hline
1st&\blu{Harvard University}&11th&\gre{University Laval}\\
2nd&\vio{University of Oxford}&12th&\red{Pantheon-Sorbonne University}\\
3rd&\red{\'Ecole polytechnique}&13th&\blu{Princeton University}\\
4th&\vio{University of Cambridge}&14th&\blu{University of California, Berkeley}\\
5th&\red{\'Ecole normale sup\'erieure}&15th&\red{Paris-Sorbonne University}\\
6th&\blu{Massachusetts Institute of Technology}&16th&\yel{Universit\'e libre de Bruxelles}\\
7th&\blu{Yale University}&17th&\gre{University of Montreal}\\
8th&\blu{Columbia University}&18th&\yel{Universit\'e catholique de Louvain}\\
9th&\blu{Stanford University}&19th&\red{Paris Nanterre University}\\
10th&\red{\'Ecole pratique des hautes \'etudes}&20th&\blu{University of Chicago}\\
\hline
\end{tabular}
}
\label{tab:top20FRWRWU}
\end{table}

This network presentation of friends in Fig.~\ref{fig:ENU20netw}
shows that the close friends are mainly located in same region;
thus MIT, Harvard, Yale form one group of east coast,
Oxford, Cambridge, Edinburgh are friends inside UK.
However, there are also inter-regional friends formed by
Princeton, UCLA, USC, UT Austin and proximity between
Stanford, Berkeley and Cornell.
The direct links shown in black play 
an important role but indirect links
shown in red are also present and 
significant like e.g. MIT being indirect friend of Stanford,
Harvard being indirect friend of Chicago and Oxford,
and Edinburgh, Princeton being indirect friend of Oxford.

We also consider countries which are friends of each regional university leader
as shown in Fig.~\ref{fig:ENWIKIGsumU20C85}. For this we consider the matrix elements
of $G_{\mathrm{rr}}+G_{\mathrm{qrnd}}$ constructed for top 20 universities of Tab.~\ref{tab:top20ENWRWU}
and 85 countries listed in Tab.~\ref{tab:WRWUcountries}. For each regional leader university we select top 4 friendliest countries.
The network of country-university friends is presented in Fig.~\ref{fig:ENWIKIGsumU20C85}
with countries marked by colors corresponding to mostly spoken language. Thus we see that countries friends of Oxford are mostly Arab and English speaking countries;
countries friends of Harvard are dominantly from English speaking countries excepting one Hebrew
 speaking country;  friends of Stanford are two Spanish speaking countries, one English and one Hebrew;
Chicago is mostly diversified having English, Hindi, Chinese and Spanish speaking friends.

\begin{figure}[t]
\centering
\resizebox{\columnwidth}{!}{%
\includegraphics{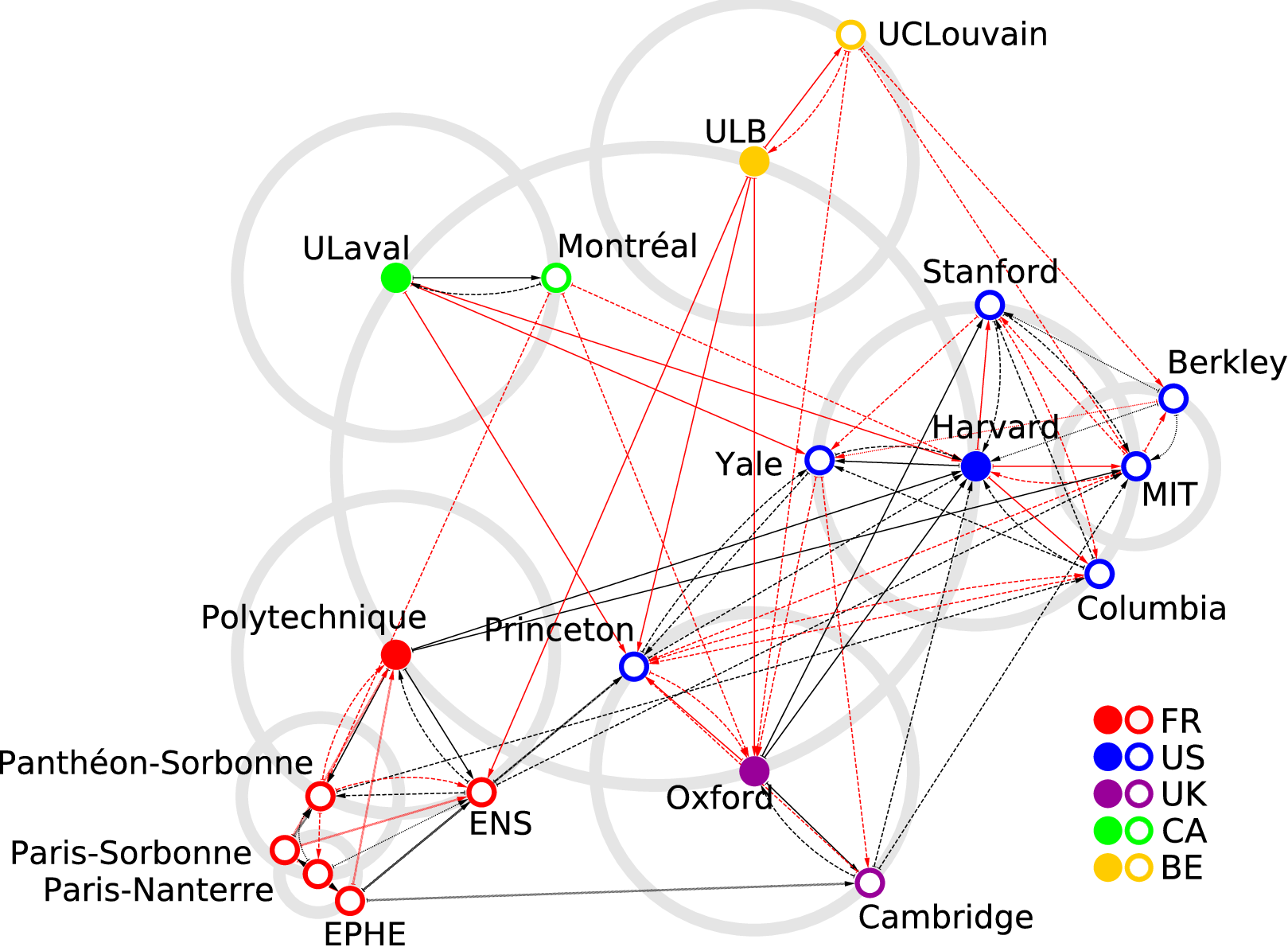}
}
\caption{Same as in Fig.~\ref{fig:ENU20netw} for PageRank top 20 universities 
of FRWIKI2017 from Tab.~\ref{tab:top20FRWRWU}.
Color filled nodes are country leaders. 
Red links are purely hidden links, 
i.e., no corresponding adjacency matrix entry.
We obtain 4 friendship levels (gray circles). 
Links originating from 1st level universities are presented by solid lines, 
from 2nd level by dashed lines, from 3rd level by doted lines, 
and from 4th level by ``\textbackslash'' symbol lines.
}
\label{fig:FRU20netw}
\end{figure}

\subsection{Top 20 universities in French Wikipedia edition}

Here we consider the cultural view of FRWIKI on top 20 universities and their
interactions. We select from FRWIKI2017 top PageRank universities listed in 
Tab.~\ref{tab:top20FRWRWU}. These universities belong to 5 countries (US, UK, FR, CA, BE) 
marked by corresponding color. Top PageRank university of each country is
considered as a country leader. Similar to the case of Fig.~\ref{fig:ENU20netw}
we obtain from the reduced Google matrix components
$G_{\mathrm{rr}}+G_{\mathrm{qrnd}}$ of these 20 universities the network of friends
shown in Fig.~\ref{fig:FRU20netw}.

The obtained network of friends of Fig.~\ref{fig:FRU20netw} shows clear cluster of universities inside 
their own countries. However, the inter-country links are well 
present and they are mainly indirect links (shown in red) pointing toward English speaking universities.
Thus Princeton is indirect friend of Oxford, ULaval and ULB;
Yale and Harvard are indirect friends of ULaval.
Since we consider FRWIKI there are 6 French universities 
being the next in number after US with 7 universities among top 20.
However, US universities are strongly linked with universities of Canada, Belgium  and UK
while French universities are weakly linked to other countries.
This FRWIKI-analysis demonstrate certain world isolation of French
universities.
Also from FRWIKI point of view, the leading English speaking universities in Fig.~\ref{fig:FRU20netw} form an invariant subset from which a random surfer cannot escape: the friends of these universities are uniquely English speaking universities.

\begin{table}[b]
\caption{List of top 20 PageRank universities of German edition WRWU2017. 
The color code corresponds to the country location of universities:
green for DE, blue for US, violet for UK, and black for AT.} 
\resizebox{\columnwidth}{!}{
\begin{tabular}{rl|rl}
\hline
Rank&University&Rank&University\\
\hline
\hline
1st&\gre{Ludwig Maximilian University of Munich}&11th&\gre{University of Freiburg}\\
2nd&\gre{Humboldt University of Berlin}&12th&\gre{University of Cologne}\\
3rd&\gre{University of G\"ottingen}&13th&\gre{University of M\"unster}\\
4th&\gre{Heidelberg University}&14th&\vio{University of Oxford}\\
5th&\gre{Free University of Berlin}&15th&\gre{University of Hamburg}\\
6th&{\bf University of Vienna}&16th&\gre{Goethe University Frankfurt}\\
7th&\gre{University of T\"ubingen}&17th&\vio{University of Cambridge}\\
8th&\blu{Harvard University}&18th&\gre{University of Marburg}\\
9th&\gre{University of Bonn}&19th&\gre{University of Kiel}\\
10th&\gre{Leipzig University}&20th&\gre{University of Jena}\\
\hline
\end{tabular}
}
\label{tab:top20DEWRWU}
\end{table}

\begin{figure}[t]
\centering
\resizebox{\columnwidth}{!}{%
\includegraphics{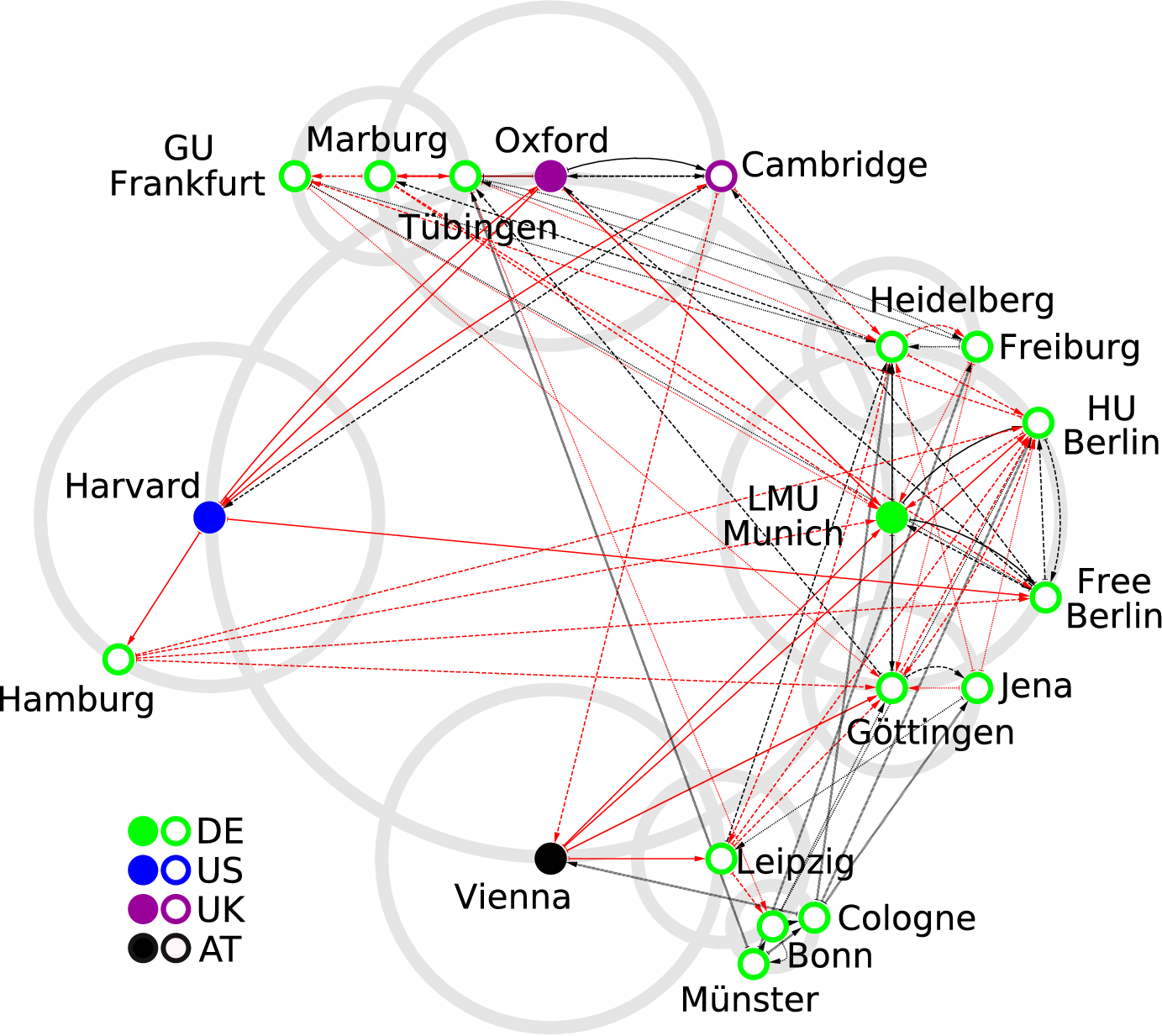}
}
\caption{Same as in Fig.~\ref{fig:ENU20netw} for PageRank top 20 universities 
of DEWIKI2017 from Tab.~\ref{tab:top20DEWRWU}.
Color filled nodes are country leaders.
Red links are purely hidden links, i.e., no corresponding adjacency matrix entry.
We obtain 4 friendship levels (gray circles). Links originating from 
1st level universities are presented by solid lines, from 2nd level by dashed lines, 
from 3rd level by doted lines, and from 4th level by ``\textbackslash'' symbol lines.
}
\label{fig:DEU20netw}
\end{figure}

\subsection{Top 20 universities in German Wikipedia edition}

Top 20 PageRank universities are given in Tab.~\ref{tab:top20DEWRWU}.
The network of friends, constructed
for these 20 nodes from the matrix components
$G_{\mathrm{rr}}+G_{\mathrm{qrnd}}$ of the reduce Google matrix,
is shown in Fig.~\ref{fig:DEU20netw}.
A specific point of these top 20 universities is that 
there are only 3 non-German-speaking universities
(Harvard, Oxford, Cambridge). This is due to the already discussed
feature of DEWIKI which gives strong preference 
to German universities (see right panel of Fig.~\ref{fig:overlap}).
The universities belong only to 4 countries AT, DE, UK, US.
The main cluster are formed around LMU Munich and Vienna
however GU Frankfurt, Marburg and Tubingen
are closely linked with Oxford and Cambridge;
Hamburg is linked with Harvard.
It should be pointed that there is a dominance of
indirect links which are also linking different countries.

\subsection{Top 20 universities in Russian Wikipedia edition}

\begin{figure}
\centering
\resizebox{\columnwidth}{!}{%
\includegraphics{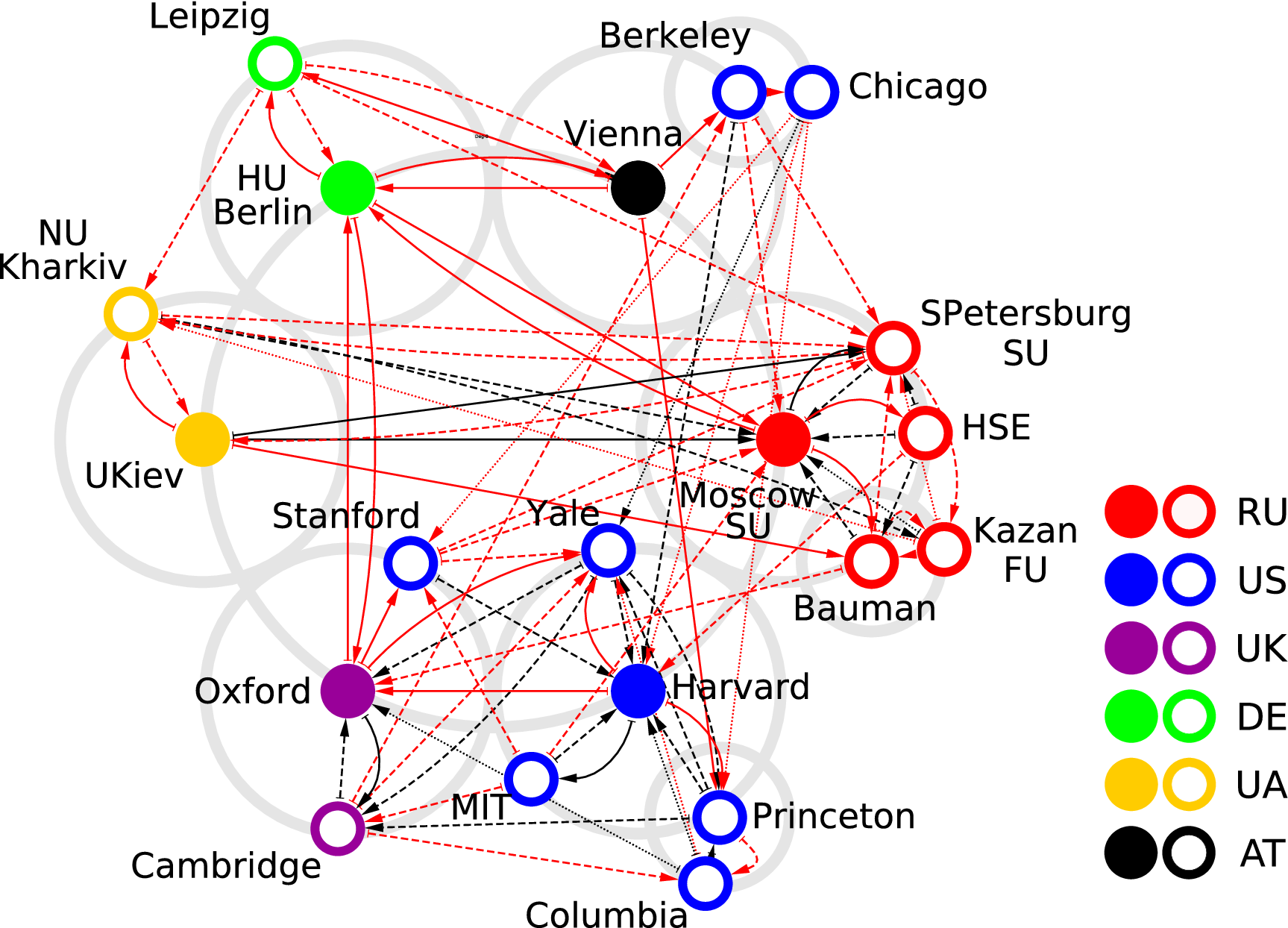}
}
\caption{Same as in Fig.~\ref{fig:ENU20netw} for PageRank top 20 universities 
of RUWIKI2017 from Tab.~\ref{tab:top20RUWRWU}.
Color filled nodes are country leaders.
Reduced network from top20 RUWRWU $G_{\mathrm{rr}}+G_{\mathrm{qrnd}}$. 
Color filled nodes are regional leaders. Red links are purely hidden links, 
i.e., no corresponding adjacency matrix entry. We obtain 3 acquaintance levels (gray circles). 
Links originating from 1st level universities are presented by solid lines,
 from 2nd level by dashed lines, and from 3rd level by doted lines.
}
\label{fig:RUU20netw}
\end{figure}

Top 20 PageRank universities are given in Tab.~\ref{tab:top20RUWRWU}.
The network of friends, constructed
for these 20 nodes from the matrix components
$G_{\mathrm{rr}}+G_{\mathrm{qrnd}}$ of the reduce Google matrix,
is shown in Fig.~\ref{fig:RUU20netw}. Among these 20 universities there 8 from US, 5 from Russia,
2 from Ukraine, 2 from Germany (its former DDR part), 2 from UK 
and 1 from Austria so that there are 6 different countries.
The clusters of universities are mainly linked with
their own countries even if there is very close proximity
between UK and US even if Berkeley and Chicago
are in the circle proximity of Vienna.
The main intercountry links are mainly indirect
(except direct links between Kiev pointing to Moscow and St. Petersburg
which were all inside former USSR). It is interesting to note that
German university, belonging to the former DDR part of Germany,
have strong links with Russian universities,
showing that the links inside Soviet block
are still significant even if Wikipedia had been created well 
after disappearance of DDR.

\begin{table}[b]
\caption{List of top 20 PageRank universities of German edition WRWU2017. 
The color code corresponds to the country location of universities:
red for RU, blue for US, violet for UK, green for DE, black for AT, and yellow for UA.} 
\resizebox{\columnwidth}{!}{
\begin{tabular}{rl|rl}
\hline
Rank&University&Rank&University\\
\hline
\hline
1st&\red{Moscow State University}&11th&\red{Kazan Federal University}\\
2nd&\red{Saint Petersburg State University}&12th&\yel{National University of Kharkiv}\\
3rd&\blu{Harvard University}&13th&\blu{Stanford University}\\
4th&\vio{University of Oxford}&14th&\blu{Princeton University}\\
5th&\vio{University of Cambridge}&15th&\blu{University of Chicago}\\
6th&\blu{Massachusetts Institute of Technology}&16th&\red{Higher School of Economics}\\
7th&\blu{Yale University}&17th&\red{Bauman Moscow State Technical University}\\
8th&\blu{Columbia University}&18th&\gre{Leipzig University}\\
9th&\yel{Kyiv University}&19th&{\bf University of Vienna}\\
10th&\gre{Humboldt University of Berlin}&20th&\blu{University of California, Berkeley}\\
\hline
\end{tabular}
}
\label{tab:top20RUWRWU}
\end{table}

\begin{figure}[t]
\centering
\resizebox{\columnwidth}{!}{%
\includegraphics{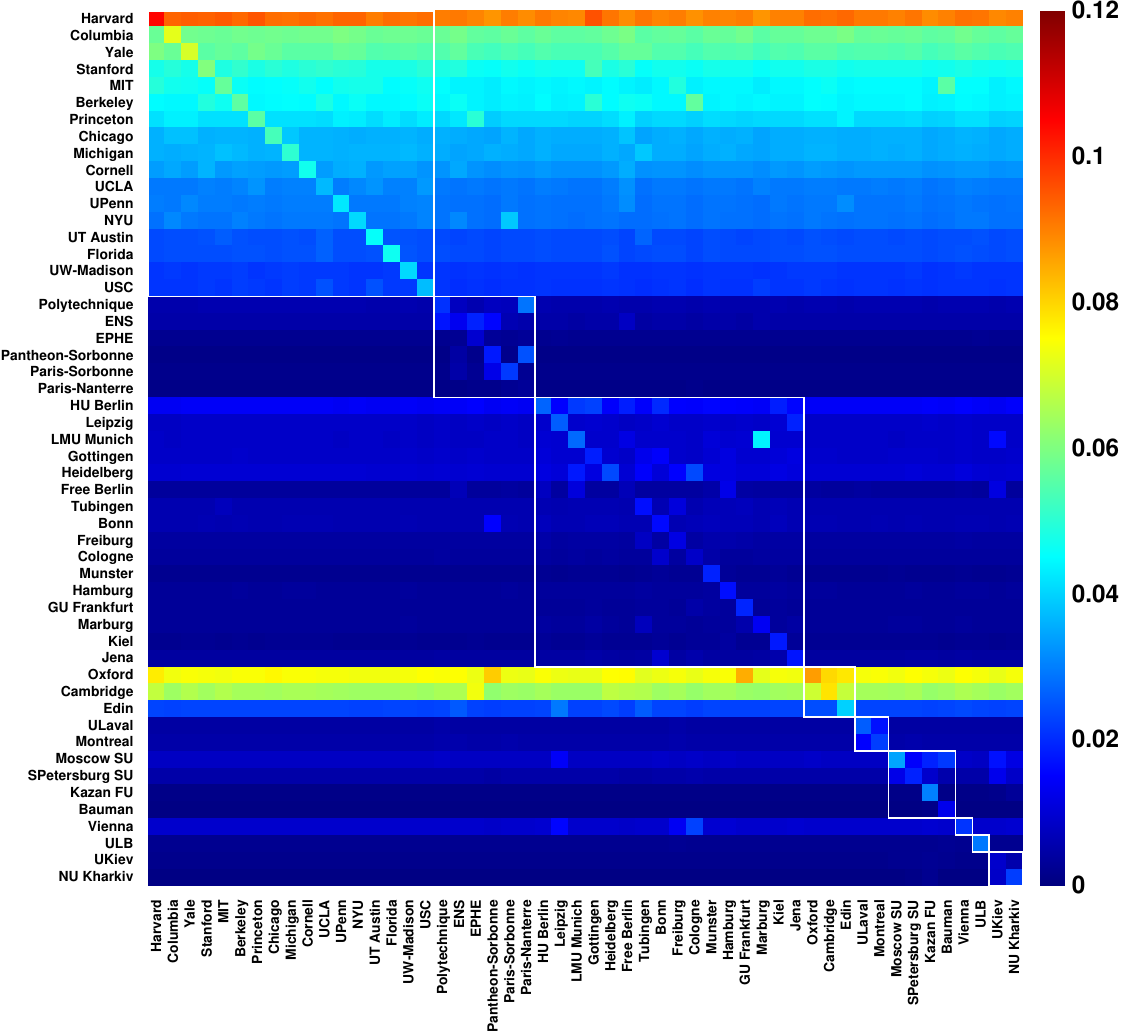}
\includegraphics{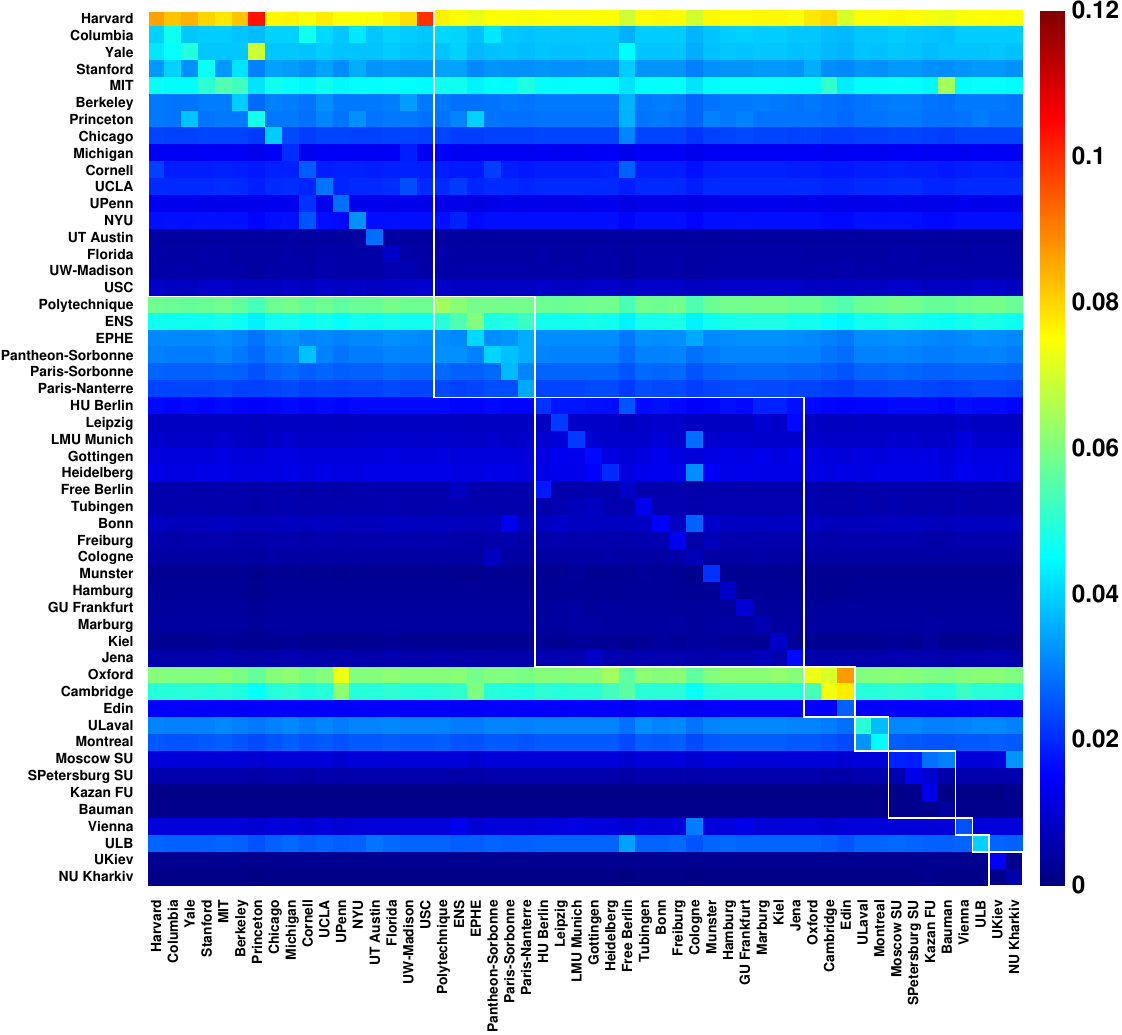}
}
\resizebox{\columnwidth}{!}{%
\includegraphics{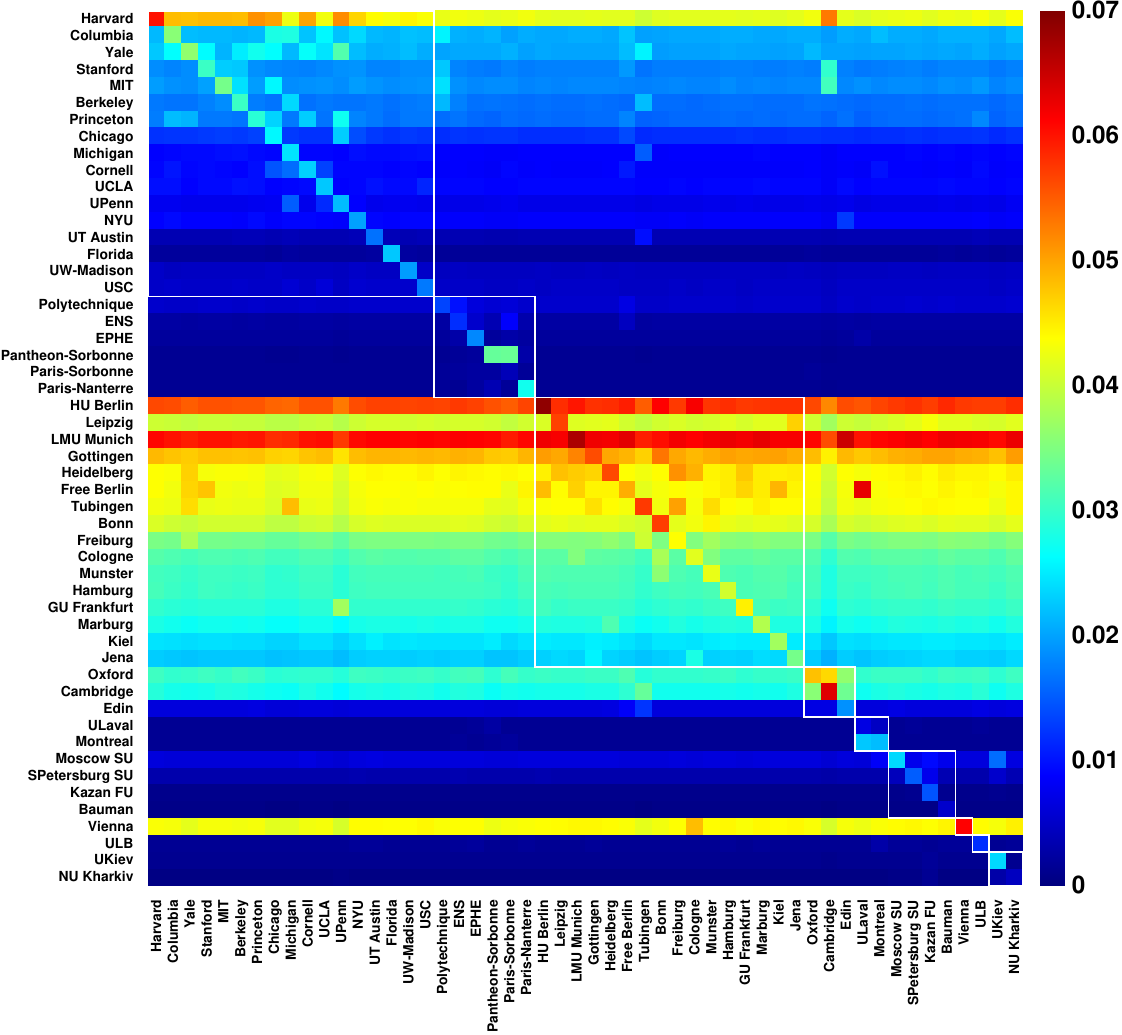}
\includegraphics{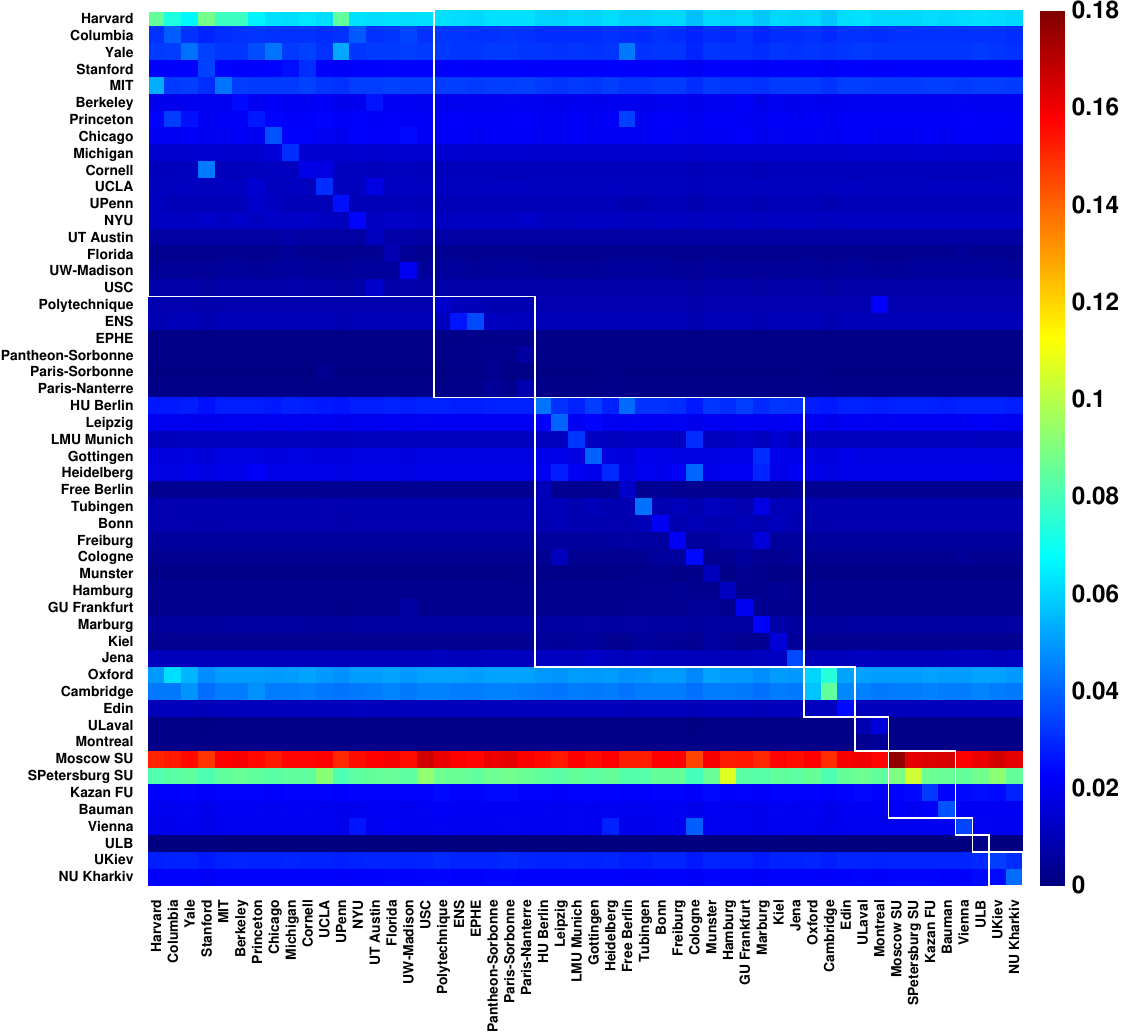}
}
\caption{Reduced Google matrix $G_{\mathrm{R}}$
  for universities listed in Tab.~\ref{tab:top52ENFRDERUWRWU}
  computed from EN (top left), FR (top right), DE (bottom left),
  and RU (bottom right) Wikipedia editions; for each edition nodes have the same
  order as in Tab.~\ref{tab:top52ENFRDERUWRWU}. The images of components
  $G_{\mathrm{pr}}$, $G_{\mathrm{rr}}$, $G_{\mathrm{qrnd}}$ are given in Figs.~\ref{fig:ENFRDERU_G_pr}, \ref{fig:ENFRDERU_G_rr}, \ref{fig:ENFRDERU_G_qr}.
}
\label{fig:ENFRDERU_G_R}
\end{figure}

\subsection{Comparison between English, French, German, and Russian Wikipedia editions}

In this subsection we perform a comparison of different cultural views of
DEWIKI, ENWIKI, FRWIKI and RU\-WIKI on top
universities. With this aim we take 20 PageRanked universities of each of these editions.
This gives us 52 different universities presented in these editions.
The list of them is given in Tab.~\ref{tab:top52ENFRDERUWRWU}. 
Each of these universities is attributed to its own foundation country
shown by colors in this Table. There are 9 different countries: US, FR, DE, UK, CA, RU, AT, BE and UA.

\begin{table}[b]
\caption{List of the 52 different universities appearing in the top20s of the EN, FR, DE, RU WRWU. Color code corresponds to country: \blu{US}, \bro{FR}, \gre{DE}, \vio{UK}, \yel{CA}, \red{RU}, \cya{AT}, {\bf BE} and \ora{UA}. Universities are ordered by countries (country groups are ordered according to PageRanking of 2017 English Wikipedia). Within country groups universities are ordered according to PageRanking of 2017 English Wikipedia.
} 
\resizebox{\columnwidth}{!}{
\begin{tabular}{rl|rl}
\hline
Rank&University&Rank&University\\
\hline
\hline
1&\blu{Harvard University}&28&\gre{Heidelberg University}\\
2&\blu{Columbia University}&29&\gre{Free University of Berlin}\\
3&\blu{Yale University}&30&\gre{University of T\"ubingen}\\
4&\blu{Stanford University}&31&\gre{University of Bonn}\\
5&\blu{Massachusetts Institute of Technology}&32&\gre{University of Freiburg}\\
6&\blu{University of California, Berkeley}&33&\gre{University of Cologne}\\
7&\blu{Princeton University}&34&\gre{University of M\"unster}\\
8&\blu{University of Chicago}&35&\gre{University of Hamburg}\\
9&\blu{University of Michigan}&36&\gre{Goethe University Frankfurt}\\
10&\blu{Cornell University}&37&\gre{University of Marburg}\\
11&\blu{University of California, Los Angeles}&38&\gre{University of Kiel}\\
12&\blu{University of Pennsylvania}&39&\gre{University of Jena}\\
13&\blu{New York University}&40&\vio{University of Oxford}\\
14&\blu{University of Texas at Austin}&41&\vio{University of Cambridge}\\
15&\blu{University of Florida}&42&\vio{University of Edinburgh}\\
16&\blu{University of Wisconsin–Madison}&43&\yel{University Laval}\\
17&\blu{University of Southern California}&44&\yel{University of Montreal}\\
18&\bro{\'Ecole polytechnique}&45&\red{Moscow State University}\\
19&\bro{\'Ecole normale sup\'erieure}&46&\red{Saint Petersburg State University}\\
20&\bro{\'Ecole pratique des hautes \'etudes}&47&\red{Kazan Federal University}\\
21&\bro{Panth\'eon-Sorbonne University}&48&\red{Bauman Moscow State Technical University}\\
22&\bro{Paris-Sorbonne University}&49&\cya{University of Vienna}\\
23&\bro{Paris Nanterre University}&50&{\bf Universit\'e Libre de Bruxelles}\\
24&\gre{Humboldt University of Berlin}&51&\ora{Kyiv University}\\
25&\gre{Leipzig University}&52&\ora{National University of Kharkiv}\\
26&\gre{Ludwig Maximilian University of Munich}&&\\
27&\gre{University of G\"ottingen}&&\\
\hline
\end{tabular}
}
\label{tab:top52ENFRDERUWRWU}
\end{table}

Then we perform the reduced Google matrix analysis for these 52 universities
for each edition constructing $G_{\mathrm{R}}$ and its 3 components.
The matrices  $G_{\mathrm{R}}$ for each edition are shown
in Fig.~\ref{fig:ENFRDERU_G_R}. The $G_{\mathrm{pr}}$, $G_{\mathrm{rr}}$, $G_{\mathrm{qrnd}}$ matrix components are available
in Figs.~\ref{fig:ENFRDERU_G_pr}, \ref{fig:ENFRDERU_G_rr}, \ref{fig:ENFRDERU_G_qr} with their weights which are similar to those
given in Fig.~\ref{fig:ENWIKIGRU20C85}; the weights of direct and indirect links
are comparable. From Fig.~\ref{fig:ENFRDERU_G_R} we see that each edition has
its own view on these 52 universities. Indeed, there is a clear tendency
that edition rank higher universities belonging to the countries
with edition language, e.g. RUWIKI places Moscow and
St. Petersburg universities on top
PageRank positions with a similar situation for DEWIKI.
Using the matrix components of $G_{\mathrm{rr}}+G_{\mathrm{qrnd}}$
we analyze the network of friend of 52 universities from the view point of
ENWIKI, FRWIKI, DEWIKI and RUWIKI.
The approach is the same as those used in 
network of friends discussed in the previous subsection. 


\begin{figure}[t]
\centering
\resizebox{\columnwidth}{!}{%
\includegraphics{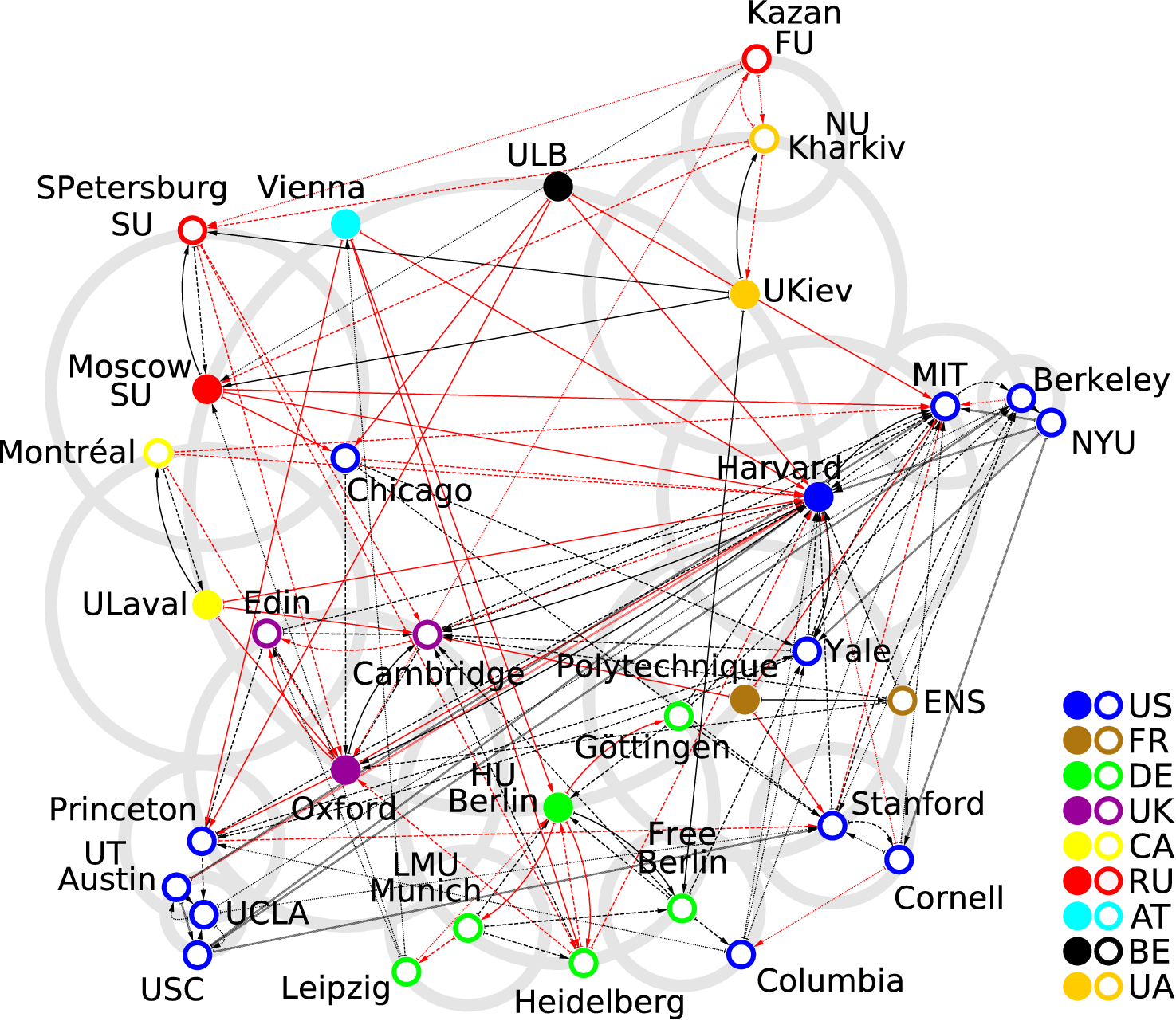}
}
\caption{Network of friends of 52 universities listed in Tab.~\ref{tab:top52ENFRDERUWRWU} 
computed from $G_{\mathrm{rr}}+G_{\mathrm{qrnd}}$ of ENWIKI. 
Color filled nodes are country leaders (same colors as in the Tab.~\ref{tab:top52ENFRDERUWRWU}).
 Red links are purely hidden links, i.e., no corresponding adjacency matrix entry.
We obtain 4 acquaintance levels (gray circles). 
Links originating from 1st level universities are presented by solid lines, 
from 2nd level by dashed lines, from 3rd level by doted lines, and 
from 4th level by ``\textbackslash'' symbol lines.
}
\label{fig:ENFRDERU_en_netw}
\end{figure}

The network of top friends for 52 universities in ENWIKI is shown in
Fig.~\ref{fig:ENFRDERU_en_netw}. We see that the majority of links
are indirect (red) comparing to the direct links (black).
As expected two clusters of English speaking universities are well visible; 
in fact as in Fig.~\ref{fig:FRU20netw}, these English speaking universities 
form an invariant subspace from which a random surfer cannot escape. 
These universities act as an attractor subset in this friendship network.
Chicago is located aside as it was already
visible in previous subsection (see Fig.~\ref{fig:ENU20netw}).
A compact cluster of German universities is also well visible.
We point that there are only 2 isolated French universities among top friends
appearing in Fig.~\ref{fig:ENFRDERU_en_netw}; no university from other countries points toward these 2 universities, and these 2 universities point exclusively toward UK/US universities.
In total this network of top friends has 13 US universities,
6 of DE, 3 of UK, 3 of RU, 2 of UA, 2 of CA, 2 of FR.
Since the network is obtained from ENWIKI
it is understandable that US universities (with UK ones)
form the majority. However, German universities
show their strength and significant influence.
Comparing to them French university group is small
and not significant being placed behind Russian universities.
This network clearly shows the weak representation and influence
of French universities that reflects a certain reality.

\begin{figure}[t]
\centering
\resizebox{\columnwidth}{!}{%
\includegraphics{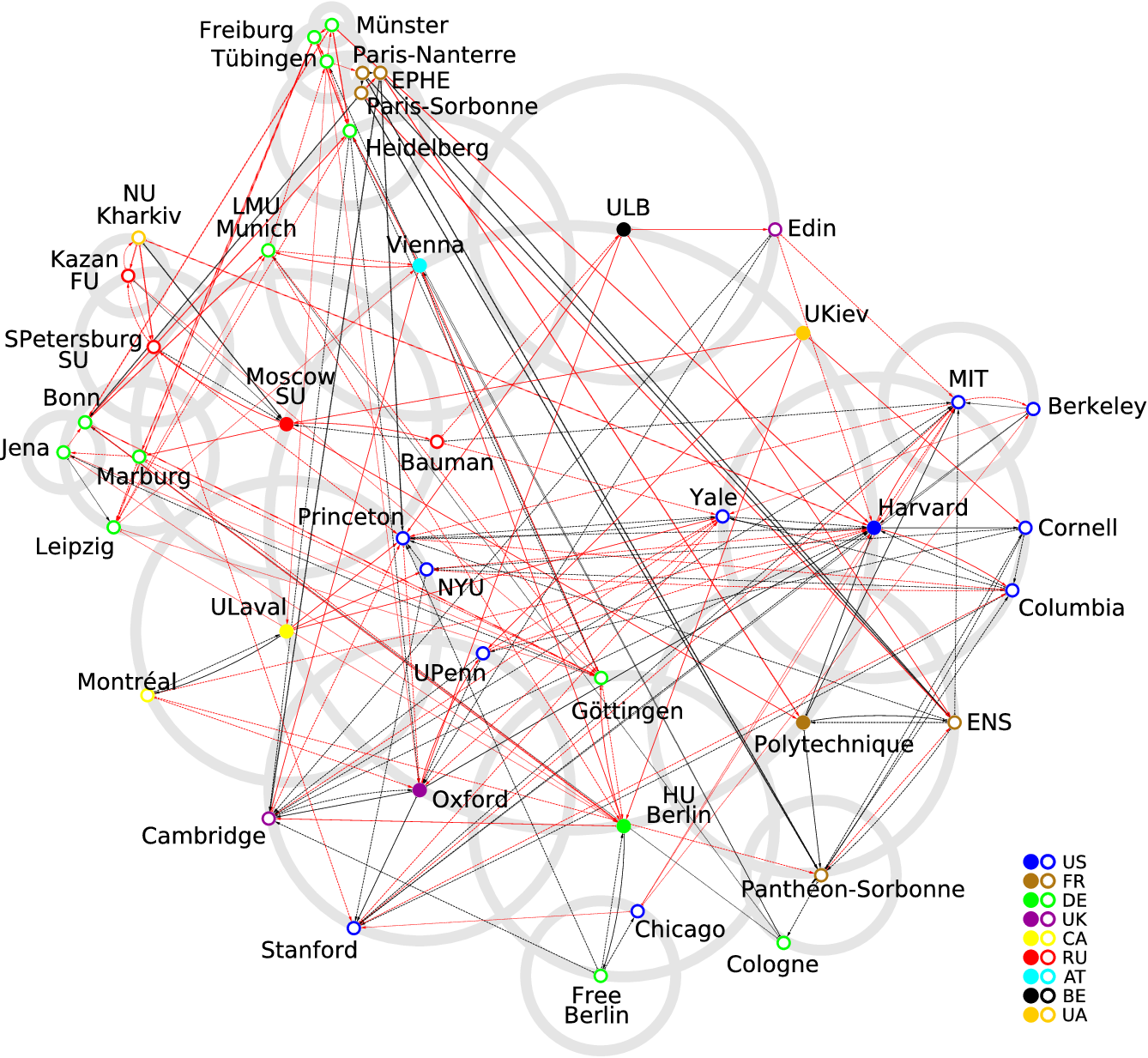}
}
\caption{Same as in Fig.~\ref{fig:ENFRDERU_en_netw} but from FRWIKI.
}
\label{fig:ENFRDERU_fr_netw}
\end{figure}

The network of top friends from FRWIKI is shown in Fig.~\ref{fig:ENFRDERU_fr_netw}.
Here we have the dominance of 13 German universities,
followed by 11 of US, 6 of France and 4 of Russia.
Still the indirect links play a dominant or
comparable role with the direct links.
In this network the cluster structure is less visible, 
however as in Figs.~\ref{fig:FRU20netw} and \ref{fig:ENFRDERU_en_netw} 
the cluster of US-UK universities is central and acts as an attractor subset.

\begin{figure}[t]
\centering
\resizebox{\columnwidth}{!}{%
\includegraphics{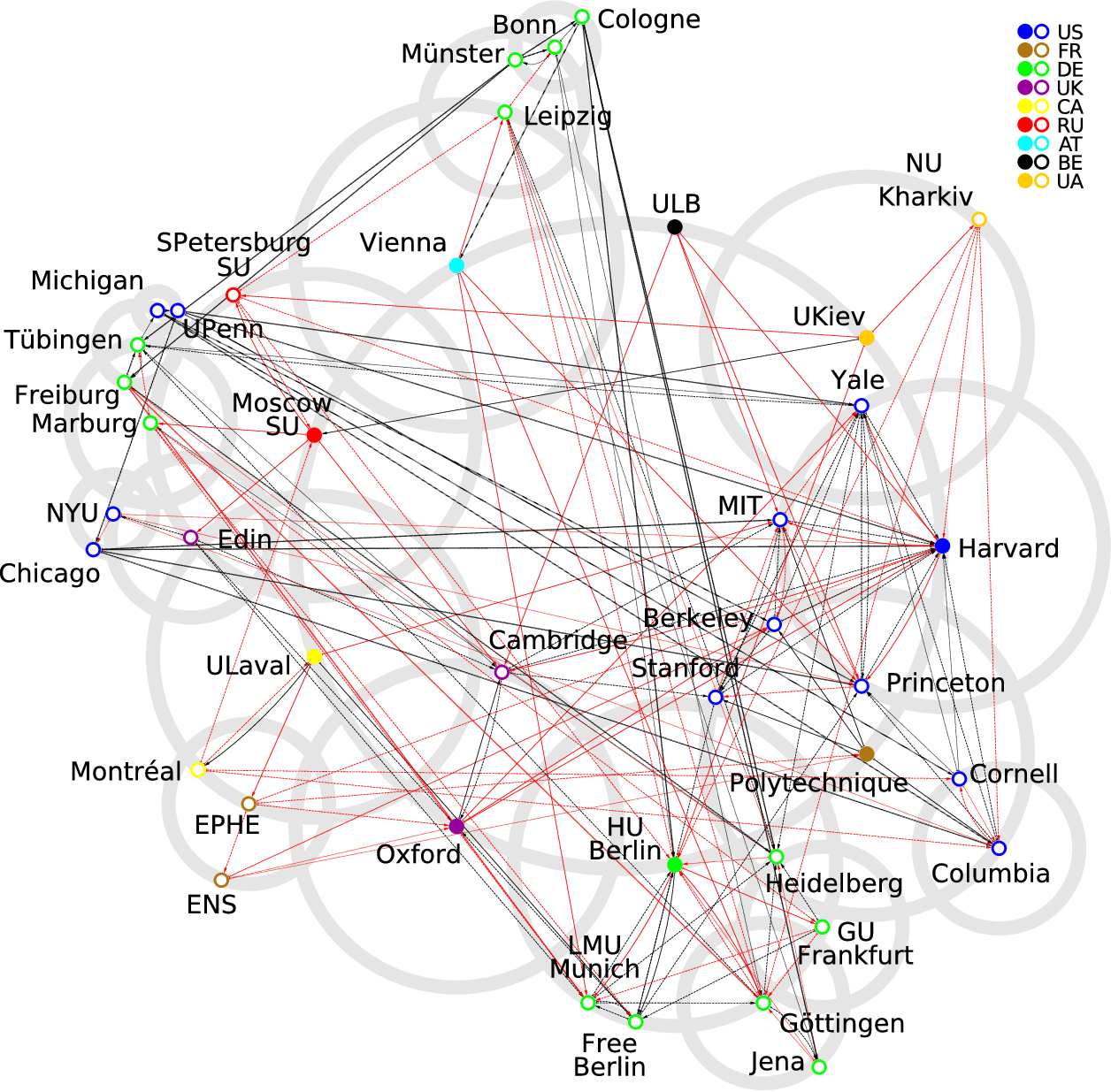}
}
\caption{Same as in Fig.~\ref{fig:ENFRDERU_en_netw} but from DEWIKI.}
\label{fig:ENFRDERU_de_netw}
\end{figure}

Fig.~\ref{fig:ENFRDERU_de_netw} shows the network of top
friends from DEWIKI. Here, naturally, the dominance 
of 14 German universities remains,
to be compared with 11 of US, 3 of France, 3 of UK and 2 of Russia.
In global DE universities are distributed over 3 clusters,
and US over 2 clusters. 

\begin{figure}[t]
\centering
\resizebox{\columnwidth}{!}{%
\includegraphics{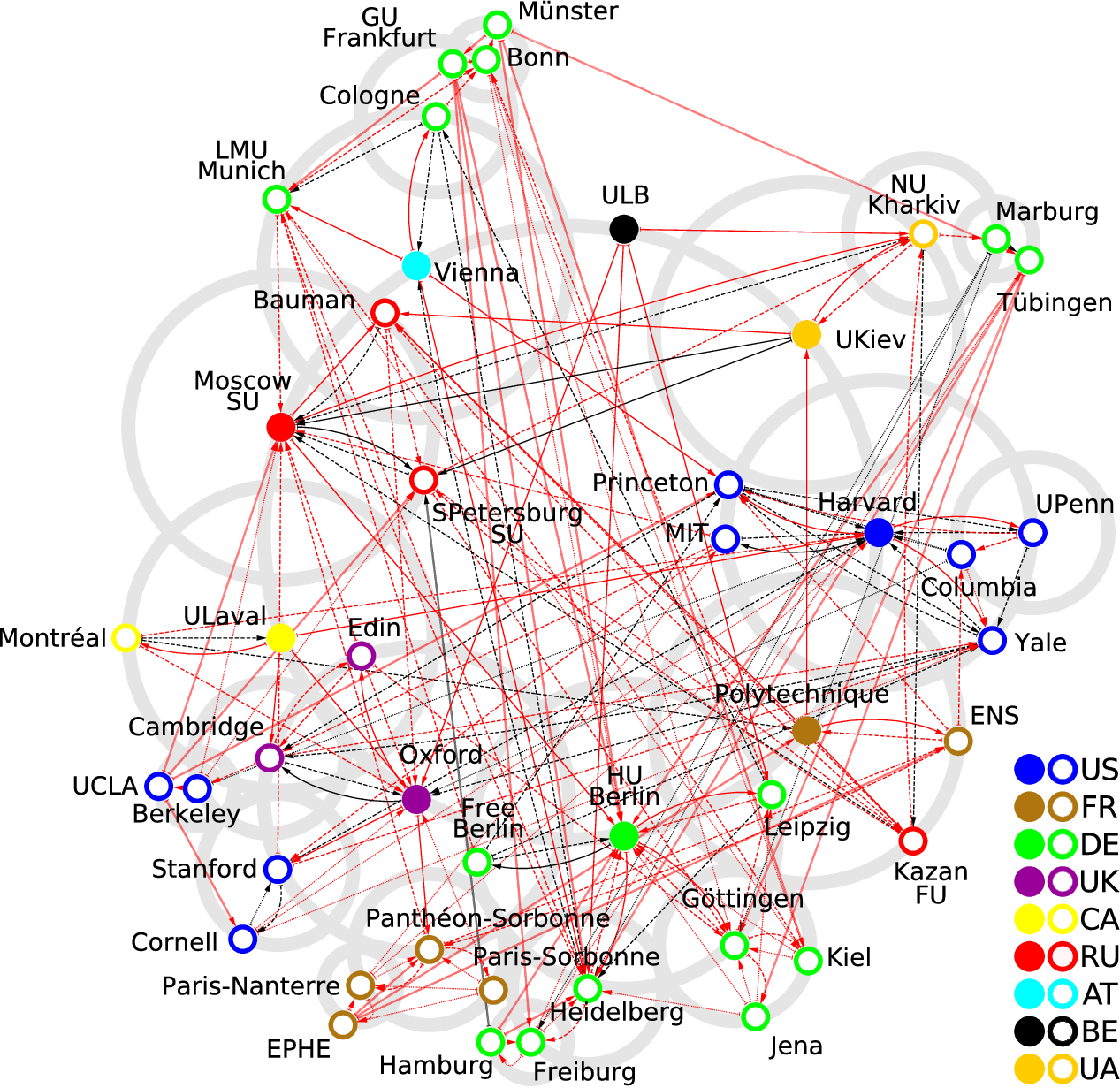}
}
\caption{Same as in Fig.~\ref{fig:ENFRDERU_en_netw} but from RUWIKI.
}
\label{fig:ENFRDERU_ru_netw}
\end{figure}

The network of top friends from RUWIKI is shown in
Fig.~\ref{fig:ENFRDERU_ru_netw}. The dominance of German
universities is well present here with 16 of them
followed by 10 of US, 4 of Russia and 3 of UK.
Three major hubs are clearly visible a German one centered around HU Berlin / Heidelberg / G\"ottingen, an US-UK one centered around Harvard and Oxford, and a Russian one centered around Moscow SU.

The analysis of this subsection allows to establish most close links
between top world universities. It also shows the dominance
of US and German universities.

\section{Reduced Google matrix averaged over 24 Wikipedia editions}

Above we have considered ranking and interactions from a view point
of a given edition. Using the reduced Google matrix approach it is possible to
perform an averaging over all 24 editions thus determining the
averaged cultural view on selected universities.
With this aim,
for each of the 24 Wikipedia editions listed in Tab.~\ref{tab:24editions},
we select the subset of articles devoted to the 100 first PageRank universities of WRWU
(see Tab.~\ref{tab:top100WRWU}). Then we compute the corresponding reduced Google matrix
$\bar{G}_\mathrm{R}$ averaged over  24 Wikipedia editions.
The averaging is defined by the relation
\begin{equation}
\bar{G}_\mathrm{R}=\displaystyle\frac{1}{24}
\sum_{E}G_\mathrm{R}^{(E)}
\label{eq:GRAVER}
\end{equation}
where $G_\mathrm{R}^{(E)}$ is the reduced Google matrix (\ref{eq:GR})
of the Wikipedia edition $E$.
Each one of the 24 reduced Google matrices is written in the same basis corresponding to
the ordered PageRank list of 100 universities.
For a given edition $E$,
reduced Google matrix entries corresponding to a link pointing toward an absent university in edition $E$ are set to 0 and 
reduced Google matrix columns corresponding
to absent universities in edition $E$ are filled with $1/100$ entries.
These contributions from absent universities
are added to the $G_\mathrm{pr}^{(E)}$ matrix components
of the full reduced Google matrices $G_\mathrm{R}^{(E)}$.

We note that the averaging of 24 $G_\mathrm{R}^{(E)}$ matrices with equal weights
gives us again the reduced Google matrix which
performs an averaging over different cultural
views of 24 editions.

The PageRank vector computed from the averaged reduced Google matrix
$\bar{G}_\mathrm{R}$ is presented in Tab.~\ref{tab:24wiki_top100_WRWU}.
We see that the rank order is changed comparing to
the $\Theta$-averaging (\ref{eq:theta})
with the top 10 PageRank universities given in Tab.~\ref{tab:top10WRWU}
(list of top 100 is given in Tab.~\ref{tab:top100WRWU}). 
We see that Harvard takes the first position instead
of the third one in Tab.~\ref{tab:top10WRWU}
and then Oxford and Cambridge are moved to second
and third positions in Tab.~\ref{tab:24wiki_top100_WRWU}.
The top ten universities of Tab.~\ref{tab:24wiki_top100_WRWU}
have overlap of 100\% with PageRank WRWU of  Tab.~\ref{tab:top10WRWU}
and 90\% with ARWU of Tab.~\ref{tab:top10ARWU}.
It can be discussed what ranking averaging
over 24 cultural view of editions is more appropriate:
with $\Theta$-averaging or with averaging of $G_\mathrm{R}^{(E)}$.
We think that both approaches are useful:
in $\Theta$-averaging all PageRanking vectors are
completely independent while averaging of $G_\mathrm{R}^{(E)}$
introduces some additional links which are not present
in certain network editions.

\begin{figure*}[h]
\centering
\resizebox{2\columnwidth}{!}{%
\includegraphics{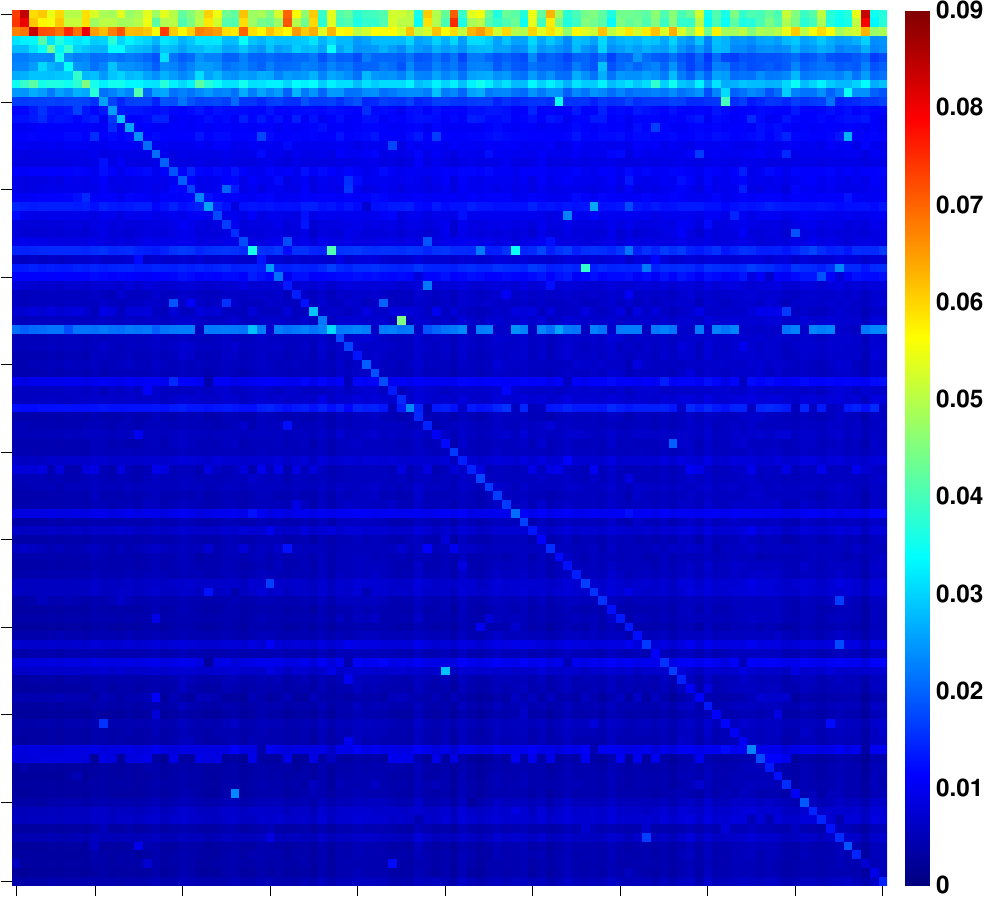}
\includegraphics{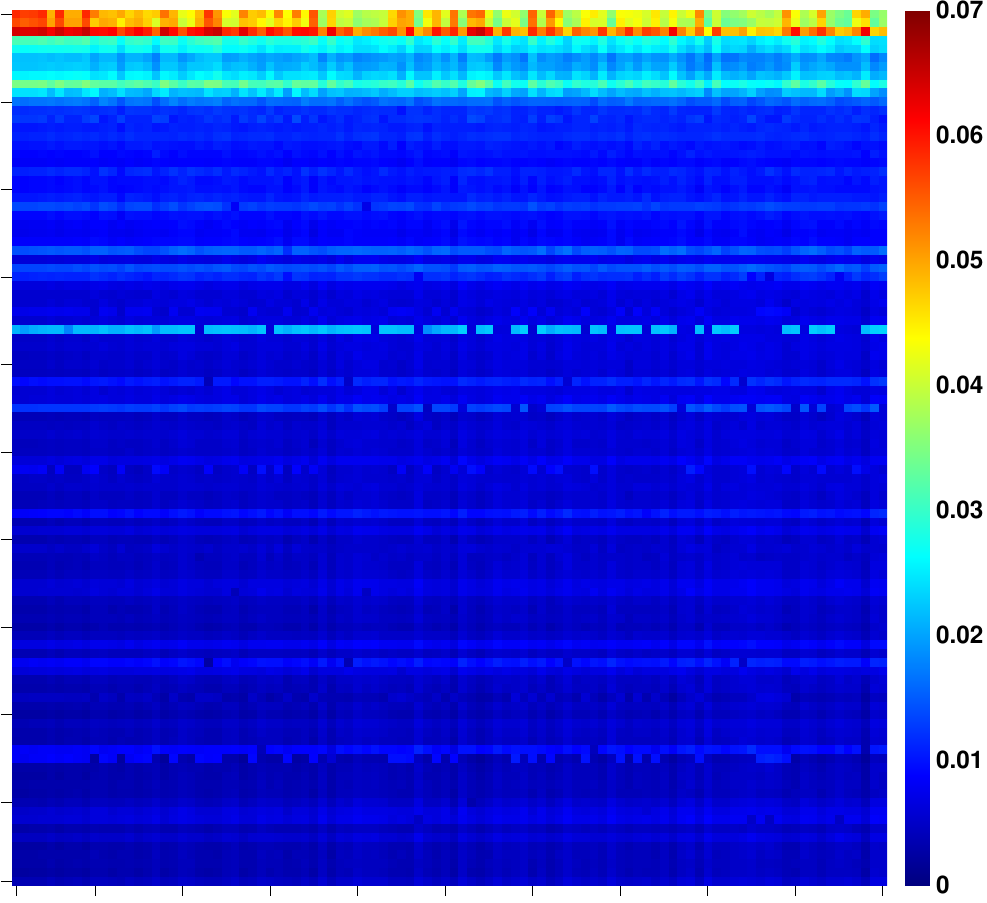}
}
\resizebox{2\columnwidth}{!}{%
\includegraphics{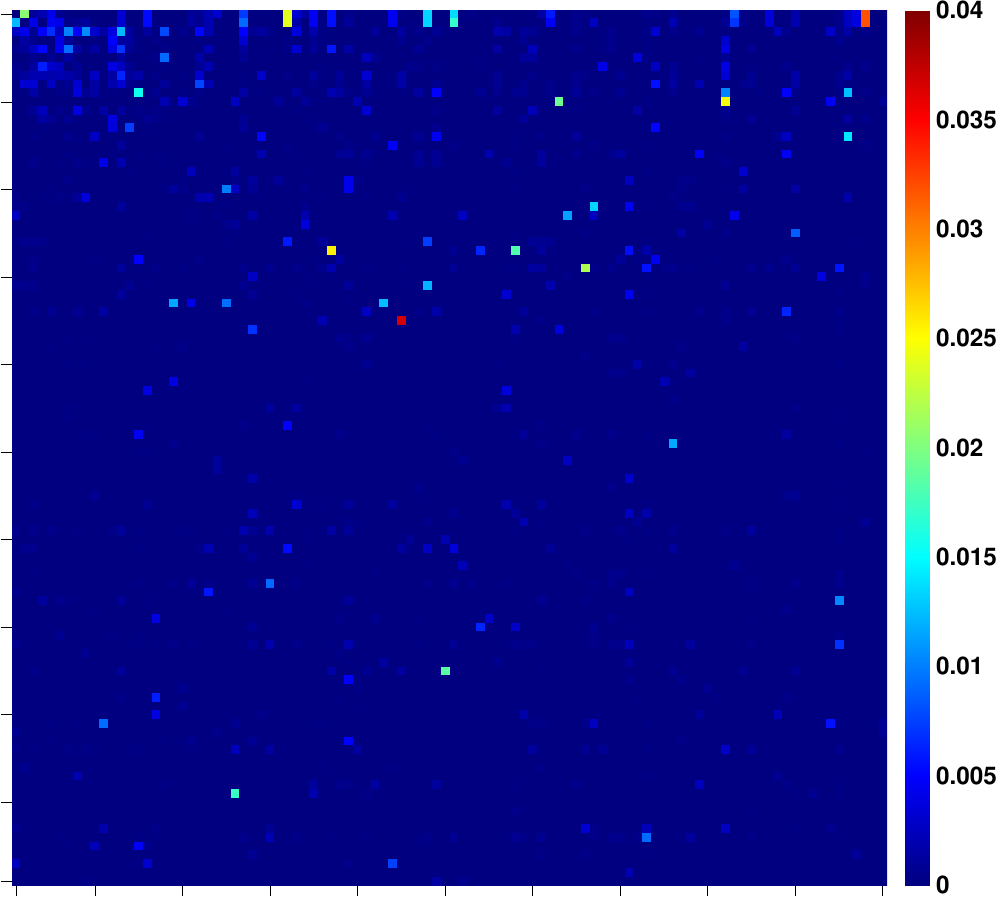}
\includegraphics{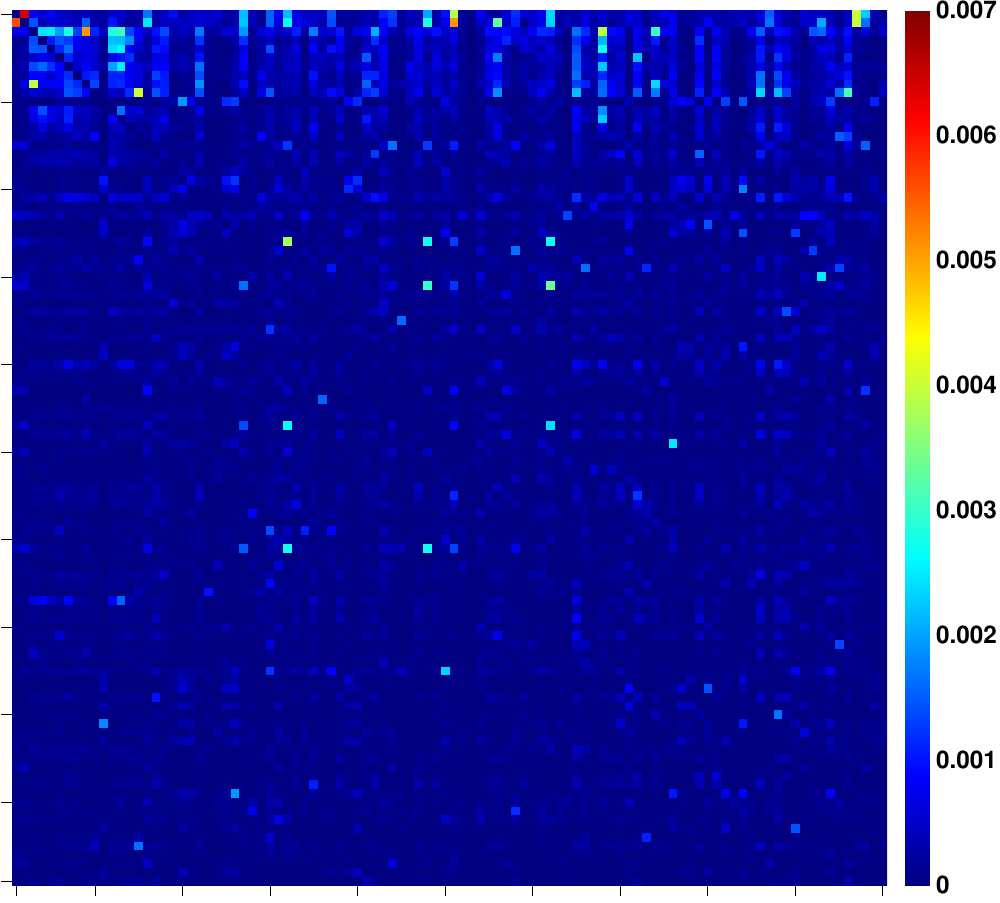}
}
\caption{Reduced Google matrix $\bar{G}_\mathrm{R}$
for  PageRank top 100 universities in WRWU (top 10 is in Tab.~\ref{tab:top10WRWU},
top 100 is in Tab.~\ref{tab:top100WRWU}) averaged over 24 Wikipedia editions.
Matrix entries correspond to universities ordered according
the top 100 WRWU PageRank order.
The full reduced Google matrix $\bar{G}_\mathrm{R}$ is presented in top left panel,
$\bar{G}_{\mathrm{pr}}$ in top right panel,
$\bar{G}_{\mathrm{rr}}$ in bottom left panel, and
$\bar{G}_{\mathrm{qrnd}}$ in bottom right panel.
The matrix weights are $W_\mathrm{R}=1$, $W_{\mathrm{pr}}=0.957$, $W_{\mathrm{rr}}=0.019$,
$W_{\mathrm{qr}}=0.024$, $W_{\mathrm{qrnd}}=0.015$. 
}
\label{fig:24wiki_top100_WRWU}
\end{figure*}

The obtained averaged Reduced Google matrix $\bar{G}_\mathrm{R}$
and its three components are shown in Fig.~\ref{fig:24wiki_top100_WRWU}.
In global the matrix structure is similar to those of individual
editions discussed above. The component $\bar{G}_{\mathrm{pr}}$
has the dominant weight but it is rather close to
the columns of PageRank vector and hence the interesting links
are determined by the components
$\bar{G}_{\mathrm{rr}}$ and $\bar{G}_{\mathrm{qr}}$
which have  comparable weights.
It is well seen that there are indirect links which are not
present between direct ones.

\begin{table}[b]
\caption{Top100 2017 WRWU ordered according to averaged reduced Google matrix.}
\resizebox{1\columnwidth}{!}{
\begin{tabular}{rll|rll}
\hline
Rank&\multicolumn{1}{p{1.5cm}}{\centering PageRank value}&University&Rank&\multicolumn{1}{p{1.5cm}}{\centering PageRank value}&University\\
\hline
\hline
1&0.0633191&Harvard University&51&0.00659655&Univ. of Colorado Boulder\\
2&0.0528587&Univ. of Oxford&52&0.00657266&Univ. of Glasgow\\
3&0.0518905&Univ. of Cambridge&53&0.00636839&Univ. of Toronto\\
4&0.0339304&MIT$^a$&54&0.0063255&Stockholm University\\
5&0.0301911&Columbia University&55&0.00624184&Univ. of T\"ubingen\\
6&0.0283041&Yale University&56&0.00609986&Univ. of Texas at Austin\\
7&0.0261455&Stanford University&57&0.00593539&Univ. of Virginia\\
8&0.024318&UC Berkeley$^b$&58&0.00584412&Imperial College London\\
9&0.0229394&Princeton University&59&0.00582829&Carnegie Mellon University\\
10&0.0215136&Univ. of Chicago&60&0.00579437&Univ. of Bonn\\
11&0.0197203&Univ. of Copenhagen&61&0.00570673&Univ. of Minnesota\\
12&0.0168679&HU Berlin$^c$&62&0.00567465&Keio University\\
13&0.0160439&Uppsala university&63&0.00557384&Univ. of Helsinki\\
14&0.0148231&Univ. of Tokyo&64&0.00548871&King's College London\\
15&0.0135633&Moscow State University&65&0.0054485&Univ. of Florida\\
16&0.0127305&Cornell University&66&0.00538279&Univ. of Zurich\\
17&0.0126064&HUJI$^d$&67&0.00536546&Univ. of Manchester\\
18&0.0125732&Univ. of Pennsylvania&68&0.00523928&McGill University\\
19&0.0120329&UCLA$^e$&69&0.00507791&Free University of Berlin\\
20&0.011732&Leiden University&70&0.00505635&Univ. of Washington\\
21&0.011246&Caltech$^f$&71&0.00505447&Univ. of Illinois U.-C.\\
22&0.0112404&New York University&72&0.00497258&Brown University\\
23&0.0112273&Univ. of Vienna&73&0.00491403&Univ. of Wisconsin-Madison\\
24&0.0104997&Univ. of Edinburgh&74&0.00485964&Northwestern University\\
25&0.0103698&Jagiellonian University&75&0.00480294&Univ. of Coimbra\\
26&0.0101557&Univ. of Bologna&76&0.00479832&Univ. of Oslo\\
27&0.0100089&Univ. of G\"ottingen&77&0.00477973&Univ. of Padua\\
28&0.00987766&Heidelberg University&78&0.00476805&Georgetown University\\
29&0.00982921&Univ. of Michigan&79&0.00475634&UNAM$^l$\\
30&0.00974263&Lund University&80&0.00468635&Boston University\\
31&0.00929623&LSE$^g$&81&0.0045985&Ohio State University\\
32&0.00918967&Johns Hopkins University&82&0.00458516&Michigan State University\\
33&0.00909002&Univ. of Warsaw&83&0.00452351&Univ. of Geneva\\
34&0.00902656&Seoul National University&84&0.00451385&Univ. of Marburg\\
35&0.00877768&Leipzig University&85&0.00433353&Univ. of Salamanca\\
36&0.00832413&Univ. of Munich$^h$&86&0.0042273&Univ. of Freiburg\\
37&0.00791791&Waseda University&87&0.00418341&Univ. of Arizona\\
38&0.0076835&Univ. College London&88&0.00417181&Univ. of Jena\\
39&0.00751886&Duke University&89&0.00415139&MLU$^m$\\
40&0.00718132&Sapienza$^i$&90&0.00401368&Univ. of St Andrews\\
41&0.00711981&ETH Zurich&91&0.00398415&TU Berlin$^n$\\
42&0.0071081&USC$^j$&92&0.00391916&UNC Chapel Hill$^o$\\
43&0.00693105&\'Ecole Polytechnique&93&0.00390789&Univ. of Tartu\\
44&0.00692597&Peking University&94&0.00388656&TU Munich$^p$\\
45&0.00682986&Al-Azhar University&95&0.00385376&Univ. of Sydney\\
46&0.00682254&\'Ecole Normale Sup\'erieure&96&0.00384341&UC San Diego$^q$\\
47&0.00680075&Kyoto University&97&0.00371085&Trinity College, Dublin\\
48&0.00666809&Charles University&98&0.00368454&Indiana University\\
49&0.00666454&SPbU$^k$&99&0.00355122& University of Notre Dame\\
50&0.00662585&Utrecht University&100&0.00353878& University of Kiel\\
\hline
\multicolumn{6}{p{1.65\columnwidth}}{
$^a$Massachusetts Institute of Technology,
$^b$University of California, Berkeley,
$^c$Humboldt University of Berlin,
$^d$Hebrew University of Jerusalem,
$^e$University of California, Los Angeles,
$^f$California Institute of Technology,
$^g$London School of Economics,
$^h$Ludwig Maximilian University of Munich,
$^i$Sapienza University of Rome,
$^j$University of Southern California,
$^k$Saint Petersburg State University,
$^l$National Autonomous University of Mexico,
$^m$Martin Luther University of Halle-Wittenberg,
$^n$Technical University of Berlin,
$^o$University of North Carolina at Chapel Hill,
$^p$Technical University of Munich,
$^q$University of California, San Diego
}
\end{tabular}
}
\label{tab:24wiki_top100_WRWU}
\end{table}

\begin{figure*}[h]
\centering
\resizebox{2\columnwidth}{!}{%
\includegraphics{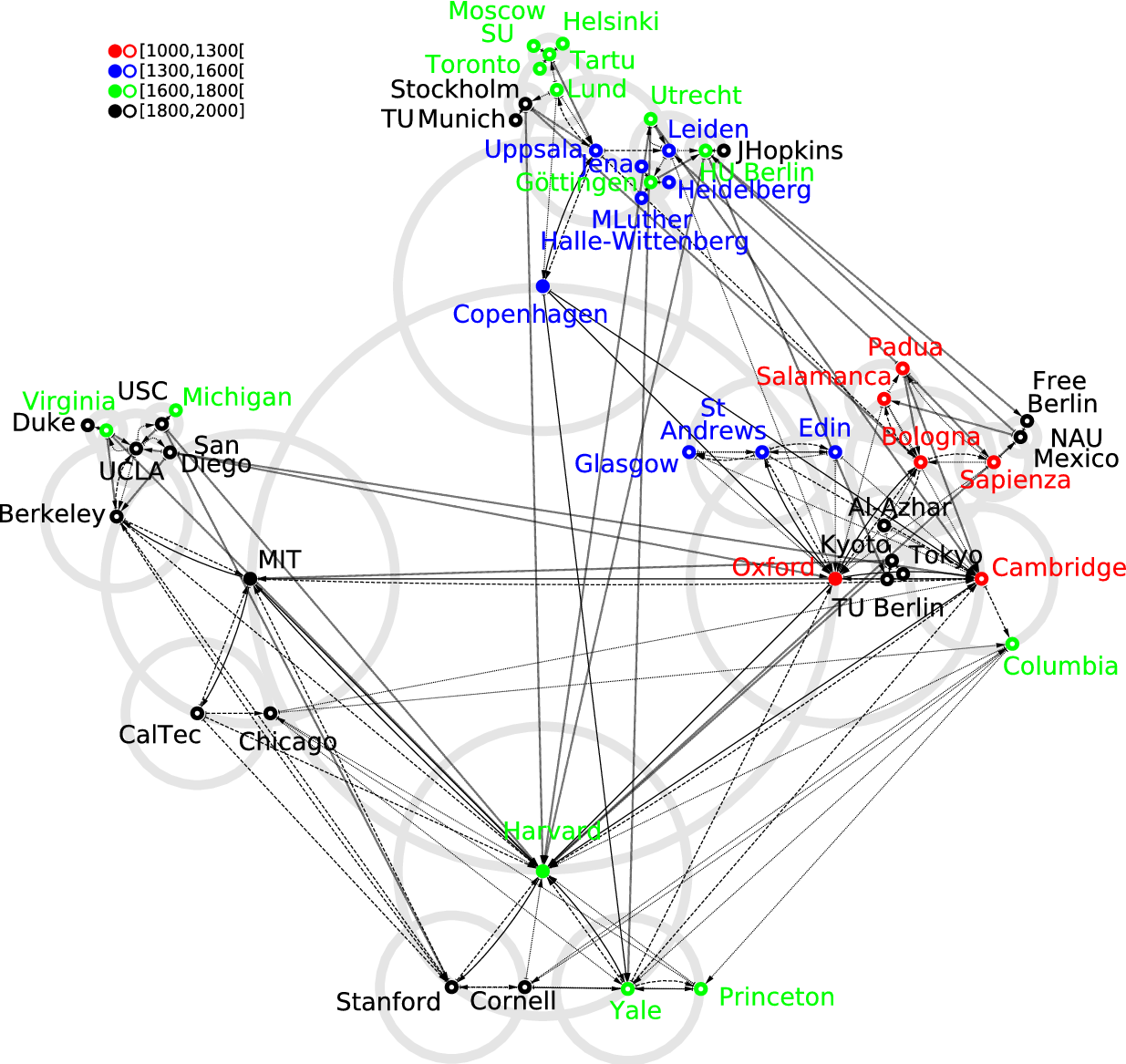}
}
\caption{Century reduced friendship network constructed for universities
  of PageRank top 100 list of WRWU (see Tab.~\ref{tab:top100WRWU},
  top 10 are in Tab.~\ref{tab:top10WRWU}), computed from
  $\bar{G}_{\mathrm{rr}} + \bar{G}_{\mathrm{qrnd}}$
  averaged over 24 Wikipedia editions.
  Color marks university founded at the same time (century) period
  given in years; color filled circles are time period leaders,
  open circles of the same color are universities from the same time period.
  We show 5 friendship levels (gray circles).
  Links originating from 1st level universities are presented by solid lines,
  from 2nd level by dashed lines, from 3rd level by doted lines,
  and from 4th or 5th level by ``\textbackslash'' symbol lines.
}
\label{fig:24wiki_top100_period_WRWU_netw}
\end{figure*}

From the matrix of direct and indirect links
$\bar{G}_{\mathrm{rr}} + \bar{G}_{\mathrm{qrnd}}$ we construct
the interaction friendship network between above considered 100 universities
divided by certain groups. Such a network takes into account
cultural views of all 24 editions. We now show all links by the same black color
since after averaging over 24 editions
there is a significant mixture of direct and indirect links.

In our first division, we mark universities
by foundation time (century) periods: red for foundation years
from 1000 to 1300 AD, blue from 1300 to 1600 AD,
green from 1600 to 1800 AD and black from 1800 to 2000 AD.
Each time period has its leader
taken as a university with highest rank position
in this period. The resulting network of friends
in shown in Fig.~\ref{fig:24wiki_top100_period_WRWU_netw}.
This network shows an interesting evolution of
interactions between universities through 10 centuries:
the cluster of universities founded in 11th-13h centuries, marked in red,
is formed mainly by UK and Italian universities (one from Spain,
group leader is Oxford). This cluster transfers its influence via
interaction and links to next 14th-16th centuries universities, marked in blue, which are mainly from northern countries including
Scotland, Denmark, Germany, Sweden and Netherlands (group leader
is Copenhagen). The influence of these universities is transfered to 17th-18th centuries
universities, marked in green, being mainly near the blue cluster
and located in the same countries with
addition of Moscow in Russia,
Tartu in Estonia, Helsinki in Finland; another group of green universities
of this time period is linked with Oxford and Cambridge and is located
mainly on US east coast (Harvard, Yale, Princeton; Columbia is directly linked to Cambridge).
The university of next centuries 19th-20th, marked in black
(MIT is group leader),
are mainly located in US but new universities of this time period
appear also in Japan (Tokyo, Kyoto), Egypt (Al-Azhar), Germany (TU Munich,
TU Berlin, Free Berlin) and Sweeden (Stockholm).
Thus the obtained friendship network
provides a compact description of
world universities development through 10 centuries
taking into account the balanced view of 24 cultures
presented by Wikipedia editions.

\begin{figure*}[h]
\centering
\resizebox{2\columnwidth}{!}{%
\includegraphics{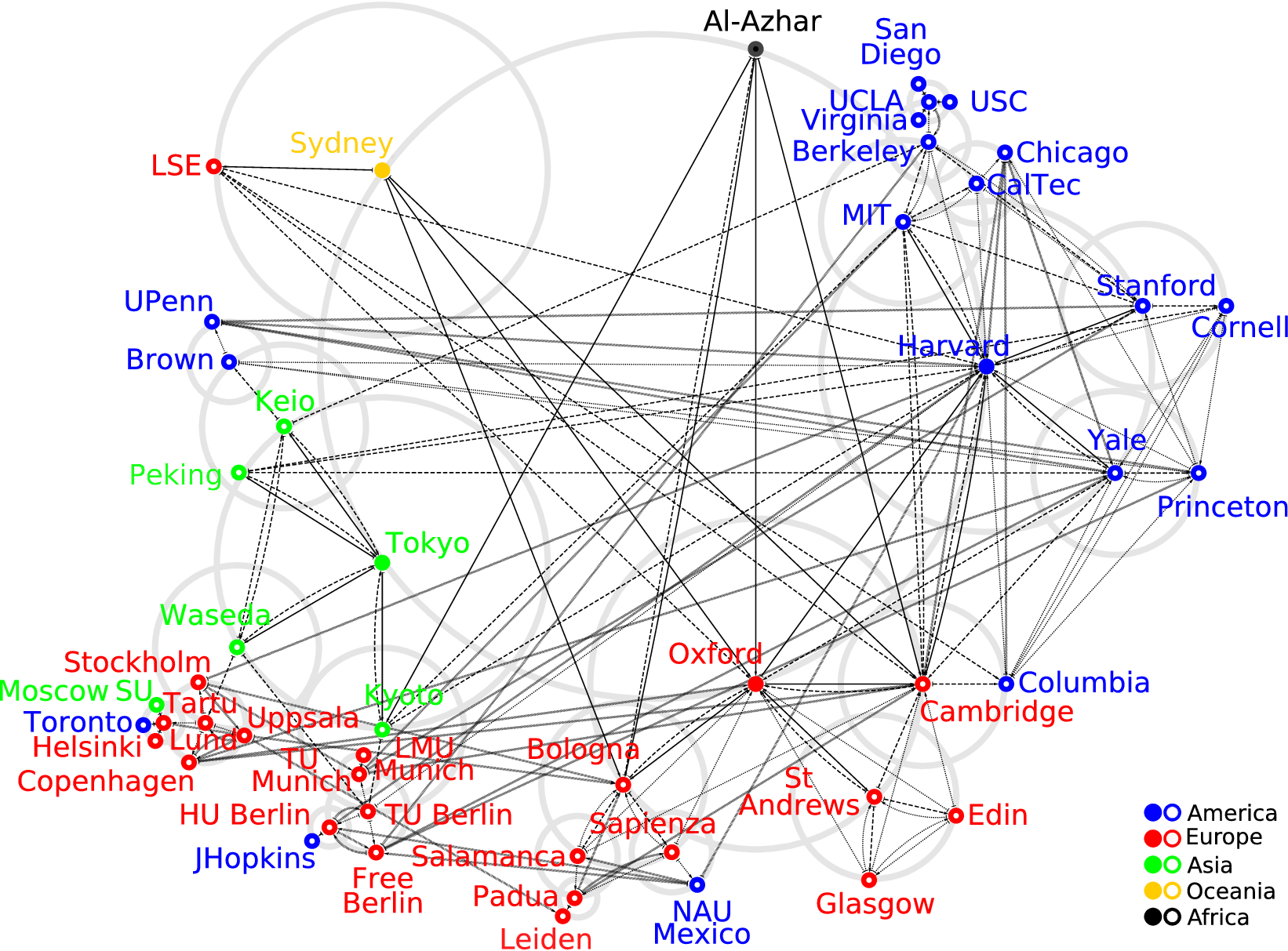}
}
\caption{Continent reduced friendship network constructed for universities
  of PageRank top 100 list of WRWU (see Tab.~\ref{tab:top100WRWU},
  top 10 are in Tab.~\ref{tab:top10WRWU}), computed from
  $\bar{G}_{\mathrm{rr}} + \bar{G}_{\mathrm{qrnd}}$
  averaged over 24 Wikipedia editions.
  Color marks university of the same continent; color filled circles are continent leaders,
  open circles of the same color are universities from the same continent.
  We obtain 5 friendship levels (gray circles). Links originating from 1st level universities are presented by solid lines, from 2nd level by dashed lines, from 3rd level by doted lines, and from 4th or 5th level by ``\textbackslash'' symbol lines.
}
\label{fig:24wiki_top100_country_WRWU_netw}
\end{figure*}

In our second division we mark universities by
continent location: blue for America, red for Europe,
green for Asia (Russia is attributed to Asia),
yellow for Oceania and black for Africa.
Again color group leaders are marked by full circles.
The friendship network obtained from
$\bar{G}_{\mathrm{rr}} + \bar{G}_{\mathrm{qrnd}}$
is shown in Fig.~\ref{fig:24wiki_top100_country_WRWU_netw}.
The network has two large clusters of European universities (red)
and US universities (blue).
The group of university in Asia (green)
is mainly linked between themselves
having secondary links with Europe and US.
Oceania (Sydney) and Africa (Al-Azhar)
are represented only by one university.
This network structure clearly shows the
influence of European and US universities
with emerging group of new group of Asian
universities with strong internal links.

\section{Discussion}

In this work we performed analysis of
ranking and interactions of world universities
from directed networks of 24 Wikipedia editions
dated by May 2017. Our results show that
obtained WRWU2017 with PageRank algorithm 
averaged over 24 editions 
gives a reliable ranking of universities
with 60\% overlap with top 100 of
ARWU2017 (Shanghai ranking) \cite{shanghai}.
At the same time WRWU2017 highlights in a stronger
way the significance of historical
path of a given university over centuries.
There are certain changes in WRWU2017 version comparing 
to WRWU2013 version demonstrating appearance
of new   universities with  time evolving
and with the increase of the number of Wikipedia articles
in the 24 selected editions.
A comparison of WRWU and ARWU ranking positions 
for specific universities (e.g. Rice University)
shows that the Wikipedia visibility can be significantly 
improved in certain cases.

We also performed an additional analysis
based on the reduced Google matrix (REGOMAX) algorithm
\cite{greduced,politwiki}. This approach allowed
us to establish direct and indirect links between 
universities and world countries.
As a result we obtain the sensitivity and influence
of specific universities on world countries 
as it is seen by Wikipedia.
The REGOMAX method allows to perform a democratic and uniform
averaging over cultural views of 24 language editions
and obtain a balanced cultural view
on the interactions of top world universities through ten centuries
of their historical development
as well as their influence over continents.

Finally we stress that the WRWU method is independent of
various personal opinions being based on purely
mathematical and statistical analysis of the 
Wikipedia database. We think that this approach
can be very complimentary to ARWU and other 
university rankings. We note that
Wikipedia articles of universities usually appear
in the top line (or lines) of Google search.
As a result the world visibility of
a given university can be publicly and 
freely broadcast all over the world
increasing visibility of certain universities.
We estimate that the improvement of Wikipedia
articles of certain universities 
(e.g. we found a low visibility of French universities)
can be an efficient way to increase their
world visibility, attractivity and influence.
Such an improvement is rather inexpensive
and can be performed by a small group of 
researchers and students having 
knowledge in languages, history,
computer and network sciences.
We think that this approach can be 
complementary to
various government projects
which aim to increase visibility of national universities
(like e.g. \cite{freducation,ru5top100}).

\begin{acknowledgement}
\textbf{Acknowledgments}\\
This work was supported
by the French ``Investissements d’Ave\-nir'' program, 
project ISITE-BFC (contract ANR-15-IDEX-0003)
and
by the Bourgogne Franche-Comt\'e Region 2017-2020 APEX project (conventions
2017Y-06426, 2017Y-06413, 2017Y-07534; 
see \url{http://perso.utinam.cnrs.fr/~lages/apex/}).
The research of DLS is supported in part by  the Programme Investissements
d'Avenir ANR-11-IDEX-0002-02, 
reference ANR-10-LABX-0037-NEXT (project THETRACOM).
\end{acknowledgement}



%
%
%


\renewcommand{\thetable}{SI\arabic{table}}
\renewcommand{\thefigure}{SI\arabic{figure}}

\setcounter{figure}{0}
\setcounter{table}{0}

\pagebreak
\onecolumn
\section*{Supplementary Information: World influence and interactions of  universities from Wikipedia networks}
\label{SI}

\begin{figure}[h]
\resizebox{\columnwidth}{!}{%
\includegraphics{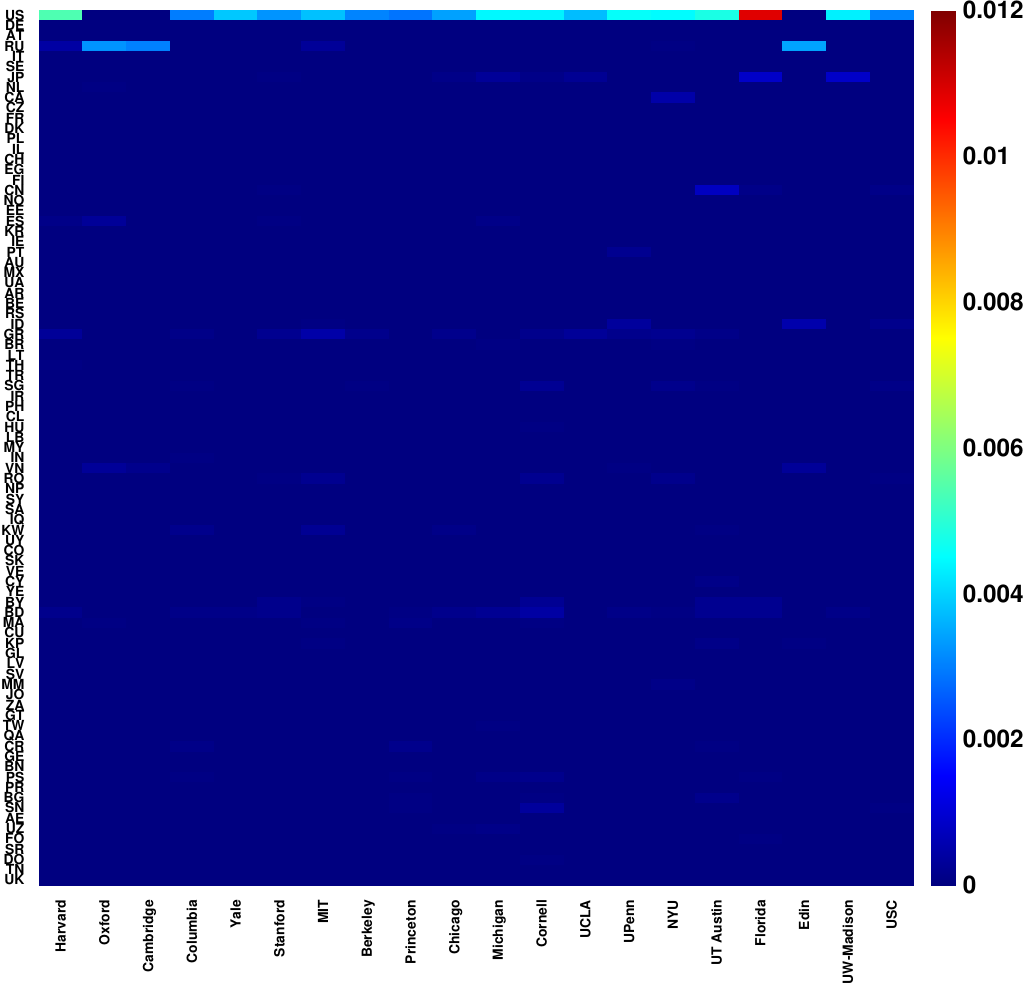}
\includegraphics{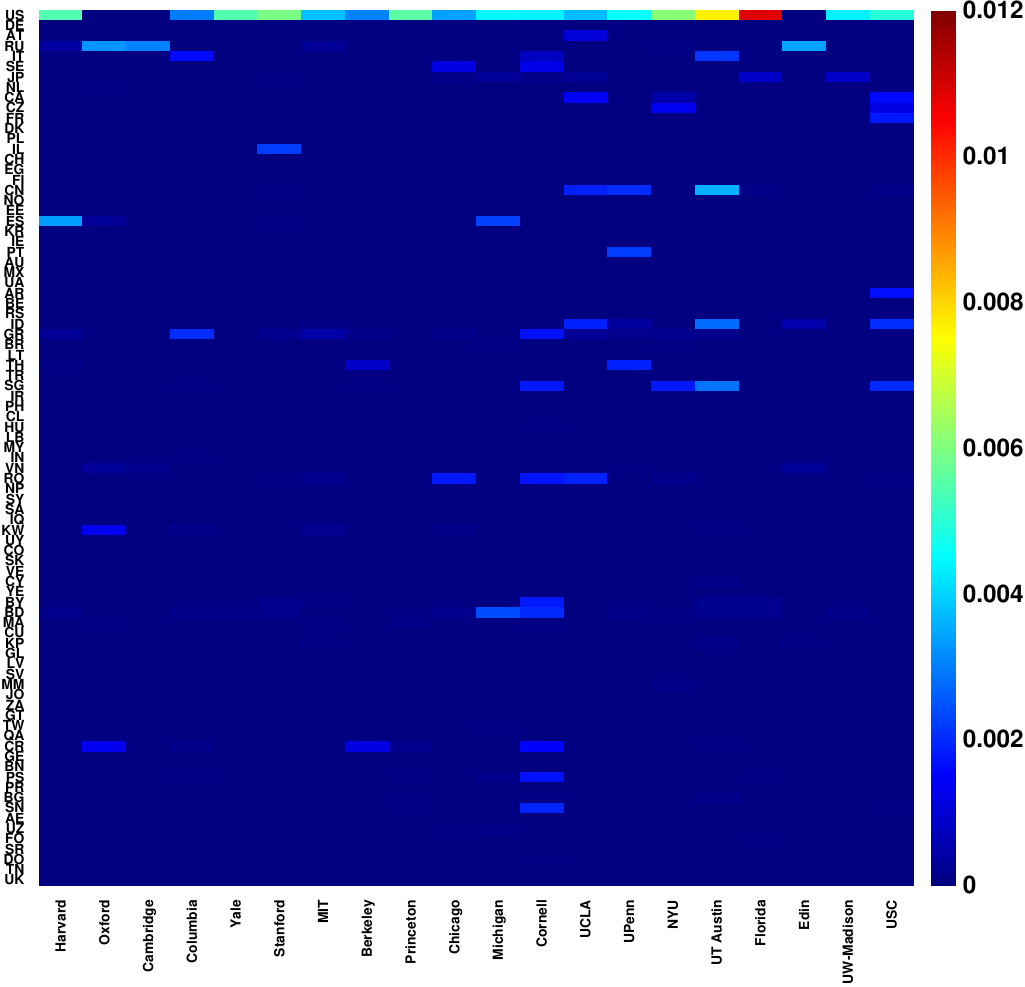}
}
\caption{University to Country sector of the $G_{\mathrm{qr}}$ matrix (left panel) and of the $G_{\mathrm{rr}}+G_{\mathrm{qrnd}}$ composite matrix (right panel). See Fig.~\ref{fig:ENWIKIGRU20C85}, bottom left (right) panel for the complete $G_{\mathrm{rr}}$ ($G_{\mathrm{qr}}$) matrix.
}
\label{fig:ENWIKIGRU20C85_UtoC}
\end{figure}

\begin{figure}[h]
\centering
\resizebox{0.6\columnwidth}{!}{%
\includegraphics{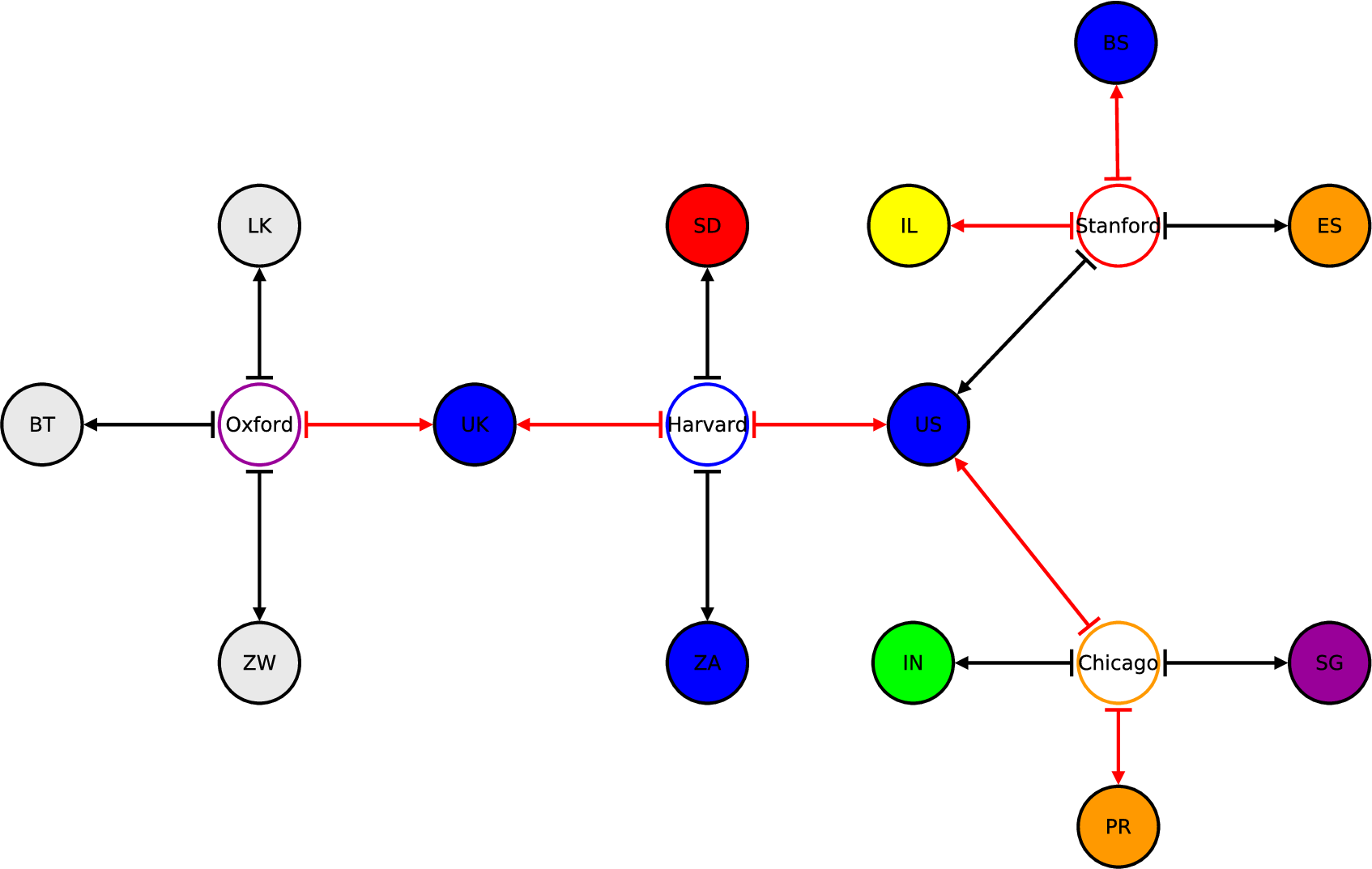}
}
\caption{Reduced network from $G_{\mathrm{rr}}+G_{\mathrm{qrnd}}$ associated to the top20 ENWRWU and 240 countries listed in Tab.~\ref{tab:CC240}.
For each regional leaders, Stanford University, University of Chicago, Harvard University, University of Oxford, the four strongest links to one of the 240 countries are presented. Universities (countries) are represented by empty (full) nodes.
The color code for countries depends on the main spoken language: \blu{blue} for English, \red{red} for Arabic, \ora{orange} for Spanish, \vio{violet} for Chinese, \gre{green} for Hindi, \yel{yellow} for Hebrew, and \gra{gray} for others. Red links are purely hidden links and black ones are at least present in the adjacency matrix.
}
\label{fig:ENWIKIGsumU20C240}
\end{figure}

\begin{figure}[h]
\resizebox{\columnwidth}{!}{%
\includegraphics{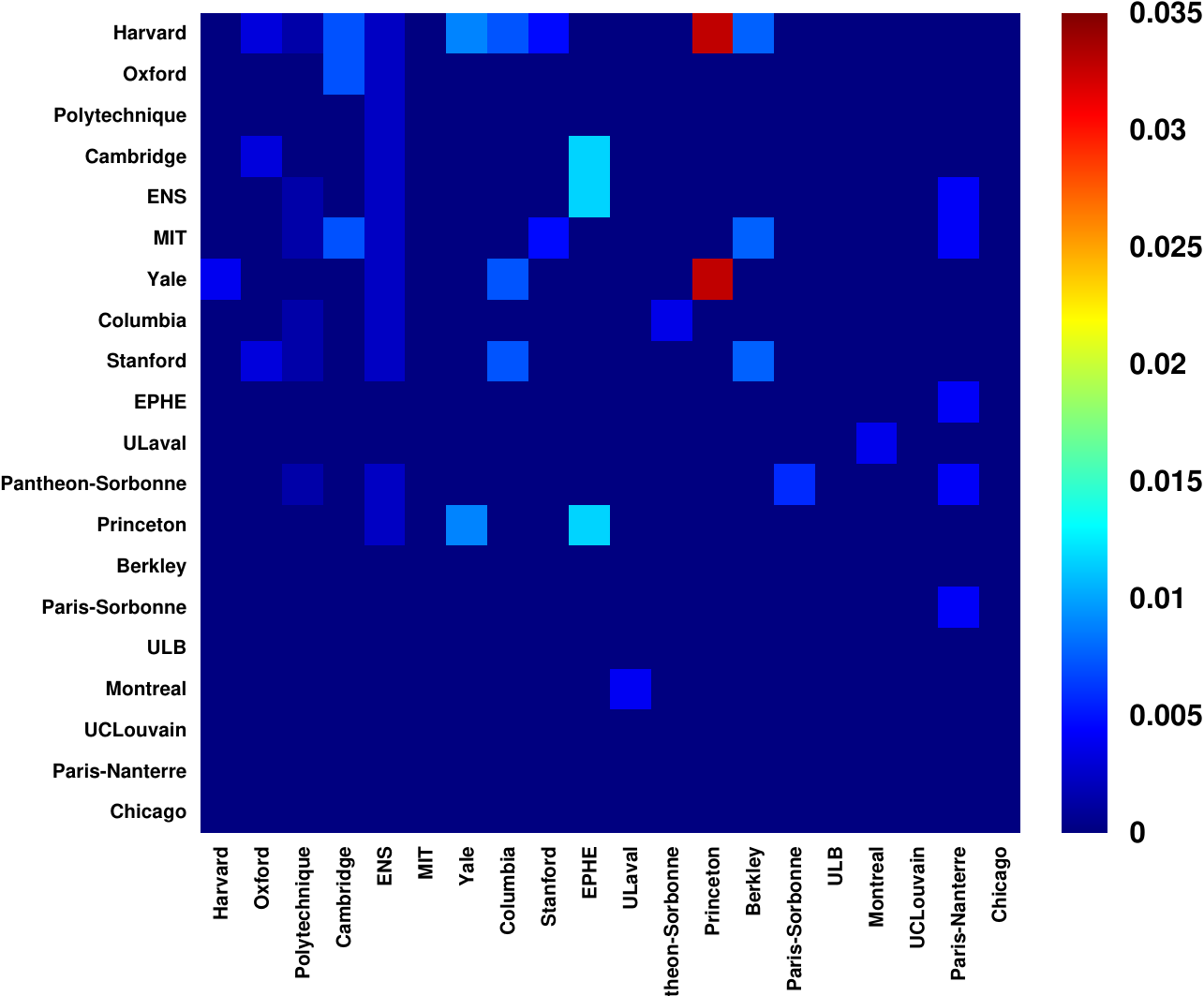}
\includegraphics{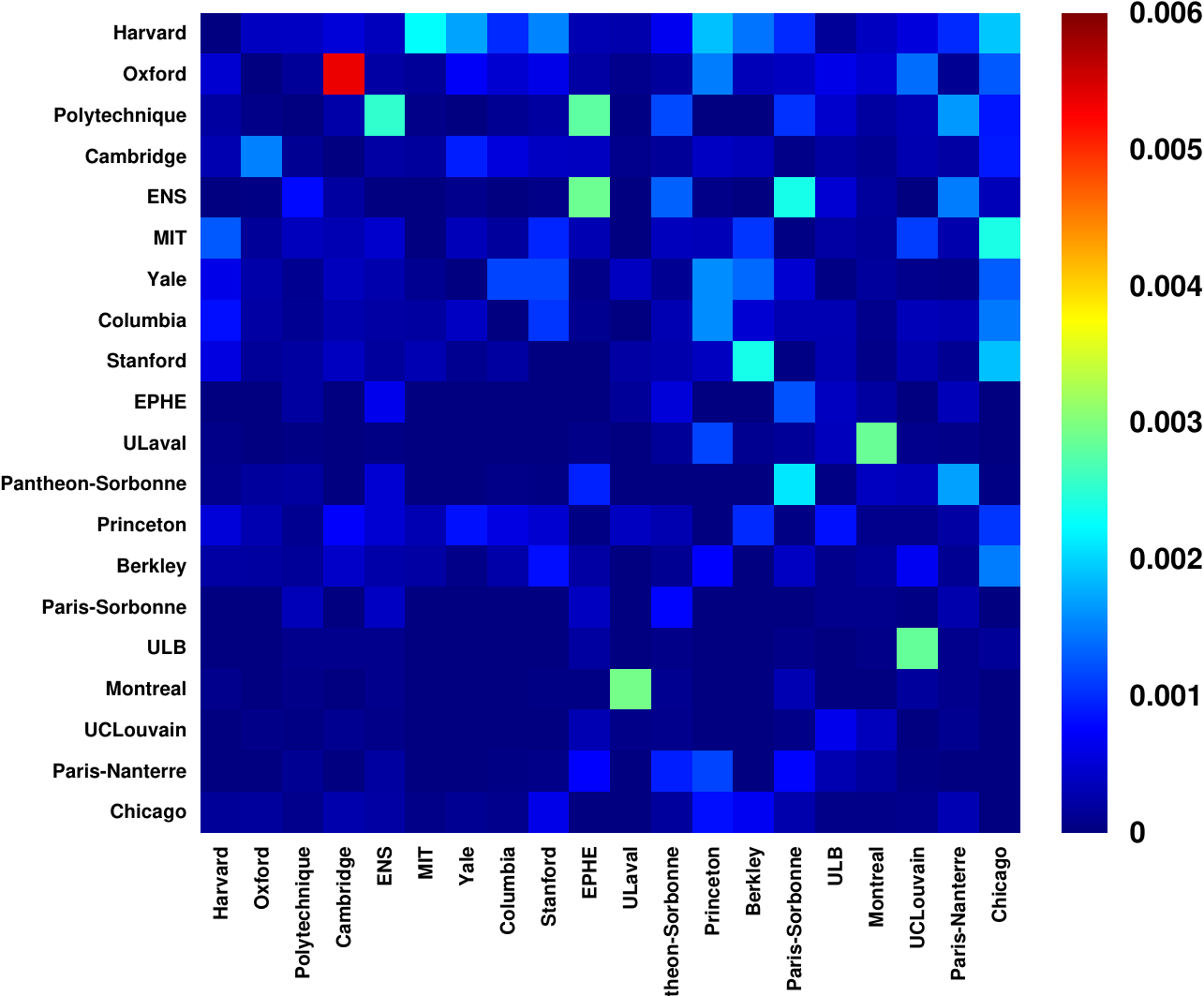}
}
\caption{Matrices $G_{\mathrm{rr}}$ (left panel) and $G_{\mathrm{qrnd}}$ (right panel) for top20 FRWRWU (Tab.~\ref{tab:top20FRWRWU}).
The matrix weights are $W_{\mathrm{rr}} = 0.01404$ and $W_{\mathrm{qrnd}} = 0.00746$.}
\label{fig:FRU20Grr_Gqr}
\end{figure}

\begin{figure}[h]
\resizebox{\columnwidth}{!}{%
\includegraphics{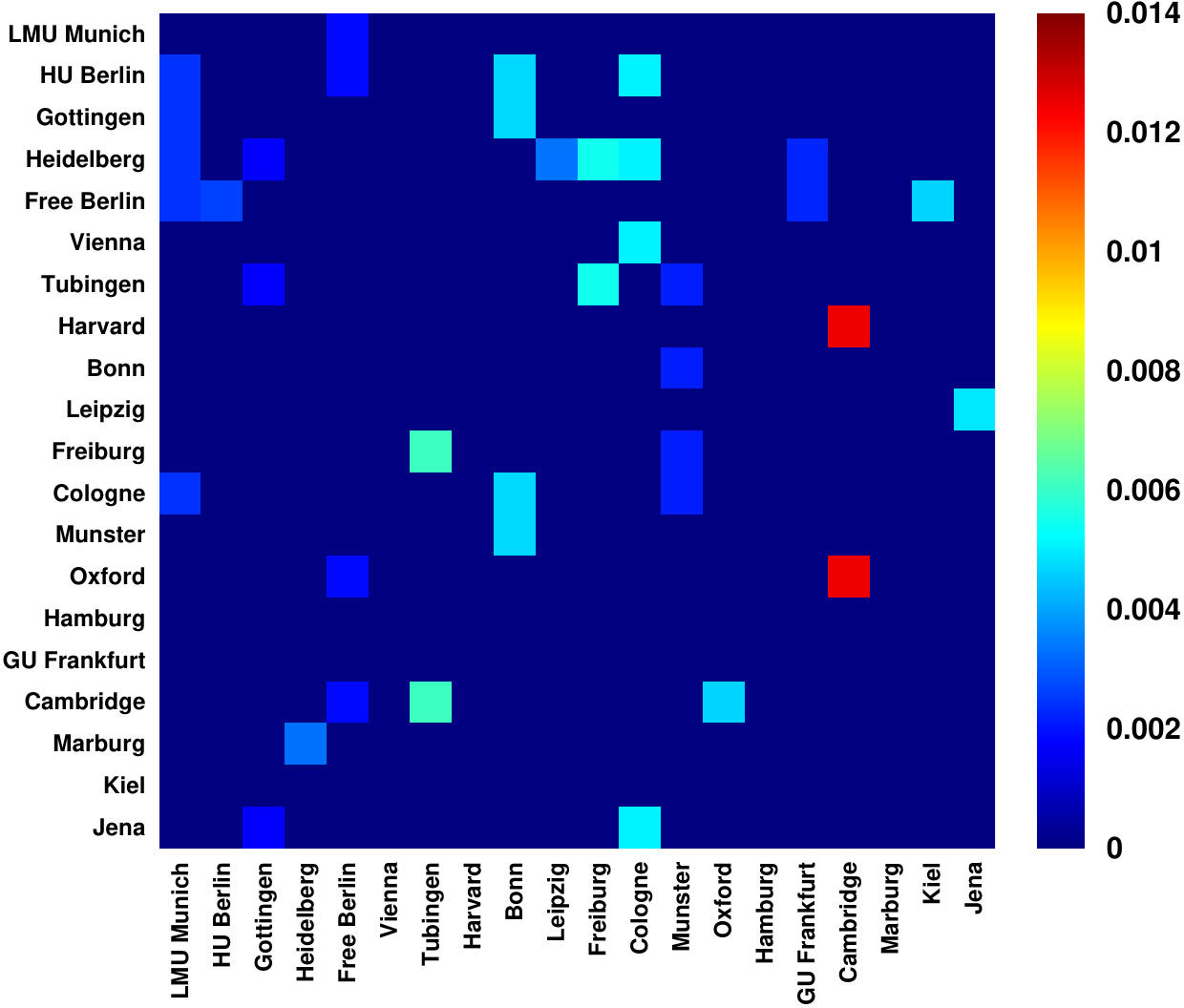}
\includegraphics{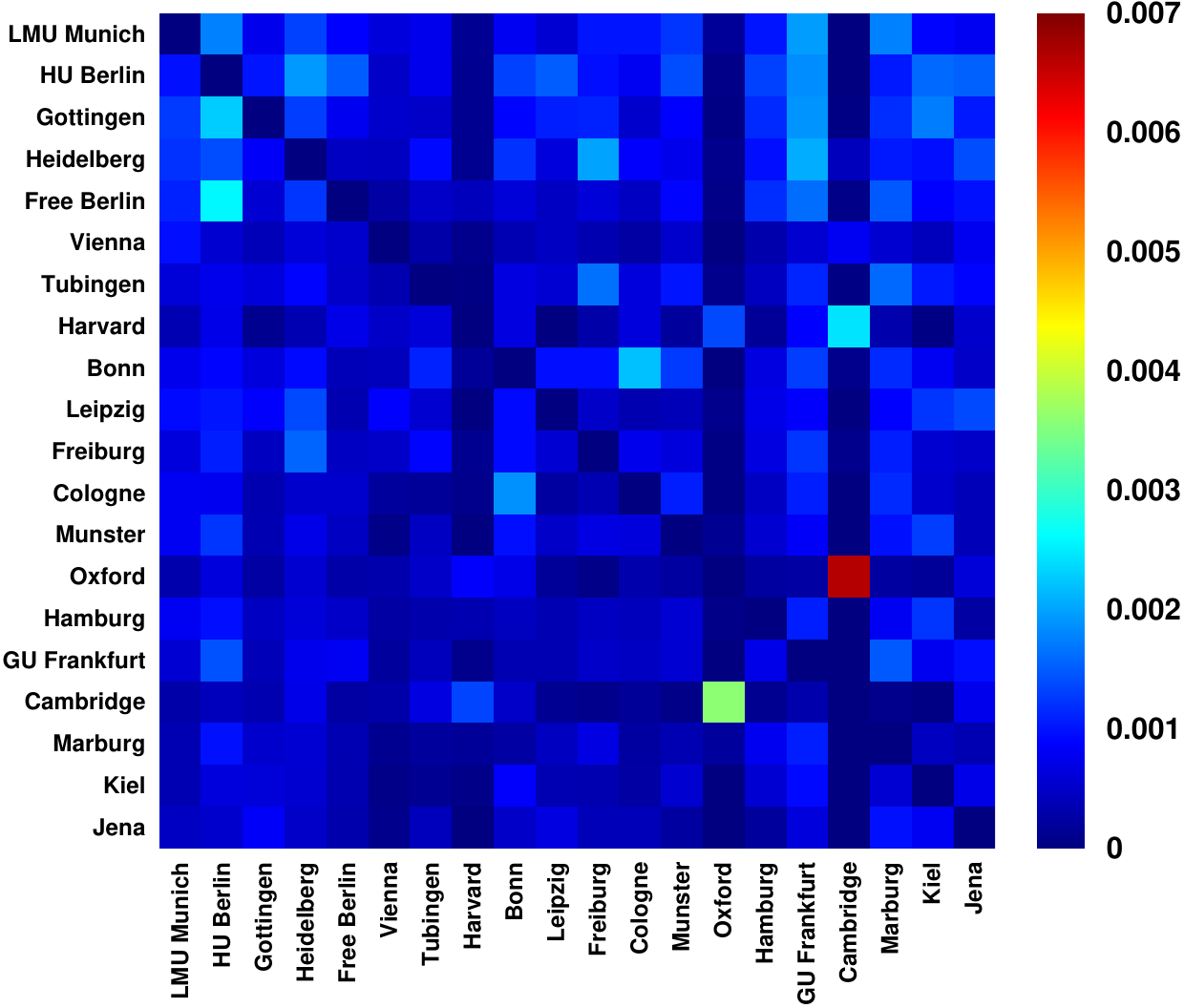}
}
\caption{Matrices $G_{\mathrm{rr}}$ (left panel) and $G_{\mathrm{qrnd}}$ (right panel) for top20 DEWRWU (Tab.~\ref{tab:top20DEWRWU}).
The matrix weights are $W_{\mathrm{rr}} = 0.00746$ and $W_{\mathrm{qrnd}} = 0.01280$.}
\label{fig:DEU20Grr_Gqr}
\end{figure}

\begin{figure}[h]
\resizebox{\columnwidth}{!}{%
\includegraphics{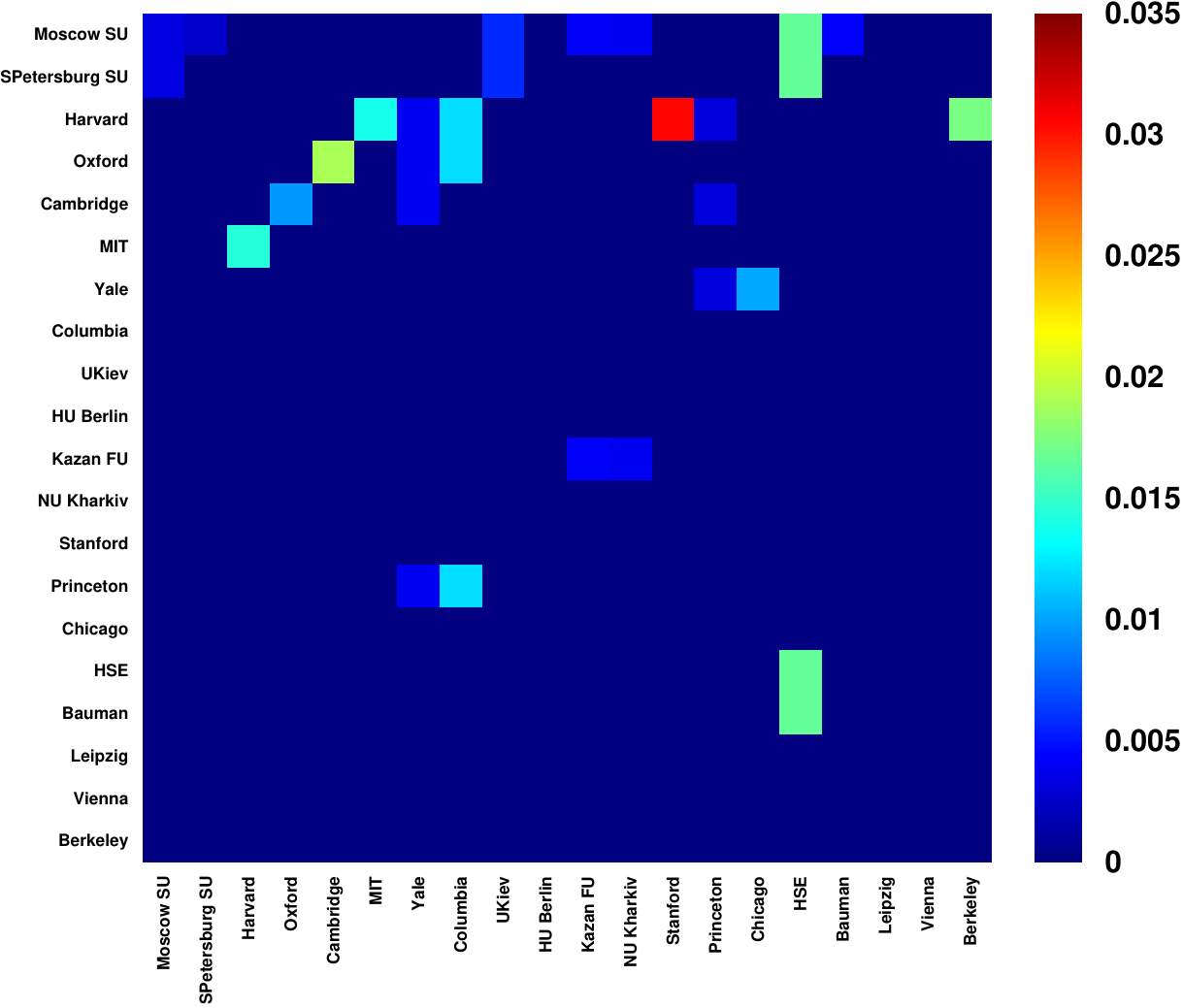}
\includegraphics{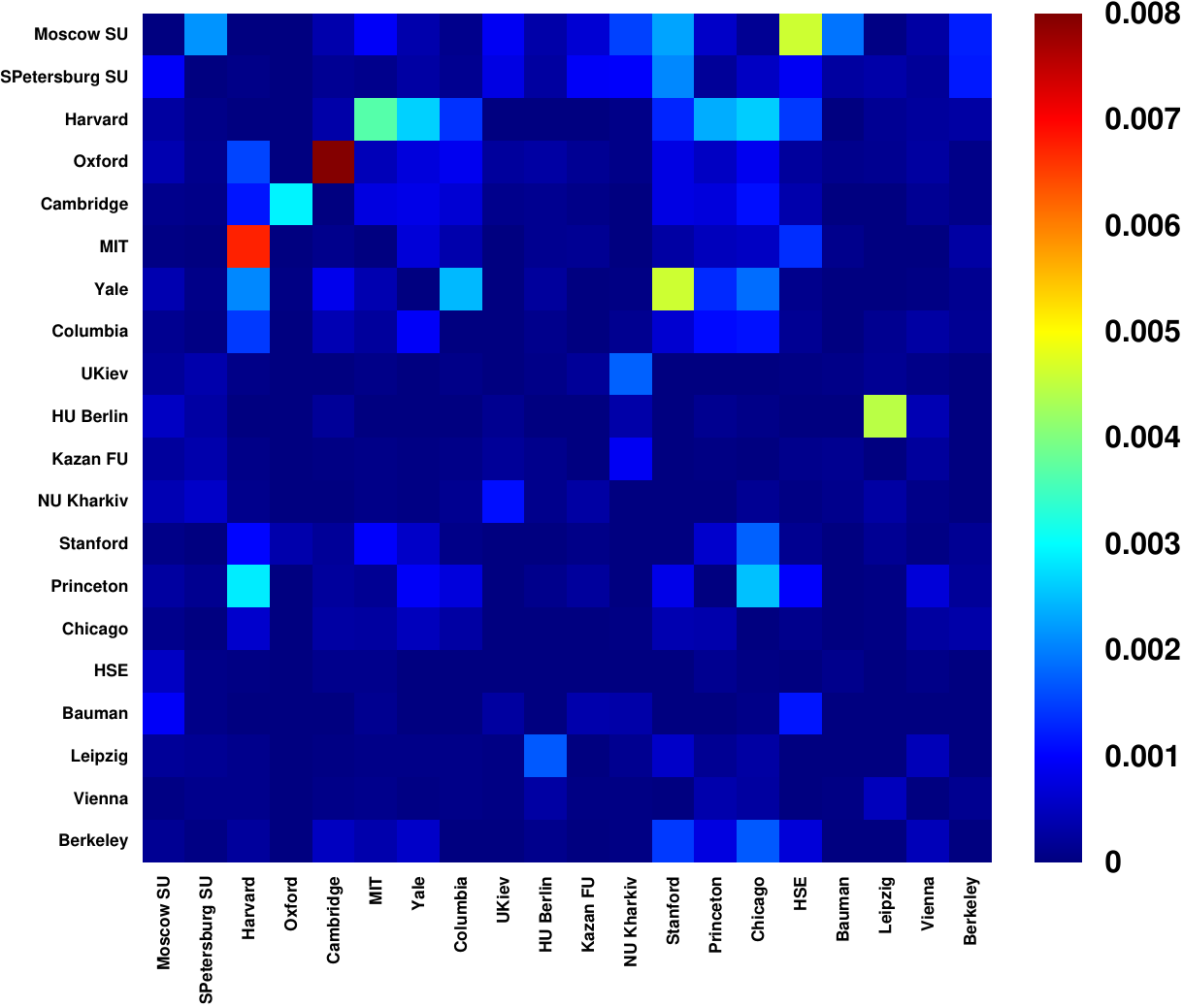}
}
\caption{Matrices $G_{\mathrm{rr}}$ (left panel) and $G_{\mathrm{qrnd}}$ (right panel) for top20 RUWRWU (Tab.~\ref{tab:top20RUWRWU}).
The matrix weights are $W_{\mathrm{rr}} = 0.01417$ and $W_{\mathrm{qrnd}} = 0.00798$.}
\label{fig:RUU20Grr_Gqr}
\end{figure}

\begin{figure}[h]
\resizebox{\columnwidth}{!}{%
\includegraphics{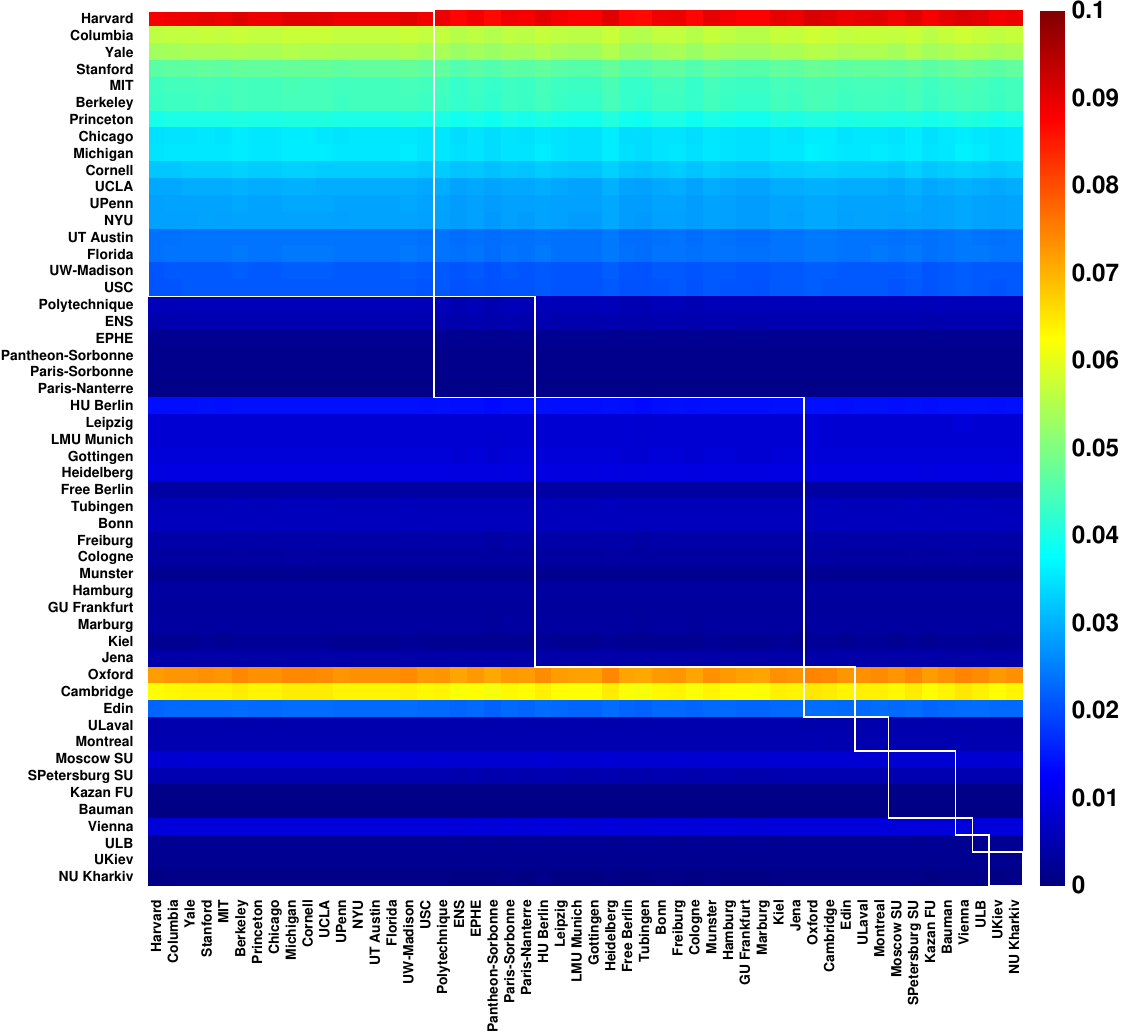}
\includegraphics{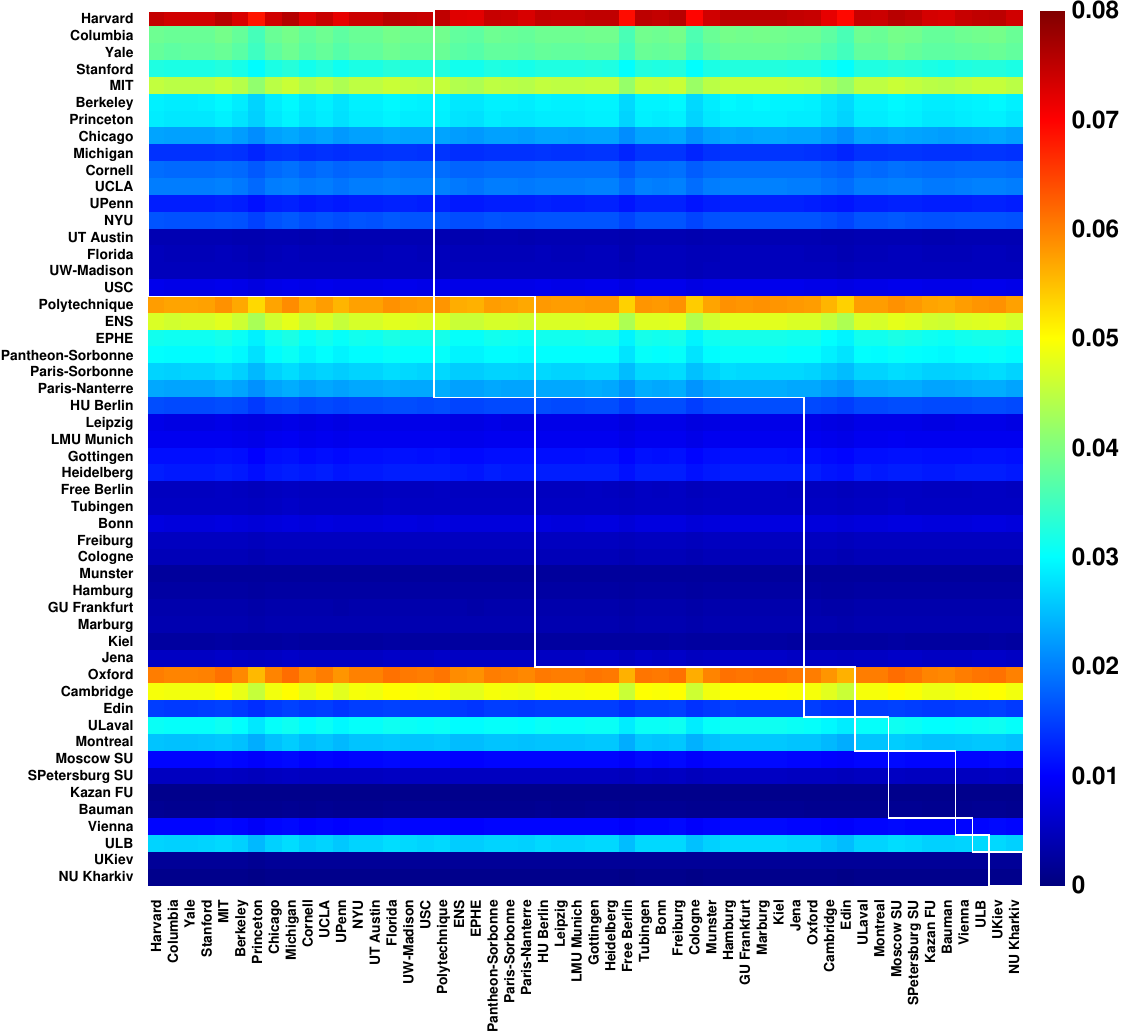}
}
\resizebox{\columnwidth}{!}{%
\includegraphics{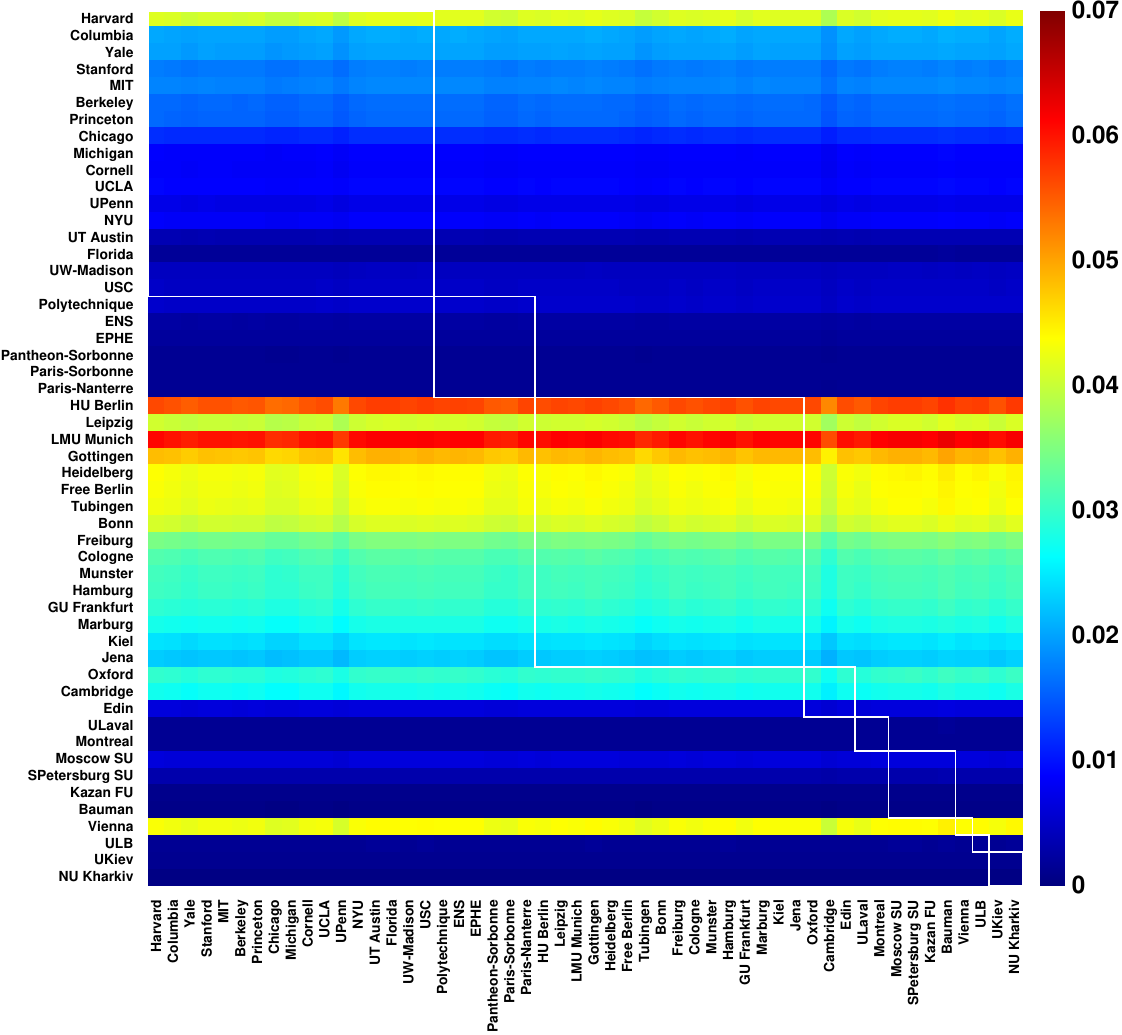}
\includegraphics{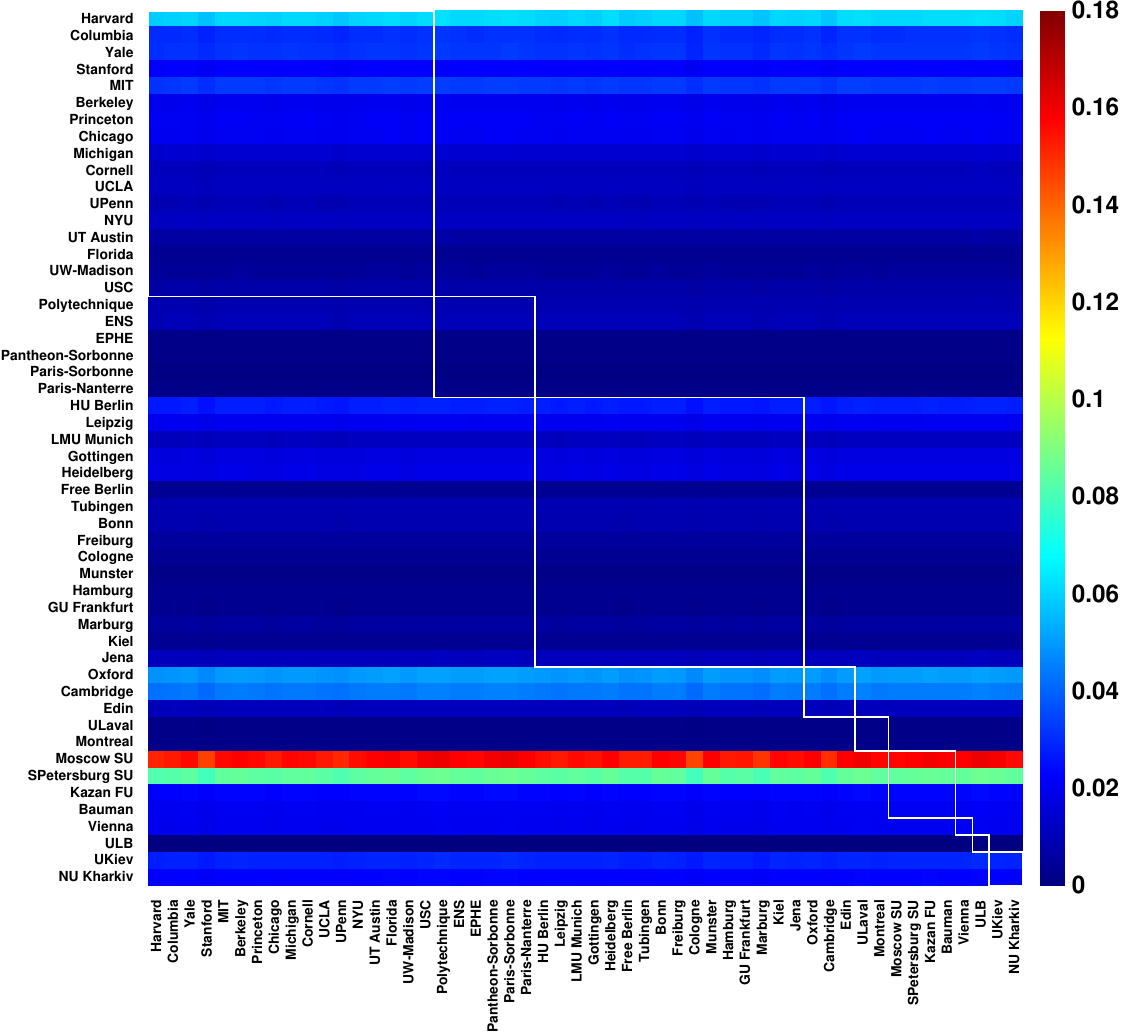}
}
\caption{$G_{\mathrm{pr}}$ part of the reduced Google matrix $G_{\mathrm{R}}$ for universities listed in Tab.~\ref{tab:top52ENFRDERUWRWU} computed from EN (top left), FR (top right), DE (bottom left), and RU (bottom right) Wikipedia editions. Matrix weights are $W_{\mathrm{pr}}\simeq0.956$ (EN), $0.966$ (FR), $0.96$ (DE) and $0.963$ (RU).
}
\label{fig:ENFRDERU_G_pr}
\end{figure}

\begin{figure}[h]
\resizebox{\columnwidth}{!}{%
\includegraphics{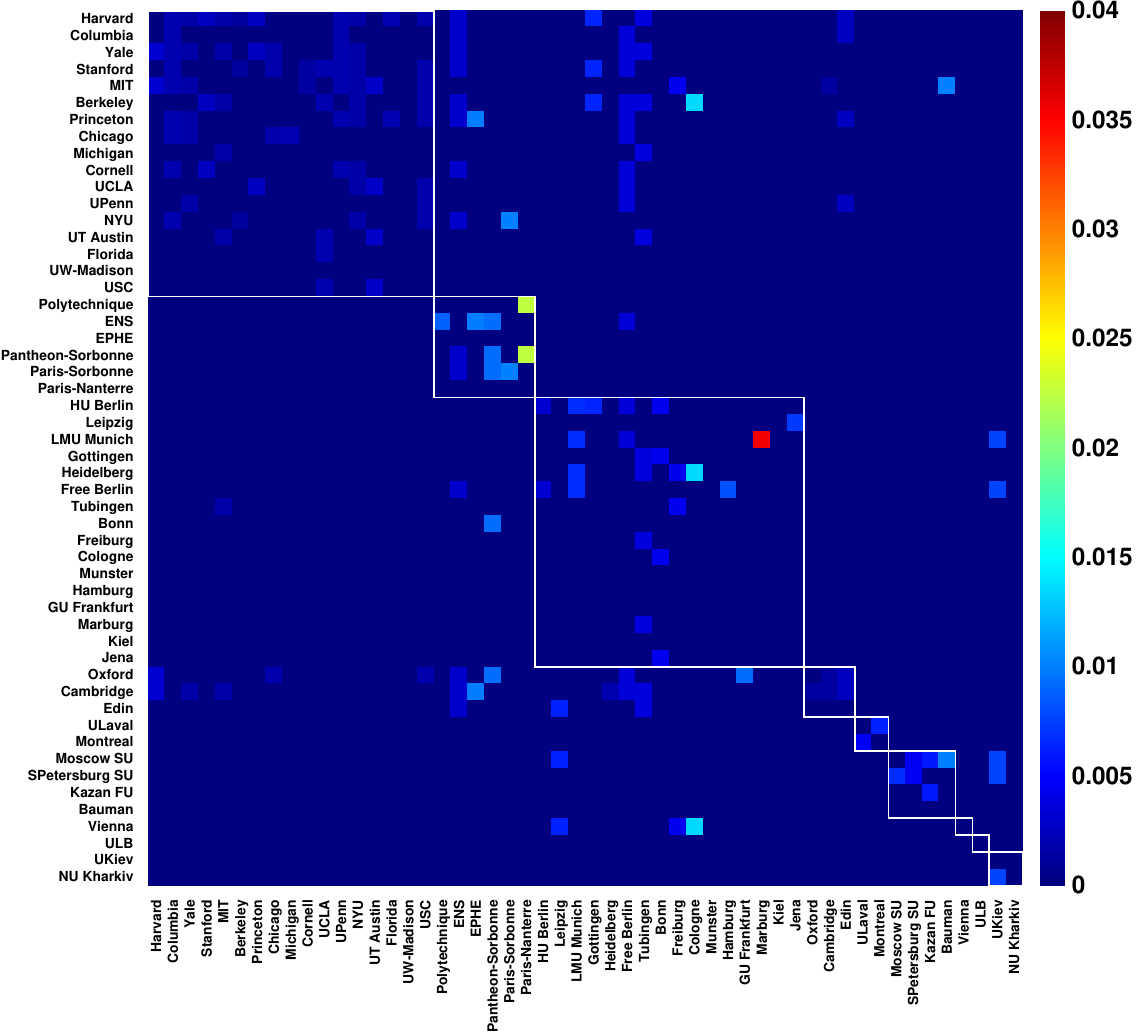}
\includegraphics{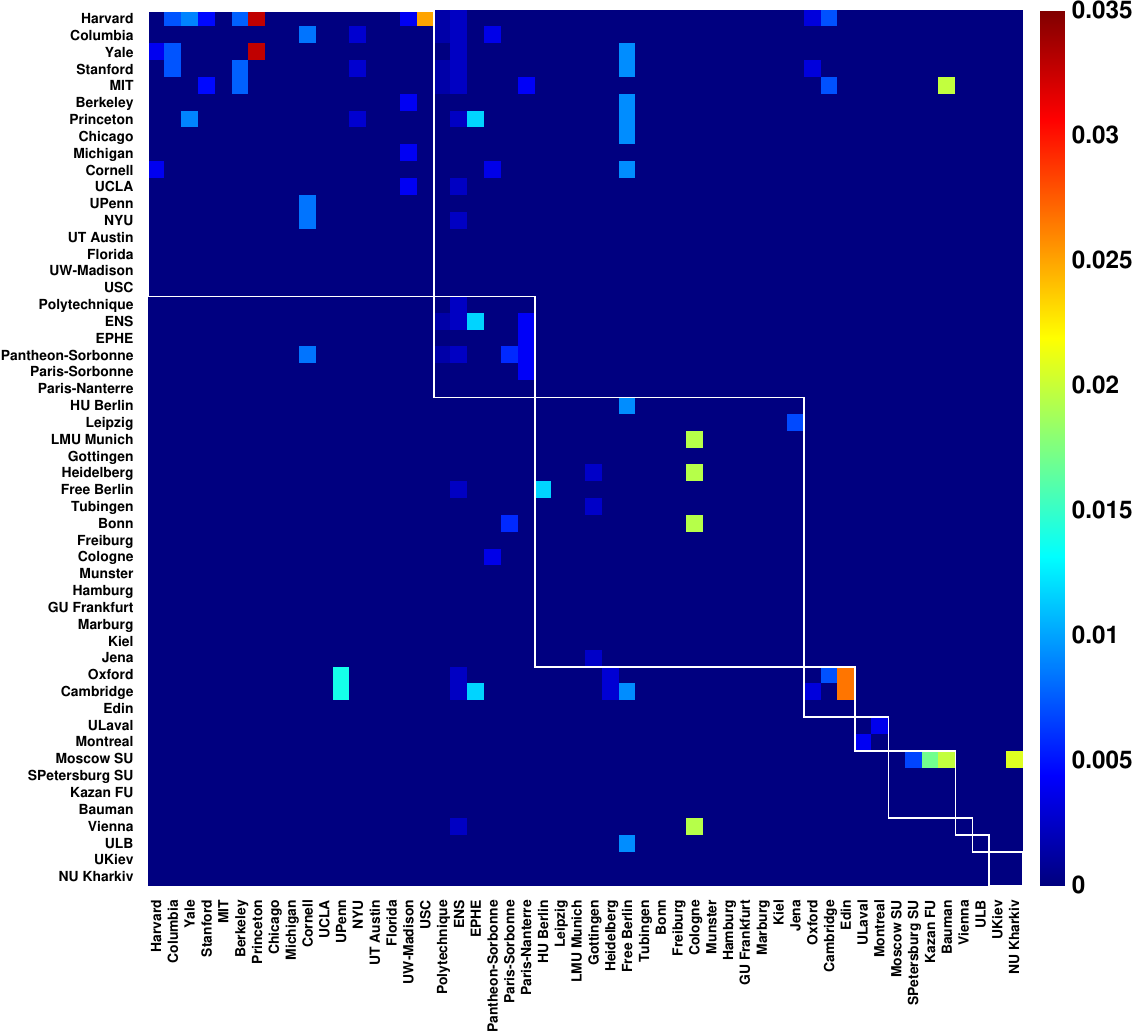}
}
\resizebox{\columnwidth}{!}{%
\includegraphics{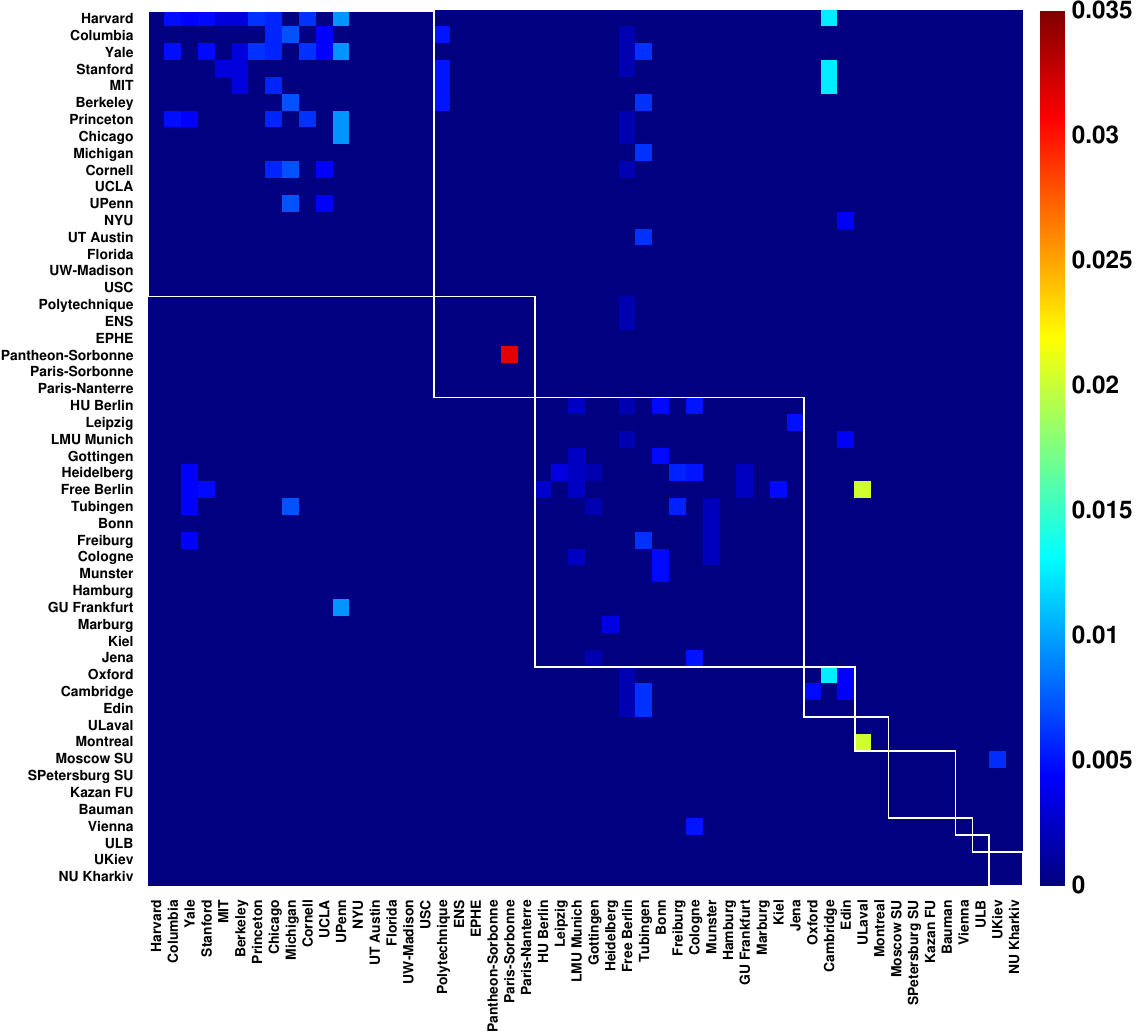}
\includegraphics{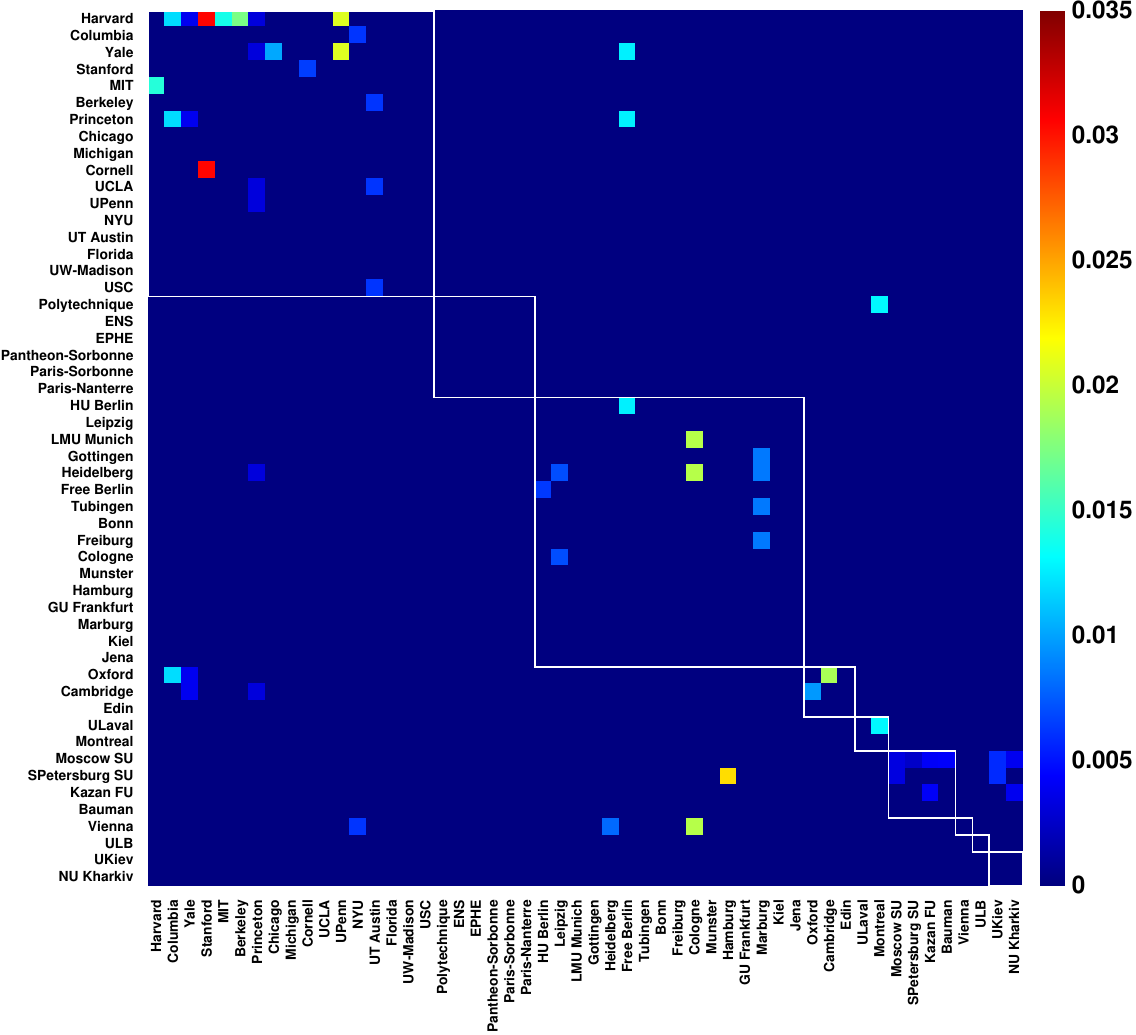}
}
\caption{$G_{\mathrm{rr}}$ part of the reduced Google matrix $G_{\mathrm{R}}$ for universities listed in Tab.~\ref{tab:top52ENFRDERUWRWU} computed from EN (top left), FR (top right), DE (bottom left), and RU (bottom right) Wikipedia editions. Matrix weights are $W_{\mathrm{rr}}\simeq0.0148$ (EN), $0.0144$ (FR), $0.0111$ (DE), $0.0105$ (RU).
}
\label{fig:ENFRDERU_G_rr}
\end{figure}

\begin{figure}[h]
\resizebox{\columnwidth}{!}{%
\includegraphics{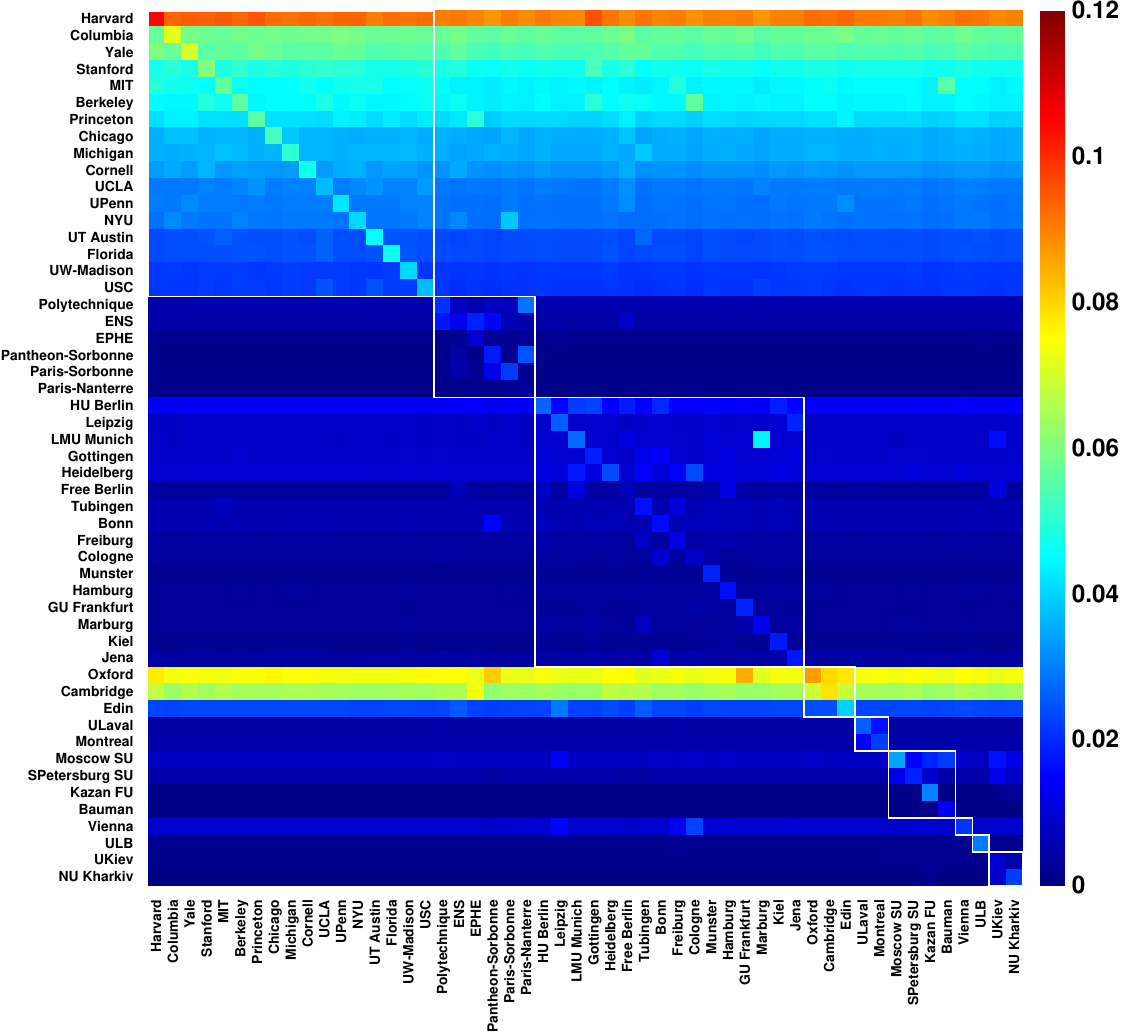}
\includegraphics{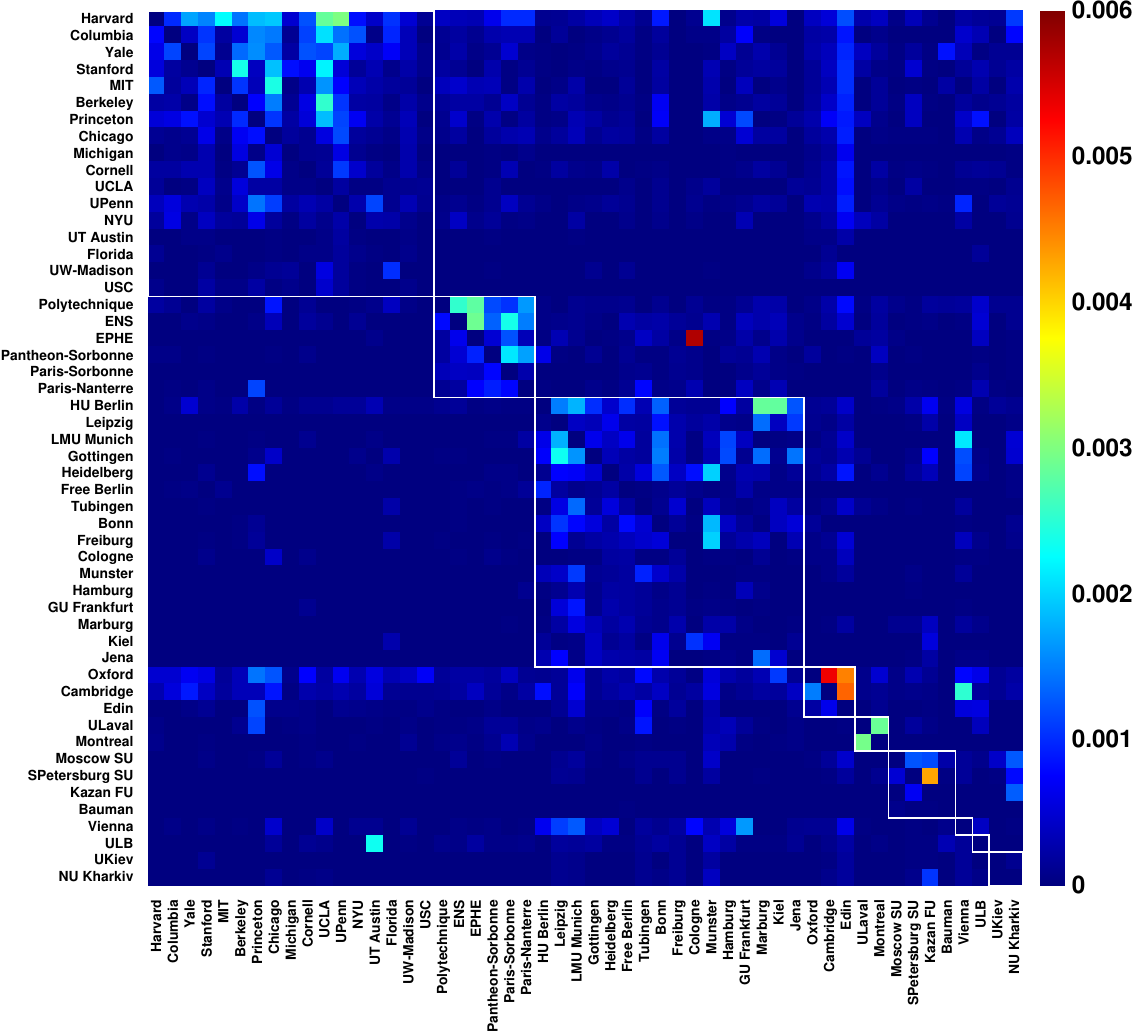}
}
\resizebox{\columnwidth}{!}{%
\includegraphics{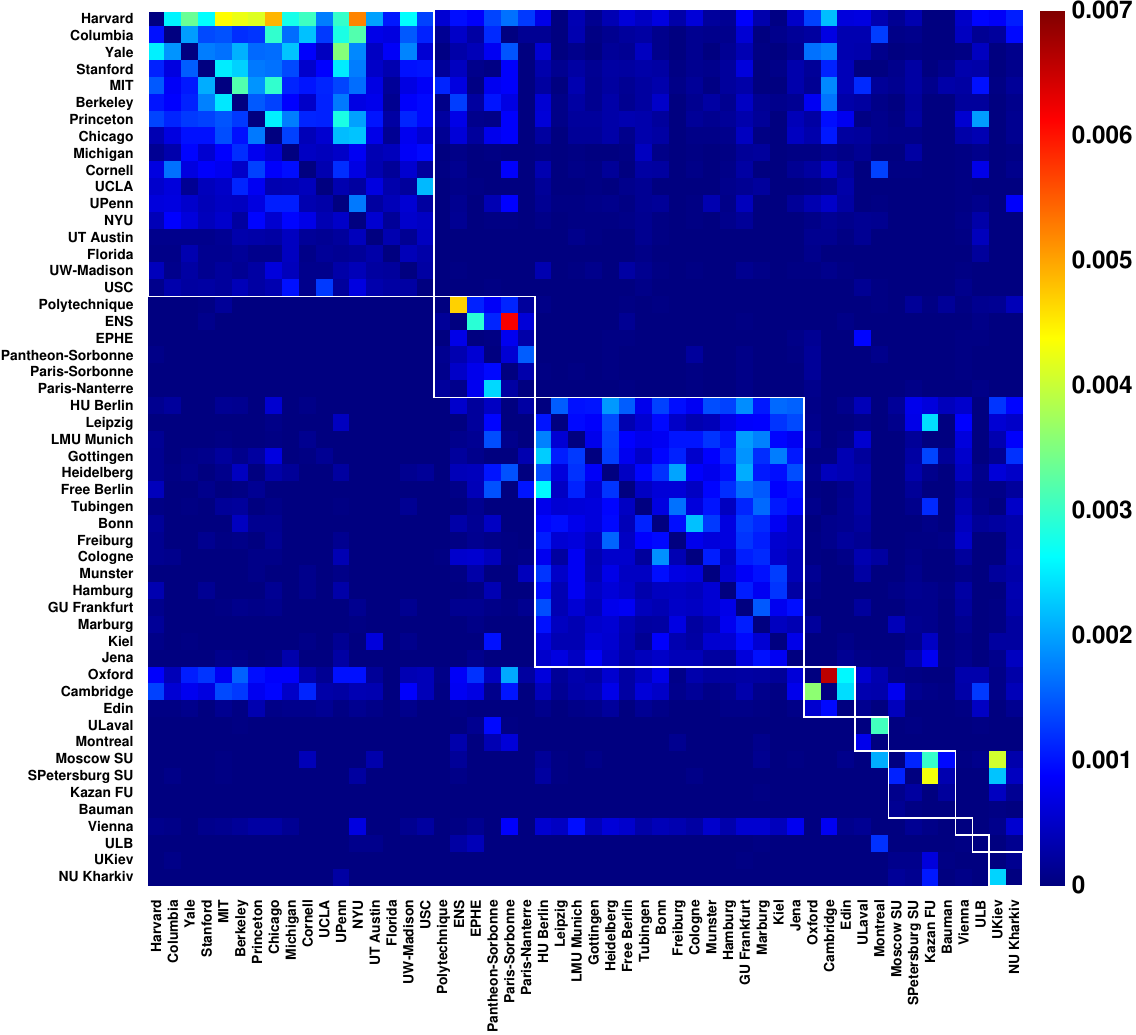}
\includegraphics{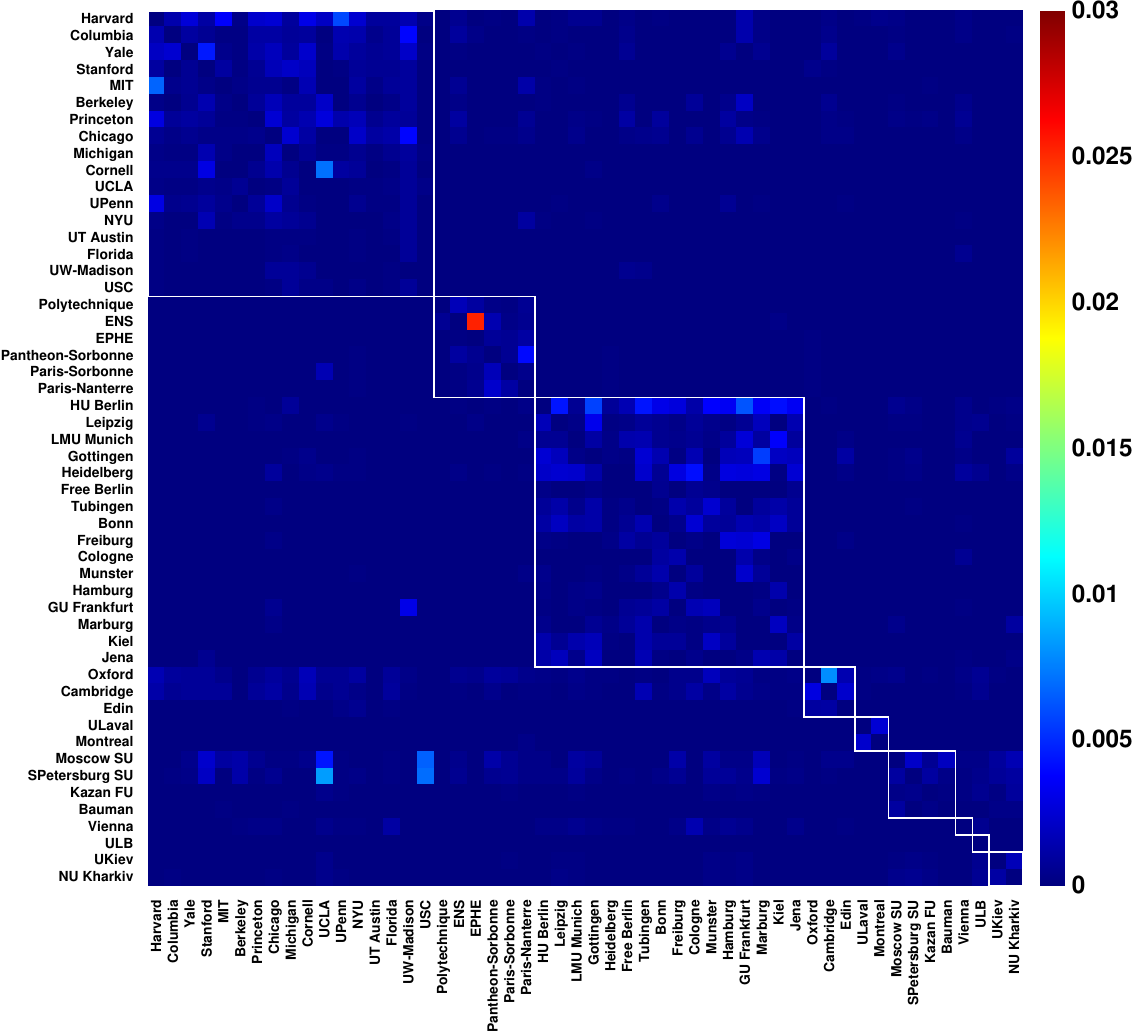}
}
\caption{$G_{\mathrm{qr}}$ part of the reduced Google matrix $G_{\mathrm{R}}$ for universities listed in Tab.~\ref{tab:top52ENFRDERUWRWU} computed from EN (top left), FR (top right), DE (bottom left), and RU (bottom right) Wikipedia editions. Matrix weights are $W_{\mathrm{qr}}\simeq0.0153$ (EN), $0.00923$ (FR), $0.015$ (DE), $0.0138$ (RU).
}
\label{fig:ENFRDERU_G_qr}
\end{figure}

\begin{table*}[h]
\caption{List of the first 100 universities of the 2017 Wikipedia Ranking of World Universities using PageRank algorithm. For a given university, the score $\Theta_{PR}$ is defined
by (\ref{eq:theta}), $N_a$ is the number of appearances in the top 100 lists of Wikipedia editions, CC is the country code, LC is the language code, and FC is the foundation century.}
\centering
\resizebox{\columnwidth}{!}{
\begin{tabular}{rrrlllr|rrrlllr}
\hline
Rank&$\Theta_{PR}$&$N_a$&University&CC&LC&FC&Rank&$\Theta_{PR}$&$N_a$&University&CC&LC&FC\\
\hline
\hline
1&2281&24&University of Oxford&UK&EN&11&51&464&13&University of Manchester&UK&EN&19\\
2&2278&24&University of Cambridge&UK&EN&13&52&461&11&Sapienza University of Rome&IT&IT&14\\
3&2277&24&Harvard University&US&EN&17&53&447&9&Al-Azhar University&EG&AR&20\\
4&2099&24&Columbia University&US&EN&18&54&437&8&University of Helsinki&FI&WR&17\\
5&1959&23&Yale University&US&EN&18&55&436&13&University of Minnesota&US&EN&19\\
6&1917&24&University of Chicago&US&EN&19&56&429&10&University of Illinois at Urbana–Champaign&US&EN&19\\
7&1858&23&Princeton University&US&EN&18&57&425&13&McGill University&CA&EN&19\\
8&1825&21&Stanford University&US&EN&19&58&391&5&Lund University&SE&SV&17\\
9&1804&21&Massachusetts Institute of Technology&US&EN&19&59&377&11&Georgetown University&US&EN&18\\
10&1693&20&University of California, Berkeley&US&EN&19&60&376&10&Peking University&CN&ZH&19\\
11&1578&22&Humboldt University of Berlin&DE&DE&19&61&351&9&University of Geneva&CH&DE&16\\
12&1534&22&Cornell University&US&EN&19&62&349&8&Imperial College London&UK&EN&20\\
13&1472&22&University of Pennsylvania&US&EN&18&63&347&7&University of Oslo&NO&WR&19\\
14&1408&23&New York University&US&EN&19&64&346&8&University of Padua&IT&IT&13\\
15&1395&21&University of California, Los Angeles&US&EN&19&65&337&13&University of Washington&US&EN&19\\
16&1303&21&University of Edinburgh&UK&EN&16&66&331&6&Kyoto University&JP&JA&19\\
17&1298&20&University of Michigan&US&EN&19&67&329&9&Saint Petersburg State University&RU&RU&18\\
18&1221&21&Johns Hopkins University&US&EN&19&68&325&9&Brown University&US&EN&18\\
19&1216&20&University of Vienna&AT&DE&14&69&308&10&University of Arizona&US&EN&19\\
20&1198&20&University of Göttingen&DE&DE&18&70&306&7&Ohio State University&US&EN&19\\
21&1189&21&Heidelberg University&DE&DE&14&71&297&7&University of Tartu&EE&WR&17\\
22&1099&18&California Institute of Technology&US&EN&19&72&290&9&Northwestern University&US&EN&19\\
23&1075&21&Moscow State University&RU&RU&18&73&289&5&Waseda University&JP&JA&19\\
24&1031&20&University of Bologna&IT&IT&11&74&281&8&Boston University&US&EN&19\\
25&998&19&Leipzig University&DE&DE&15&75&273&5&University of Warsaw&PL&PL&19\\
26&993&19&Ludwig Maximilian University of Munich&DE&DE&15&76&263&6&ETH Zurich&CH&DE&19\\
27&975&17&London School of Economics&UK&EN&19&77&261&6&University of Marburg&DE&DE&16\\
28&891&16&Uppsala University&SE&SV&15&78&247&8&Martin Luther University of Halle-Wittenberg&DE&DE&16\\
29&874&20&University of Southern California&US&EN&19&79&246&6&Michigan State University&US&EN&19\\
30&851&18&University of Tokyo&JP&JA&19&80&241&7&University of Jena&DE&DE&16\\
31&770&15&Leiden University&NL&NL&16&81&235&7&University of Notre Dame&US&EN&19\\
32&746&13&University College London&UK&EN&19&82&231&6&Free University of Berlin&DE&DE&20\\
33&653&15&University of Toronto&CA&EN&19&83&226&7&University of Salamanca&ES&ES&12\\
34&651&14&Charles University&CZ&CS&14&84&225&6&University of Freiburg&DE&DE&15\\
35&638&17&Duke University&US&EN&19&85&221&4&Seoul National University&KR&KO&20\\
36&600&13&École normale sup\'erieure&FR&FR&18&86&218&5&University of Colorado Boulder&US&EN&19\\
37&585&15&University of Copenhagen&DK&DA&15&87&215&5&Trinity College, Dublin&IE&EN&16\\
38&565&11&University of Texas at Austin&US&EN&19&88&209&8&Indiana University&US&EN&19\\
39&555&14&University of Tübingen&DE&DE&15&89&207&7&Technical University of Munich&DE&DE&19\\
40&531&13&University of Bonn&DE&DE&19&90&207&7&University of North Carolina at Chapel Hill&US&EN&18\\
41&526&15&Carnegie Mellon University&US&EN&19&91&206&5&University of Coimbra&PT&PT&13\\
42&516&10&University of Florida&US&EN&19&92&204&5&Stockholm University&SE&SV&19\\
43&511&12&Jagiellonian University&PL&PL&14&93&204&5&Utrecht University&NL&NL&17\\
44&505&11&University of Glasgow&UK&EN&15&94&203&4&Technical University of Berlin&DE&DE&20\\
45&501&11&École polytechnique&FR&FR&18&95&199&3&Keio University&JP&JA&19\\
46&501&13&Hebrew University of Jerusalem&IL&HE&20&96&199&6&University of California, San Diego&US&EN&20\\
47&499&13&University of Wisconsin–Madison&US&EN&19&97&196&5&University of St Andrews&UK&EN&15\\
48&489&9&King's College London&UK&EN&16&98&196&4&University of Sydney&AU&EN&19\\
49&482&14&University of Virginia&US&EN&19&99&195&4&University of Kiel&DE&DE&17\\
50&473&15&University of Zurich&CH&DE&16&100&193&3&National Autonomous University of Mexico&MX&ES&20\\
\hline
\end{tabular}
}
\label{tab:top100WRWU}
\end{table*}

\begin{table*}[h]
\caption{List of the first 100 universities of the 2017 Academic Ranking of World Universities (ARWU). CC is the country code, LC is the language code, and FC is the foundation century.}
\centering
\resizebox{\columnwidth}{!}{
\begin{tabular}{rlccc|rlccc}
\hline
Rank&University&CC&LC&FC&Rank&University&CC&LC&FC\\
\hline
\hline
1&Harvard University&US&EN&17&51&University of Texas at Austin&US&EN&19\\
2&Stanford University&US&EN&19&52&Vanderbilt University&US&EN&19\\
3&University of Cambridge&UK&EN&13&53&University of Maryland, College Park&US&EN&19\\
4&Massachusetts Institute of Technology&US&EN&19&54&University of Southern California&US&EN&19\\
5&University of California, Berkeley&US&EN&19&55&University of Queensland&AU&EN&20\\
6&Princeton University&US&EN&18&56&University of Helsinki&FI&WR&17\\
7&University of Oxford&UK&EN&18&57&Ludwig Maximilian University of Munich&DE&DE&15\\
8&Columbia University&US&EN&18&58&University of Zurich&CH&DE&16\\
9&California Institute of Technology&US&EN&19&59&University of Groningen&NL&DE&17\\
10&University of Chicago&US&EN&19&60&University of Geneva&CH&FR&16\\
11&Yale University&US&EN&18&61&University of Bristol&UK&EN&20\\
12&University of California, Los Angeles&US&EN&19&62&University of Oslo&NO&WR&19\\
13&University of Washington&US&EN&19&63&Uppsala University&SE&SV&15\\
14&Cornell University&US&EN&19&64&University of California, Irvine&US&EN&20\\
15&University of California, San Diego&US&EN&20&65&Aarhus University&DK&DA&20\\
16&University College London&UK&EN&19&66&McMaster University&CA&EN&19\\
17&University of Pennsylvania&US&EN&18&67&McGill University&CA&EN&19\\
18&Johns Hopkins University&US&EN&19&68&University of Pittsburgh&US&EN&18\\
19&ETH Zurich&CH&DE&19&69&\'Ecole Normale Sup\'erieure&FR&FR&18\\
20&Washington University in St. Louis&US&EN&19&70&Ghent University&BE&FR&19\\
21&University of California, San Francisco&US&EN&19&71&Mayo Medical School&US&EN&20\\
22&Northwestern University&US&EN&19&72&Peking University&CN&ZH&19\\
23&University of Toronto&CA&EN&19&73&Erasmus University Rotterdam&NL&NL&20\\
24&University of Michigan&US&EN&19&74&Rice University&US&EN&20\\
25&University of Tokyo&JP&JA&19&75&Stockholm University&SE&SV&19\\
26&Duke University&US&EN&19&76&\'Ecole Polytechnique F\'ed\'erale de Lausanne&CH&ZH&20\\
27&Imperial College London&UK&EN&20&77&Purdue University&US&EN&19\\
28&University of Wisconsin-Madison&US&EN&19&78&Monash University&AU&EN&20\\
29&New York University&US&EN&19&79&Rutgers University&US&EN&18\\
30&University of Copenhagen&DK&DA&15&80&Boston University&US&EN&19\\
31&University of British Columbia&CA&EN&20&81&Carnegie Mellon University&US&EN&19\\
32&University of Edinburgh&UK&EN&16&82&Ohio State University&US&EN&19\\
33&University of North Carolina at Chapel Hill&US&EN&18&83&University of Sydney&AU&EN&19\\
34&University of Minnesota&US&EN&19&84&Nagoya University&JP&JA&19\\
35&Kyoto University&JP&JA&19&85&Georgia Institute of Technology&US&EN&19\\
36&Rockefeller University&US&EN&20&86&Pennsylvania State University&US&EN&19\\
37&University of Illinois at Urbana-Champaign&US&EN&19&87&University of California, Davis&US&EN&20\\
38&University of Manchester&UK&EN&14&88&Leiden University&NL&NL&16\\
39&University of Melbourne&AU&EN&19&89&University of Florida&US&EN&19\\
40&Pierre and Marie Curie University&FR&FR&20&90&KU Leuven&BE&FR&15\\
41&University of Paris-Sud&FR&FR&20&91&National University of Singapore&SG&ZH&20\\
42&Heidelberg University&DE&DE&14&92&The University of Western Australia&AU&EN&20\\
43&University of Colorado Boulder&US&EN&19&93&Moscow State University&RU&RU&18\\
44&Karolinska Institute&SE&SV&19&94&Technion - Israel Institute of Technology&IL&HE&20\\
45&University of California, Santa Barbara&US&EN&19&95&University of Basel&CH&ZH&15\\
46&King's College London&UK&EN&16&96&University of G\"ottingen&DE&DE&18\\
47&Utrecht University&NL&NL&17&97&Australian National University&AU&EN&20\\
48&The University of Texas Southwestern Medical Center at Dallas&US&EN&20&98&University of California, Santa Cruz&US&EN&20\\
49&Tsinghua University&CN&ZH&20&99&Cardiff University&UK&EN&19\\
50&Technical University of Munich&DE&DE&19&100&University of Arizona&US&EN&19\\
\hline
\end{tabular}
}
\label{tab:top100ARWU}
\end{table*}

\begin{figure}[h]
\resizebox{\columnwidth}{!}{%
\includegraphics{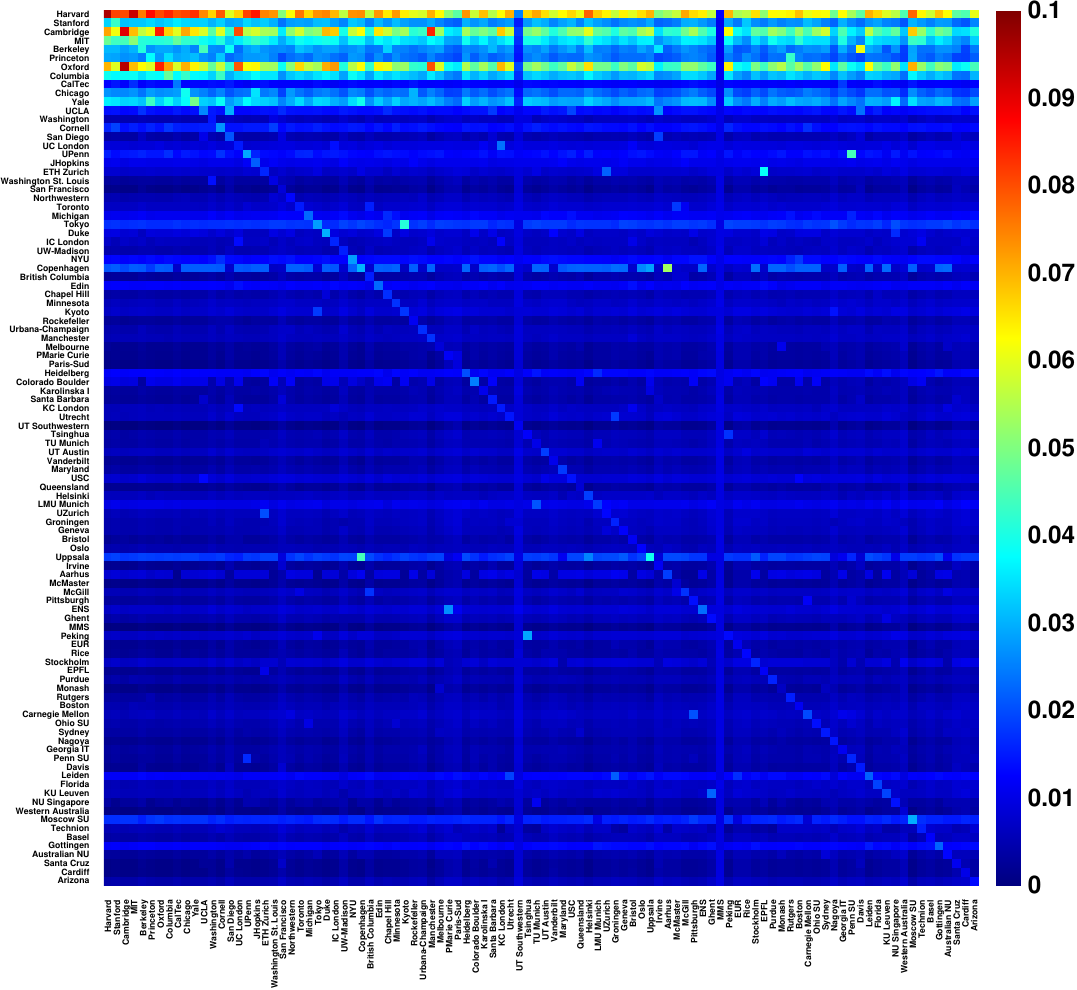}
\includegraphics{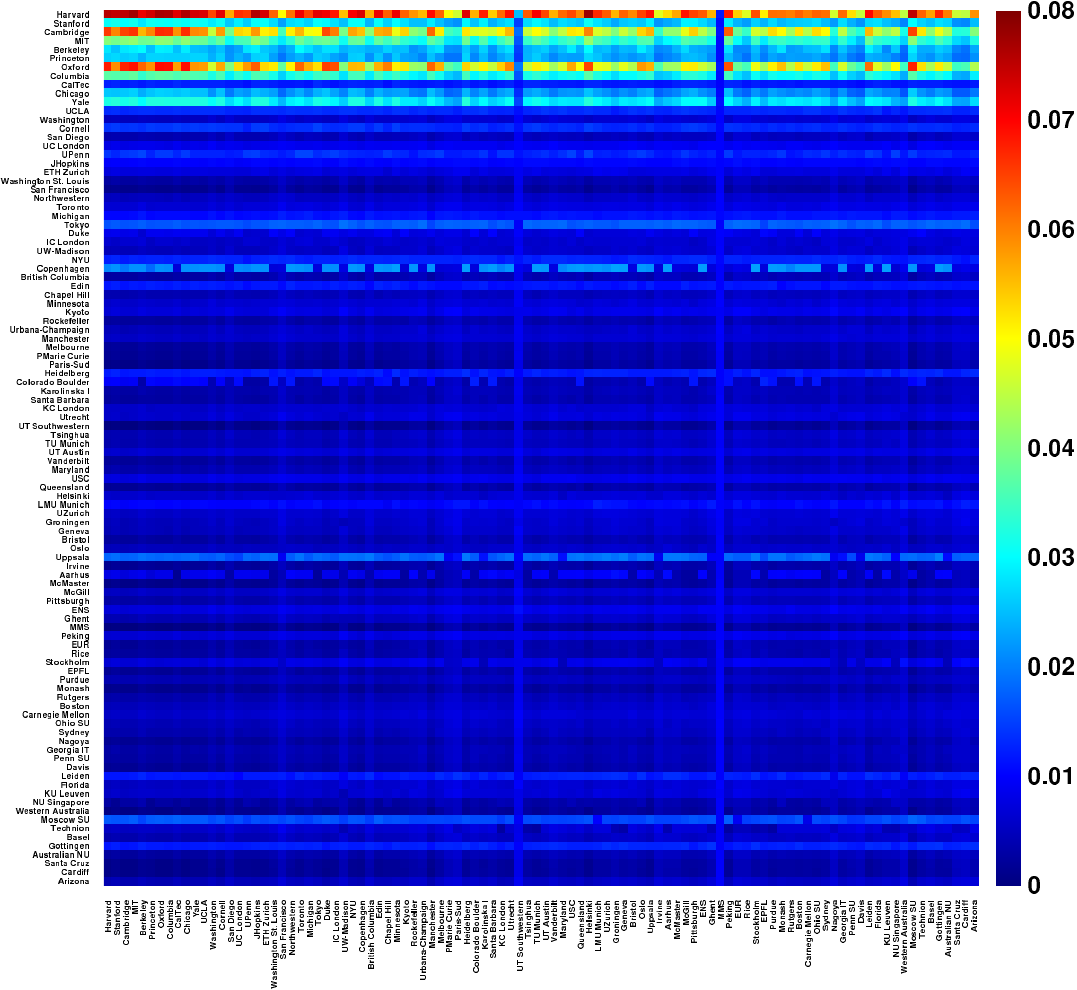}
}
\resizebox{\columnwidth}{!}{%
\includegraphics{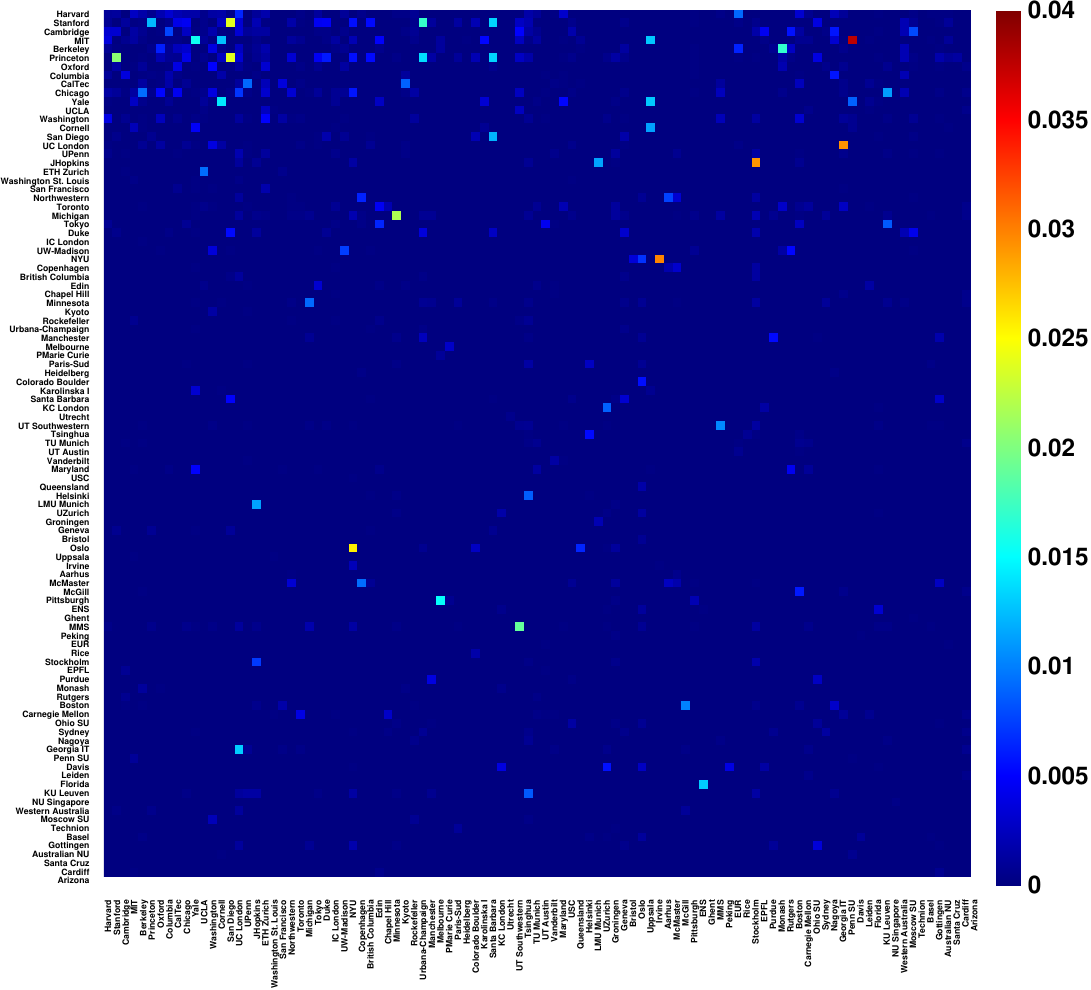}
\includegraphics{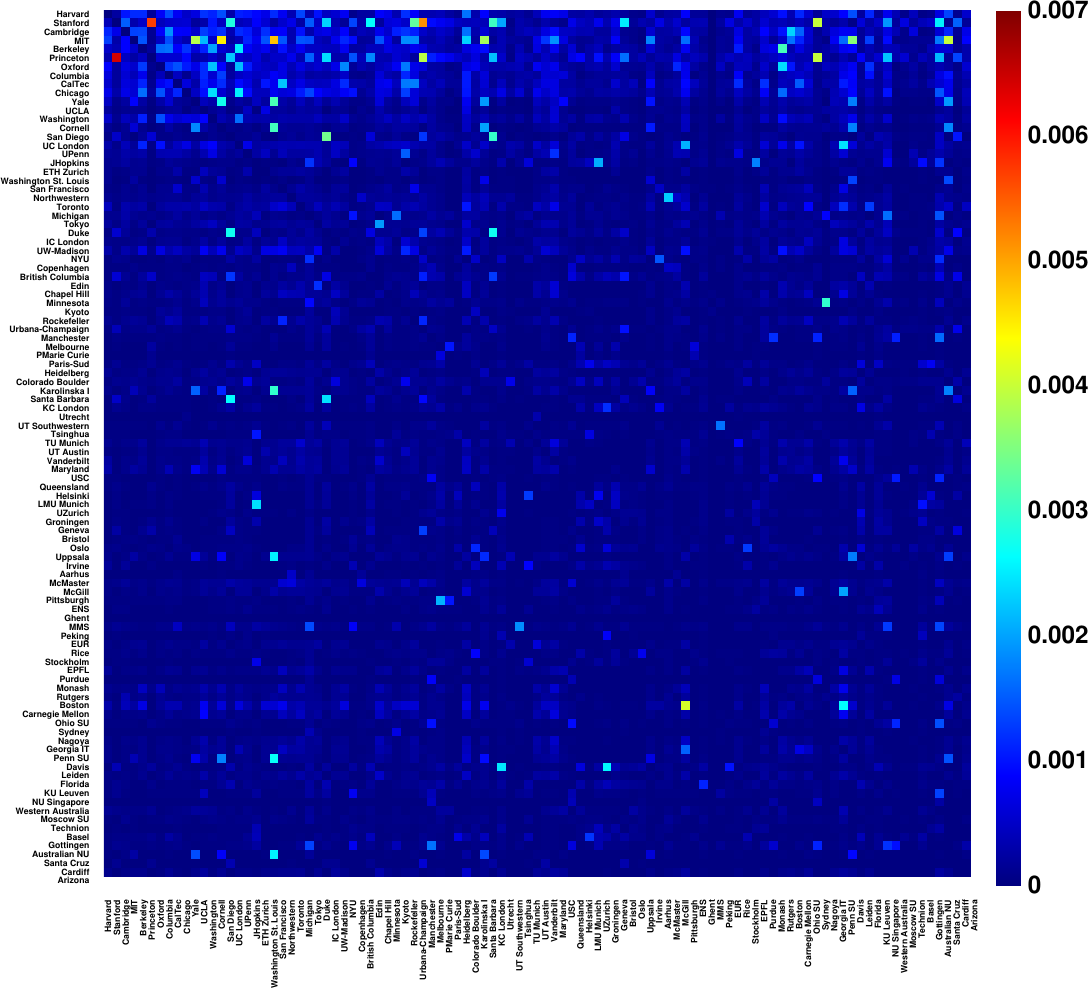}
}
\caption{Reduced Google matrix $G_\mathrm{R}$ for top100 universities in ARWU (Tab.~\ref{tab:top100WRWU}) averaged over 24 Wikipedia editions. The full reduced Google matrix $G_\mathrm{R}$ is presented in top left panel, $G_{\mathrm{pr}}$ in top right panel, $G_{\mathrm{rr}}$ in bottom left panel, and $G_{\mathrm{qrnd}}$ in bottom right panel. The matrices weights are $W_\mathrm{R}=1$, $W_{\mathrm{pr}}=0.956$, $W_{\mathrm{rr}}=0.018$, and $W_{\mathrm{qr}}=0.026$.
}
\label{fig:24wiki_top100_ARWU}
\end{figure}

\begin{table}[h]
\caption{
List of countries with corresponding country codes (CC) and language codes (LC).
Only countries appearing in the top 100 universities of the 24 considered 2017 Wikipedia editions using PageRank, CheiRank, and 2DRank algorithms are listed here. LC is determined by the most spoken language in the given country. Country codes (CC) follow ISO 3166-1 alpha-2 standard \cite{isowiki}.
Language codes are based on language edition codes of Wikipedia; 
WR represents all languages other than the considered 24 languages.}
\centering
\resizebox{\columnwidth}{!}{
\begin{tabular}{lll|lll|lll}
\hline
CC&Country&LC&CC&Country&LC&CC&Country&LC\\
\hline\hline
AE & United Arab Emirates & AR & GE & Georgia & WR & NP & Nepal & WR\\
AF & Afghanistan & FA & GL & Greenland & DA & NZ & New Zealand & EN\\
AL & Albania & WR & GR & Greece & EL & PE & Peru & ES\\
AM & Armenia & WR & GT & Guatemala & ES & PG & Papua New Guinea & EN\\
AO & Angola & PT & HK & Hong Kong & ZH & PH & Philippines & EN\\
AR & Argentina & ES & HN & Honduras & ES & PK & Pakistan & HI\\
AT & Austria & DE & HR & Croatia & WR & PL & Poland & PL\\
AU & Australia & EN & HU & Hungary & HU & PR & Puerto Rico & ES\\
AZ & Azerbaijan & TR & ID & Indonesia & WR & PS & State of Palestine & AR\\
BA & Bosnia and Herzegovina & WR & IE & Ireland & EN & PT & Portugal & PT\\
BD & Bangladesh & WR & IL & Israel & HE & QA & Qatar & AR\\
BE & Belgium & NL & IN & India & HI & RO & Romania & WR\\
BG & Bulgaria & WR & IQ & Iraq & AR & RS & Serbia & WR\\
BI & Burundi & WR & IR & Iran & FA & RU & Russia & RU\\
BJ & Benin & FR & IS & Iceland & WR & SA & Saudi Arabia & AR\\
BN & Brunei & MS & IT & Italy & IT & SD & Sudan & AR\\
BR & Brazil & PT & JM & Jamaica & EN & SE & Sweden & SV\\
BY & Belarus & RU & JO & Jordan & AR & SG & Singapore & ZH\\
CA & Canada & EN & JP & Japan & JA & SK & Slovakia & WR\\
CD & Dem. Rep. of Congo & FR & KE & Kenya & EN & SN & Senegal & FR\\
CH & Switzerland & DE & KG & Kyrgyzstan & WR & SR & Suriname & NL\\
CI & Ivory Coast & FR & KH & Cambodia & WR & SV & El Salvador & ES\\
CL & Chile & ES & KP & North Korea & KO & SY & Syria & AR\\
CN & China & ZH & KR & South Korea & KO & TH & Thailand & TH\\
CO & Colombia & ES & KW & Kuwait & AR & TL & Timor-Leste & PT\\
CR & Costa Rica & ES & KZ & Kazakhstan & WR & TN & Tunisia & AR\\
CU & Cuba & ES & LA & Laos & WR & TR & Turkey & TR\\
CV & Cape Verde & PT & LB & Lebanon & AR & TW & Taiwan & ZH\\
CY & Cyprus & EL & LT & Lithuania & WR & UA & Ukraine & WR\\
CZ & Czech Republic & WR & LV & Latvia & WR & UG & Uganda & EN\\
DE & Germany & DE & LY & Libya & AR & UK & United Kingdom & EN\\
DK & Denmark & DA & MA & Morocco & AR & US & United States & EN\\
DO & Dominican Republic & ES & MK & Macedonia & WR & UY & Uruguay & ES\\
DZ & Algeria & AR & MM & Myanmar & WR & UZ & Uzbekistan & WR\\
EC & Ecuador & ES & MT & Malta & EN & VA & Holy See & IT\\
EE & Estonia & WR & MX & Mexico & ES & VE & Venezuela & ES\\
EG & Egypt & AR & MY & Malaysia & MS & VN & Vietnam & VI\\
ES & Spain & ES & NE & Niger & FR & YE & Yemen & AR\\
FI & Finland & WR & NG & Nigeria & EN & ZA & South Africa & WR\\
FO & Faroe Islands & DA & NL & Netherlands & NL & ZW & Zimbabwe & EN\\
FR & France & FR & NO & Norway & WR & &&\\
\hline
\end{tabular}
}
\label{tab:CC}
\end{table}

\begin{table}[h]
\caption{
List of 240 countries and territories ranked from 2017 Wikipedia English edition using PageRank algorithm.
Country codes (CC) follow ISO 3166-1 alpha-2 standard \cite{isowiki}.
}
\centering\scriptsize
\resizebox{\columnwidth}{!}{
\begin{tabular}{lll|lll|lll}
\hline
Rank&Country&CC&Rank&Country&CC&Rank&Country&CC\\
\hline\hline
1&United States&US&81&Madagascar&MG&161&Niger&NE\\
2&France&FR&82&Armenia&AM&162&Gabon&GA\\
3&Germany&DE&83&Lebanon&LB&163&Brunei&BN\\
4&United Kingdom&UK&84&Cyprus&CY&164&Belize&BZ\\
5&Iran&IR&85&Antarctica&AQ&165&Guinea&GN\\
6&India&IN&86&Kazakhstan&KZ&166&Chad&TD\\
7&Canada&CA&87&Latvia&LV&167&Malawi&MW\\
8&Australia&AU&88&Panama&PA&168&Togo&TG\\
9&China&CN&89&Belarus&BY&169&Liechtenstein&LI\\
10&Italy&IT&90&Albania&AL&170&United States Virgin Islands&VI\\
11&Japan&JP&91&Papua New Guinea&PG&171&Samoa&WS\\
12&Russia&RU&92&Luxembourg&LU&172&Burundi&BI\\
13&Brazil&BR&93&Ghana&GH&173&South Sudan&SS\\
14&Spain&ES&94&United Arab Emirates&AE&174&Republic of the Congo&CG\\
15&Netherlands&NL&95&Uruguay&UY&175&East Timor&TL\\
16&Poland&PL&96&North Korea&KP&176&Cape Verde&CV\\
17&Sweden&SE&97&Yemen&YE&177&Jersey&JE\\
18&Mexico&MX&98&Costa Rica&CR&178&Eritrea&ER\\
19&Turkey&TR&99&Malta&MT&179&Mauritania&MR\\
20&Romania&RO&100&Tunisia&TN&180&Central African Republic&CF\\
21&New Zealand&NZ&101&Jamaica&JM&181&Maldives&MV\\
22&South Africa&ZA&102&Zimbabwe&ZW&182&Tonga&TO\\
23&Norway&NO&103&Cambodia&KH&183&Andorra&AD\\
24&Switzerland&CH&104&Cameroon&CM&184&Vanuatu&VU\\
25&Philippines&PH&105&Mongolia&MN&185&State of Palestine&PS\\
26&Austria&AT&106&Burkina Faso&BF&186&Lesotho&LS\\
27&Belgium&BE&107&Jordan&JO&187&Cook Islands&CK\\
28&Pakistan&PK&108&Sudan&SD&188&The Gambia&GM\\
29&Argentina&AR&109&Uganda&UG&189&French Polynesia&PF\\
30&Indonesia&ID&110&Guatemala&GT&190&Swaziland&SZ\\
31&Greece&GR&111&Libya&LY&191&American Samoa&AS\\
32&Denmark&DK&112&Greenland&GL&192&Falkland Islands&FK\\
33&South Korea&KR&113&Dominican Republic&DO&193&Seychelles&SC\\
34&Israel&IL&114&Haiti&HT&194&Grenada&GD\\
35&Hungary&HU&115&Moldova&MD&195&San Marino&SM\\
36&Finland&FI&116&Somalia&SO&196&Northern Mariana Islands&MP\\
37&Egypt&EG&117&Ivory Coast&CI&197&Saint Lucia&LC\\
38&Portugal&PT&118&Namibia&NA&198&Palau&PW\\
39&Taiwan&TW&119&Paraguay&PY&199&Marshall Islands&MH\\
40&Ukraine&UA&120&Angola&AO&200&Guernsey&GG\\
41&Sri Lanka&LK&121&Uzbekistan&UZ&201&Equatorial Guinea&GQ\\
42&Czech Republic&CZ&122&Montenegro&ME&202&Dominica&DM\\
43&Malaysia&MY&123&Kuwait&KW&203&Cayman Islands&KY\\
44&Peru&PE&124&Laos&LA&204&Aruba&AW\\
45&Thailand&TH&125&Mozambique&MZ&205&Guinea-Bissau&GW\\
46&Hong Kong&HK&126&Nicaragua&NI&206&Curaçao&CW\\
47&Colombia&CO&127&Qatar&QA&207&Comoros&KM\\
48&Bulgaria&BG&128&Senegal&SN&208&Djibouti&DJ\\
49&Chile&CL&129&Fiji&FJ&209&Tuvalu&TV\\
50&Republic of Ireland&IE&130&Mali&ML&210&Niue&NU\\
51&Singapore&SG&131&Honduras&HN&211&Western Sahara&EH\\
52&Serbia&RS&132&Macau&MO&212&British Virgin Islands&VG\\
53&Azerbaijan&AZ&133&Isle of Man&IM&213&Nauru&NR\\
54&Vietnam&VN&134&Zambia&ZM&214&Saint Vincent \& Grenadines&VC\\
55&Nepal&NP&135&El Salvador&SV&215&Kiribati&KI\\
56&Estonia&EE&136&Bermuda&BM&216&Saint Helena&SH\\
57&Croatia&HR&137&Kyrgyzstan&KG&217&Saint Kitts and Nevis&KN\\
58&Nigeria&NG&138&Kosovo&XK&218&Mayotte&YT\\
59&Afghanistan&AF&139&Guyana&GY&219&Micronesia&FM\\
60&Iraq&IQ&140&Trinidad and Tobago&TT&220&Antigua and Barbuda&AG\\
61&Bangladesh&BD&141&Mauritius&MU&221&Netherlands Antilles&AN\\
62&Syria&SY&142&Guam&GU&222&Montserrat&MS\\
63&Myanmar&MM&143&Tajikistan&TJ&223&São Tomé and Príncipe&ST\\
64&Kenya&KE&144&Monaco&MC&224&Saint Pierre and Miquelon&PM\\
65&Slovakia&SK&145&New Caledonia&NC&225&Wallis and Futuna&WF\\
66&Venezuela&VE&146&Oman&OM&226&Turks and Caicos Islands&TC\\
67&Slovenia&SI&147&Suriname&SR&227&Sint Maarten&SX\\
68&Morocco&MA&148&Liberia&LR&228&Saint Barthélemy&BL\\
69&Cuba&CU&149&Solomon Islands&SB&229&Anguilla&AI\\
70&Algeria&DZ&150&Sierra Leone&SL&230&Tokelau&TK\\
71&Bosnia and Herzegovina&BA&151&Bhutan&BT&231&Pitcairn Islands&PN\\
72&Ecuador&EC&152&The Bahamas&BS&232&Christmas Island&CX\\
73&Puerto Rico&PR&153&Bahrain&BH&233&Cocos (Keeling) Islands&CC\\
74&Saudi Arabia&SA&154&Barbados&BB&234&Saint Martin&MF\\
75&Lithuania&LT&155&Botswana&BW&235&British Indian Ocean Territory&IO\\
76&Iceland&IS&156&Rwanda&RW&236&Svalbard and Jan Mayen&SJ\\
77&Bolivia&BO&157&Gibraltar&GI&237&Georgia&GE\\
78&Tanzania&TZ&158&Faroe Islands&FO&238&Macedonia&MK\\
79&Ethiopia&ET&159&Turkmenistan&TM&239&Vatican&VA\\
80&Dem. Rep. of Congo&CD&160&Benin&BJ&240&Reunion&RE\\
\hline
\end{tabular}
}
\label{tab:CC240}
\end{table}

\begin{table*}[h]
\caption{Universities from top100 2017 ARWU ordered according to PageRank algorithm applied to the reduced Google matrix averaged over 24 Wikipedia editions.}
\resizebox{\columnwidth}{!}{
\begin{tabular}{rll|rll}
\hline
Rank&PageRank&University&Rank&PageRank&University\\
\hline
\hline
1&0.0745472&Harvard University&51&0.00583189&University of Washington\\
2&0.0621785&University of Oxford&52&0.00560402&Northwestern University\\
3&0.0609665&University of Cambridge&53&0.00556462&University of Groningen\\
4&0.0397826&MIT$^a$&54&0.00545821&University of Wisconsin-Madison\\
5&0.0355868&Columbia University&55&0.00545373&U of I, Illinois$^f$\\
6&0.0333098&Yale University&56&0.00541549&Boston University\\
7&0.0306724&Stanford University&57&0.00539776&University of Oslo\\
8&0.0282621&UC Berkeley$^b$&58&0.00530861&University of Geneva\\
9&0.0266544&Princeton University&59&0.00516361&Ohio State University\\
10&0.0250087&University of Chicago&60&0.00502027&Tsinghua University\\
11&0.0195607&University of Copenhagen&61&0.00491138&Ghent University\\
12&0.0189782&Uppsala University&62&0.0049034&University of Arizona\\
13&0.0176773&University of Tokyo&63&0.00473034&University of California, San Diego\\
14&0.0165834&Moscow State University&64&0.00468699&Purdue University\\
15&0.0148824&Cornell University&65&0.00465378&University of Basel\\
16&0.0145334&University of Pennsylvania&66&0.0046221&UNC Chapel Hill$^g$\\
17&0.0141382&UCLA$^c$&67&0.00455207&Technical University of Munich\\
18&0.0131496&New York University&68&0.00444405&University of Sydney\\
19&0.0131307&California Institute of Technology&69&0.00437431&University of Maryland, College Park\\
20&0.0125402&University G\"ottingen&70&0.00430464&Rutgers University\\
21&0.0124228&Leiden University&71&0.00404074&Karolinska Institute\\
22&0.0124135&Heidelberg University&72&0.0039799&University of Pittsburgh\\
23&0.012155&University of Edinburgh&73&0.00396299&Georgia Institute of Technology\\
24&0.0115044&University of Michigan&74&0.00385092&Penn State University\\
25&0.0107339&Johns Hopkins University&75&0.00369518&Washington University in St. Louis\\
26&0.0105176&University of Munich$^d$&76&0.00364923&University of British Columbia\\
27&0.00917197&University College London&77&0.00360777&Australian National University\\
28&0.00872849&Duke University&78&0.00355039&University of California, Santa Barbara\\
29&0.00819352&University of Southern California&79&0.00352429&University of Bristol\\
30&0.00814739&ETH Zurich&80&0.00337135&National University of Singapore\\
31&0.00802171&Kyoto University&81&0.00326108&Rockefeller University\\
32&0.00797102&Aarhus University&82&0.00320465&Rice University\\
33&0.00795232&University of Colorado Boulder&83&0.00311399&University of Melbourne\\
34&0.00783698&\'Ecole Normale Sup\'erieure&84&0.00307731&Vanderbilt University\\
35&0.00766772&Peking University&85&0.00290042&Erasmus University Rotterdam\\
36&0.0076523&Stockholm University&86&0.00284488&University of California, Davis\\
37&0.00744185&University Toronto&87&0.00279911&EPFL$^h$\\
38&0.0070079&University of Texas at Austin&88&0.00273436&Pierre and Marie Curie University\\
39&0.00685167&Imperial College London&89&0.00272264&Nagoya University\\
40&0.00681529&Utrecht University&90&0.00265046&University of California, Irvine\\
41&0.00658785&University of Minnesota&91&0.00263142&University of Queensland\\
42&0.00654586&Carnegie Mellon University&92&0.00210331&University of California, San Francisco\\
43&0.00653689&University of Helsinki&93&0.0021006&Monash University\\
44&0.00651107&KU Leuven&94&0.00198778&University of California, Santa Cruz\\
45&0.0064442&King's College London&95&0.00189623&Cardiff University\\
46&0.00639753&University of Zurich&96&0.00186566&McMaster University\\
47&0.00623053&McGill University&97&0.00186277&University of Paris-Sud\\
48&0.00621359&University of Manchester&98&0.00157434&The University of Western Australia\\
49&0.00602113&University of Florida&99&0.0011675&UT Southwestern$^i$\\
50&0.00600767&Technion$^e$&100&0.00102057&Mayo Medical School\\
\hline
\multicolumn{6}{p{\columnwidth}}{$^a$Massachusetts Institute of Technology, $^b$University of California, Berkeley, $^c$University of California, Los Angeles, $^d$Ludwig Maximilian University of Munich, $^e$Technion - Israel Institute of Technology, $^f$University of Illinois Urbana-Champaign, $^g$University of North Carolina at Chapel Hill, $^h$\'Ecole Polytechnique F\'ed\'erale de Lausanne, $^i$The University of Texas Southwestern Medical Center at Dallas}
\end{tabular}
}
\label{tab:24wiki_top100_ARWU}
\end{table*}

\begin{figure}[h]
\resizebox{\columnwidth}{!}{%
\includegraphics{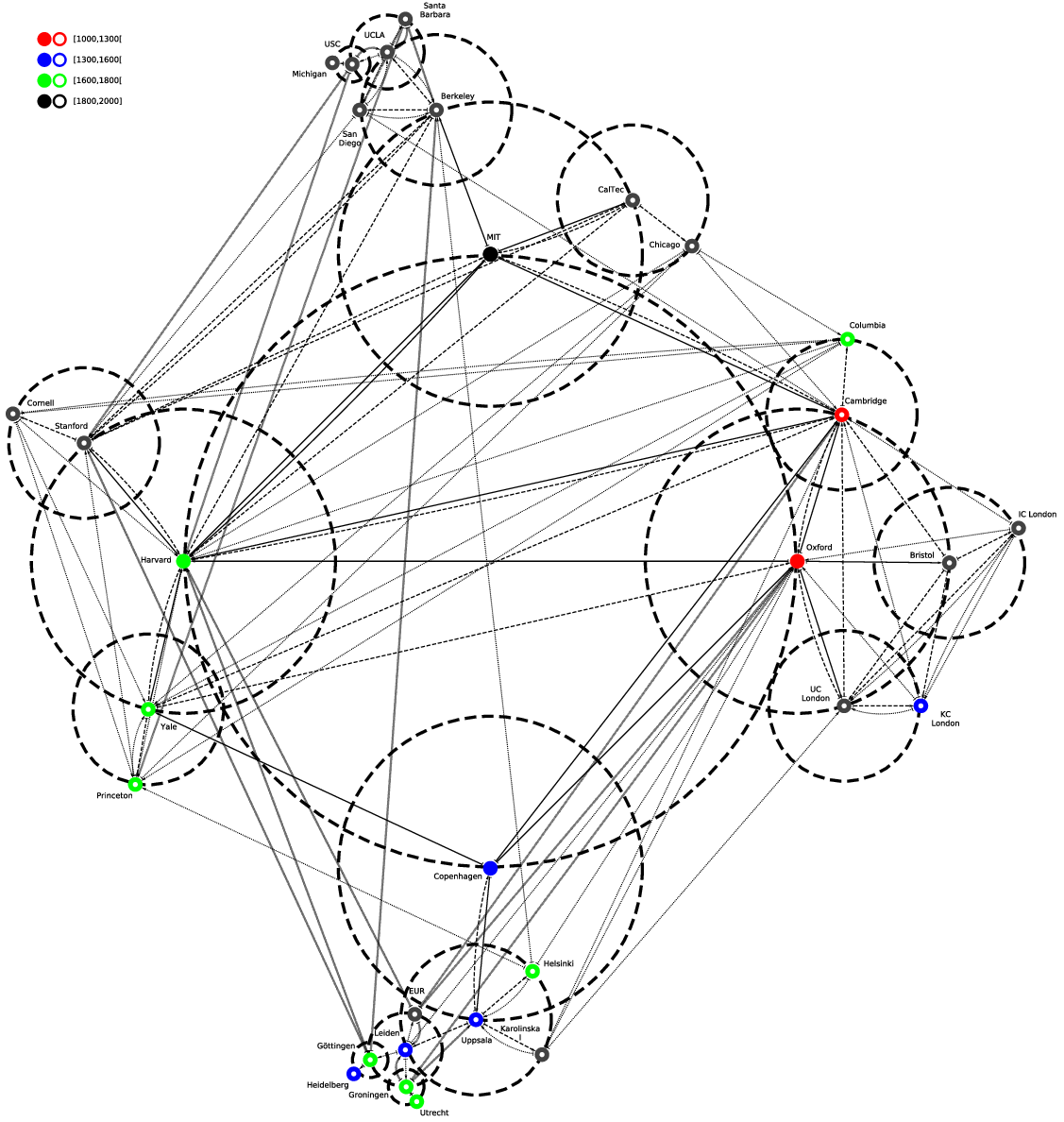}
}
\caption{Reduced network constructed for universities listed in Tab.~\ref{tab:24wiki_top100_ARWU} computed from $G_{\mathrm{rr}}+G_{\mathrm{qrnd}}$ averaged over 24 Wikipedia editions. Color filled nodes are time period leaders.
We obtain 4 friendship levels with corresponding links: black solid lines (level 1), dashed lines (level 2), doted lines (level 3) and ``\textbackslash'' symbol lines (4+).
}
\label{fig:24wiki_top100_period_ARWU_netw}
\end{figure}

\begin{figure}[h]
\resizebox{\columnwidth}{!}{%
\includegraphics{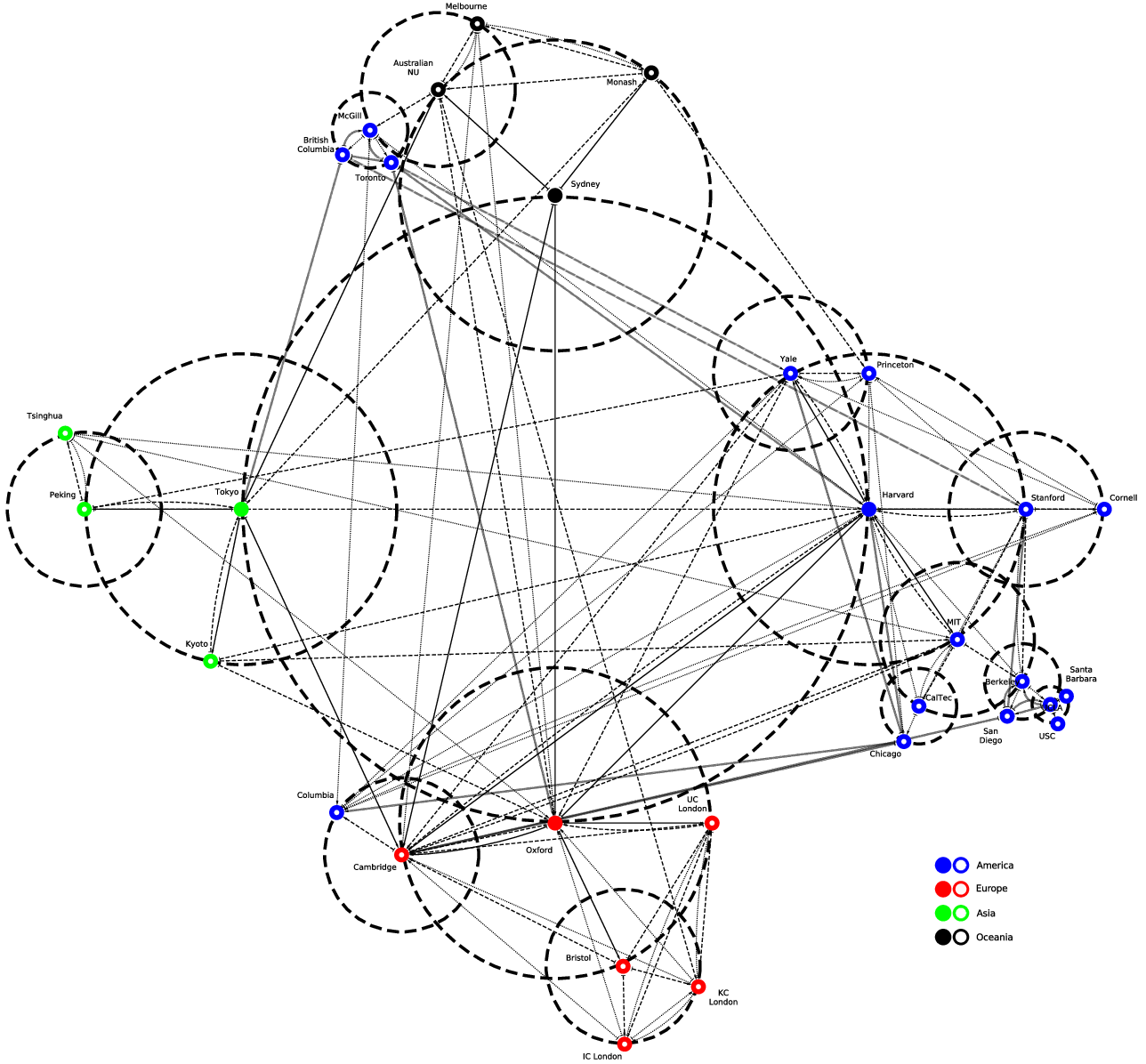}
}
\caption{Reduced network constructed for universities listed in Tab.~\ref{tab:24wiki_top100_ARWU} computed from $G_{\mathrm{rr}}+G_{\mathrm{qrnd}}$ averaged over 24 Wikipedia editions. Color filled nodes are continent leaders.
We obtain 4 friendship levels with corresponding links: black solid lines (level 1), dashed lines (level 2), doted lines (level 3) and ``\textbackslash'' symbol lines (4+).
}
\label{fig:24wiki_top100_country_ARWU_netw}
\end{figure}

\end{document}